\documentclass{iopart}  
\pdfoutput=1

\usepackage{iopams}
\usepackage{graphicx}
\usepackage{epsfig}
\usepackage{mathrsfs}
\usepackage{slashbox}

\newcommand{\cross}{\raisebox{-0.4\height}{\includegraphics[height=0.8cm]{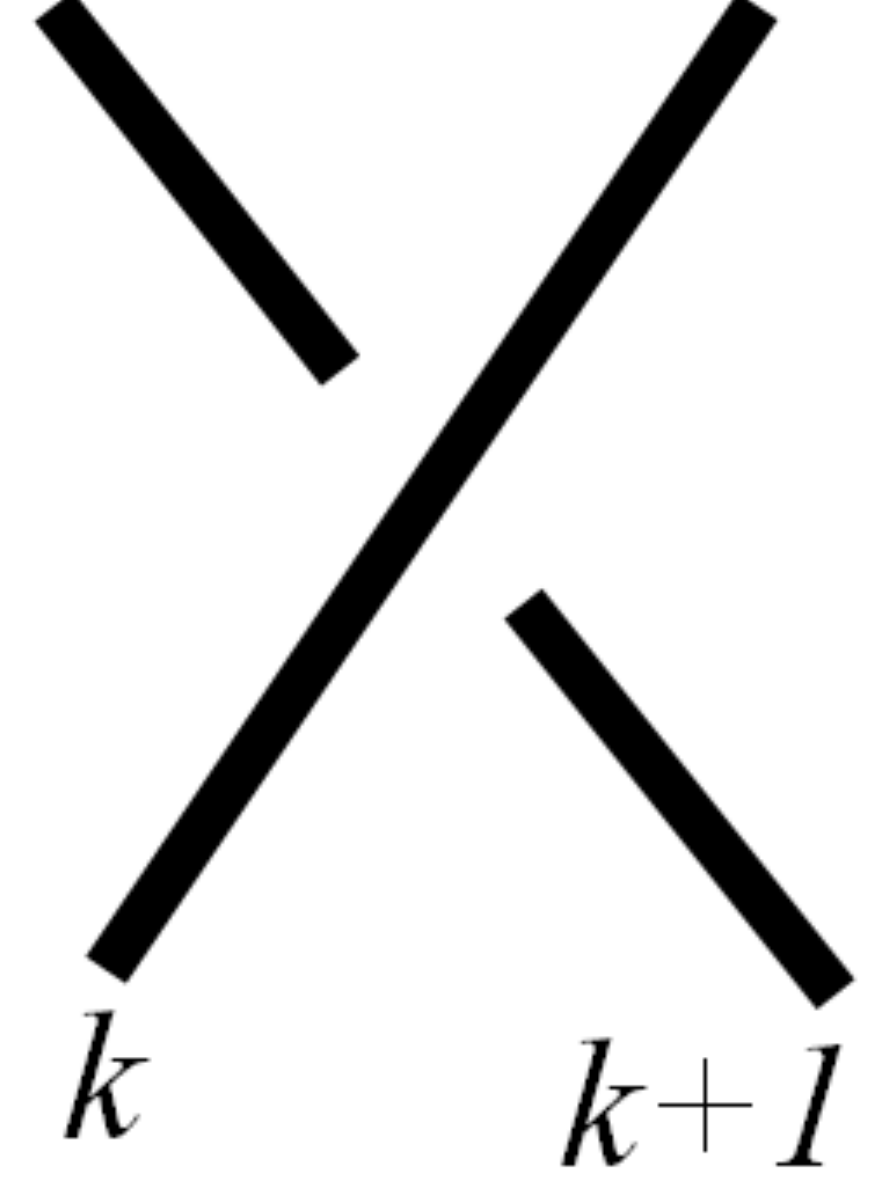}}}
\newcommand{\cros}{\raisebox{-0.4\height}{\includegraphics[height=0.8cm,angle=90]{cross}}}
\newcommand{\Un}{\raisebox{-0.4\height}{\includegraphics[height=0.8cm]{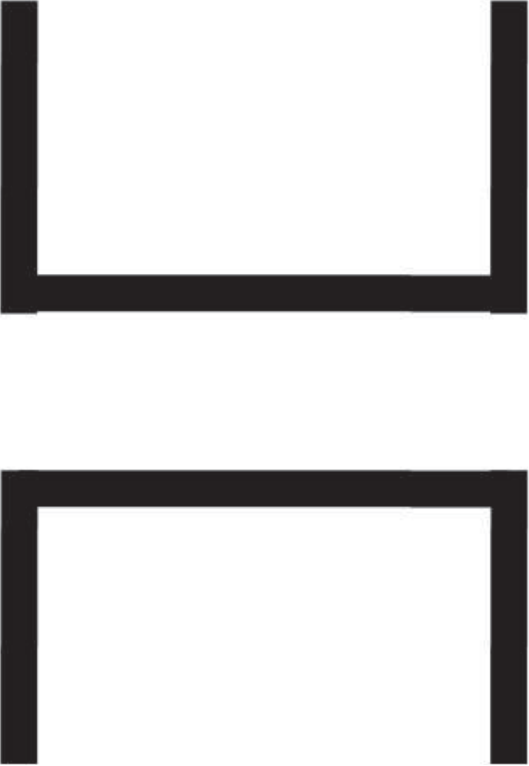}}}
\newcommand{\SSSSijk}{\raisebox{-0.4\height}{\includegraphics[height=2cm]{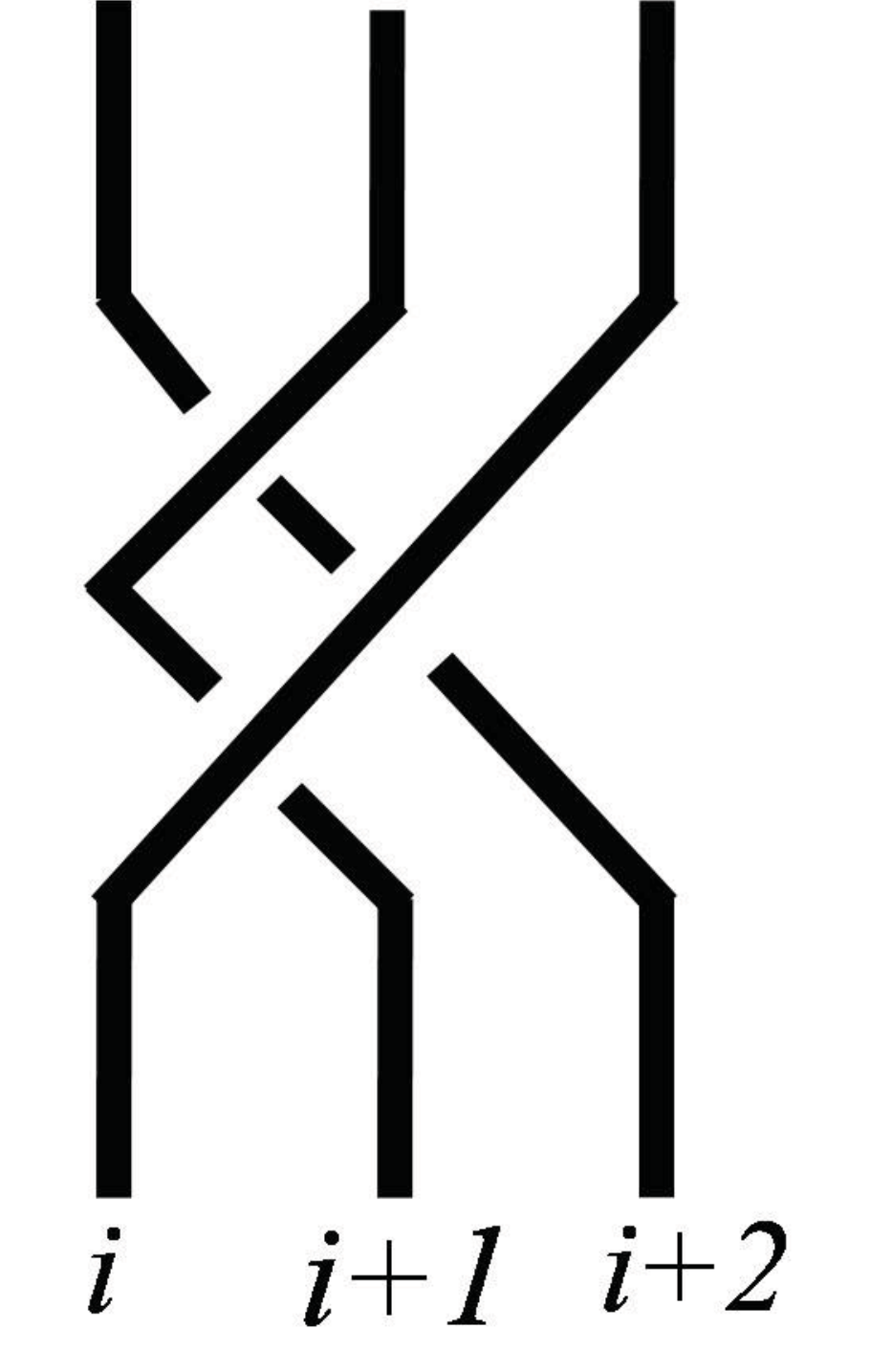}}}
\newcommand{\sssijk}{\raisebox{-0.4\height}{\includegraphics[height=2cm]{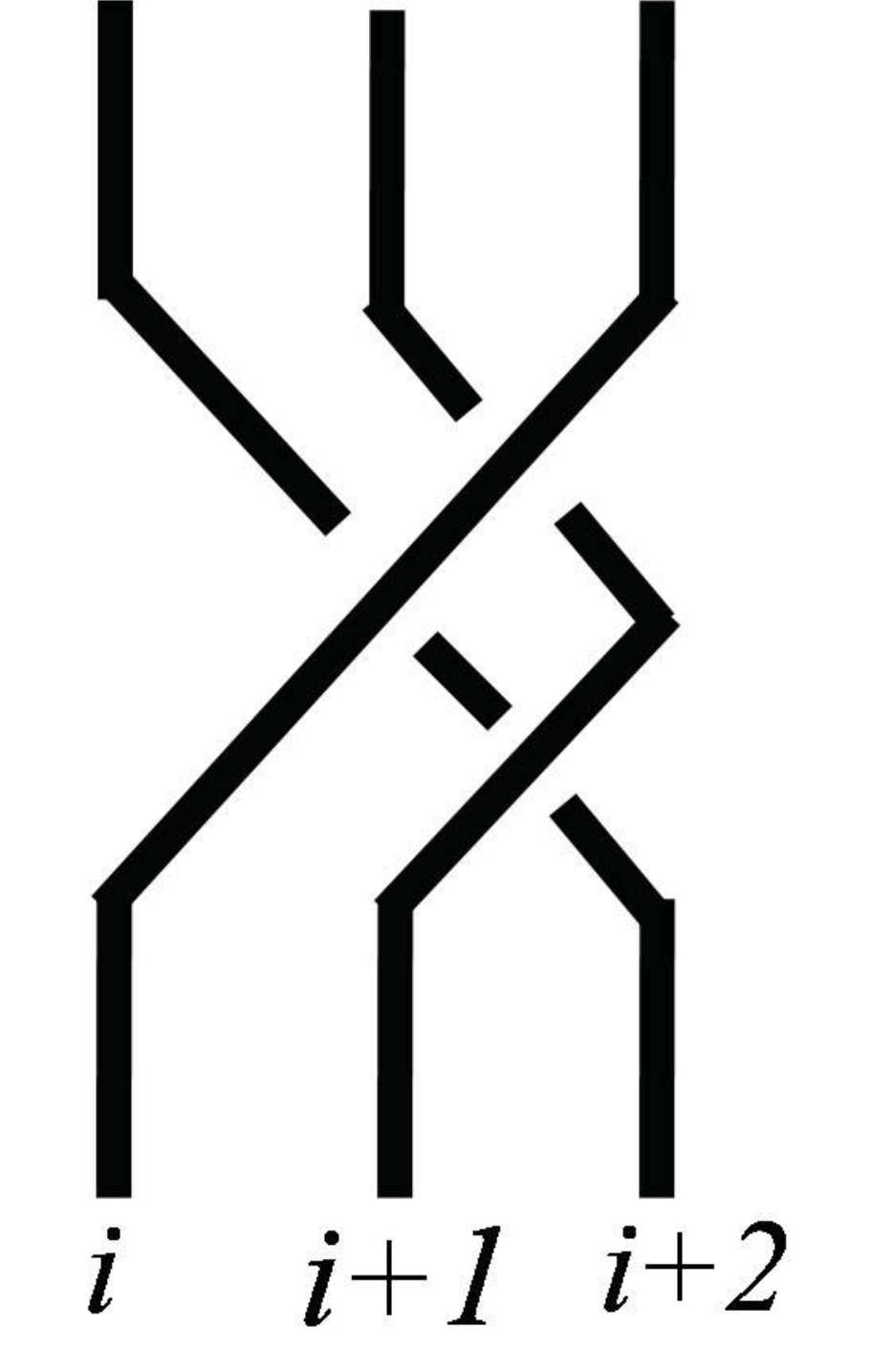}}}
\newcommand{\Uniil}{\raisebox{-0.4\height}{\includegraphics[height=1cm]{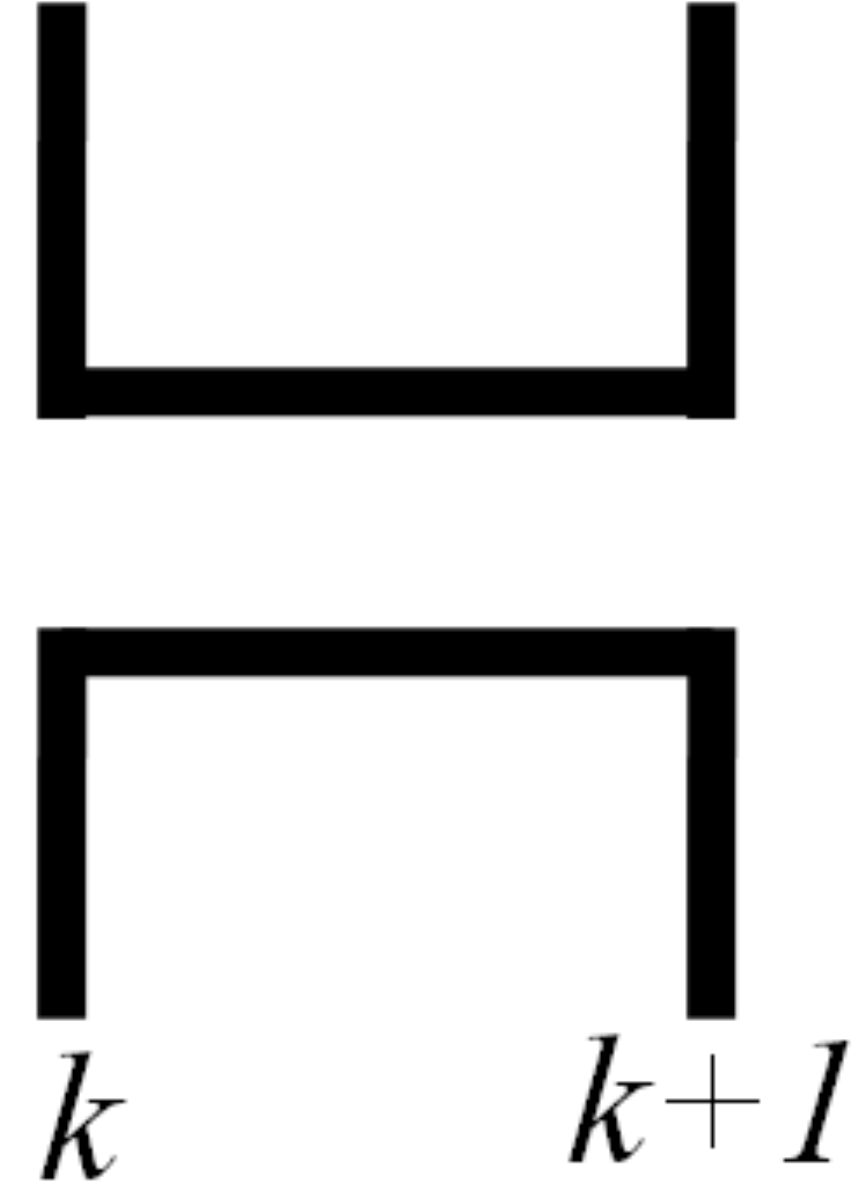}}}
\newcommand{\complexijk}{\raisebox{-0.4\height}{\includegraphics[height=1.6cm]{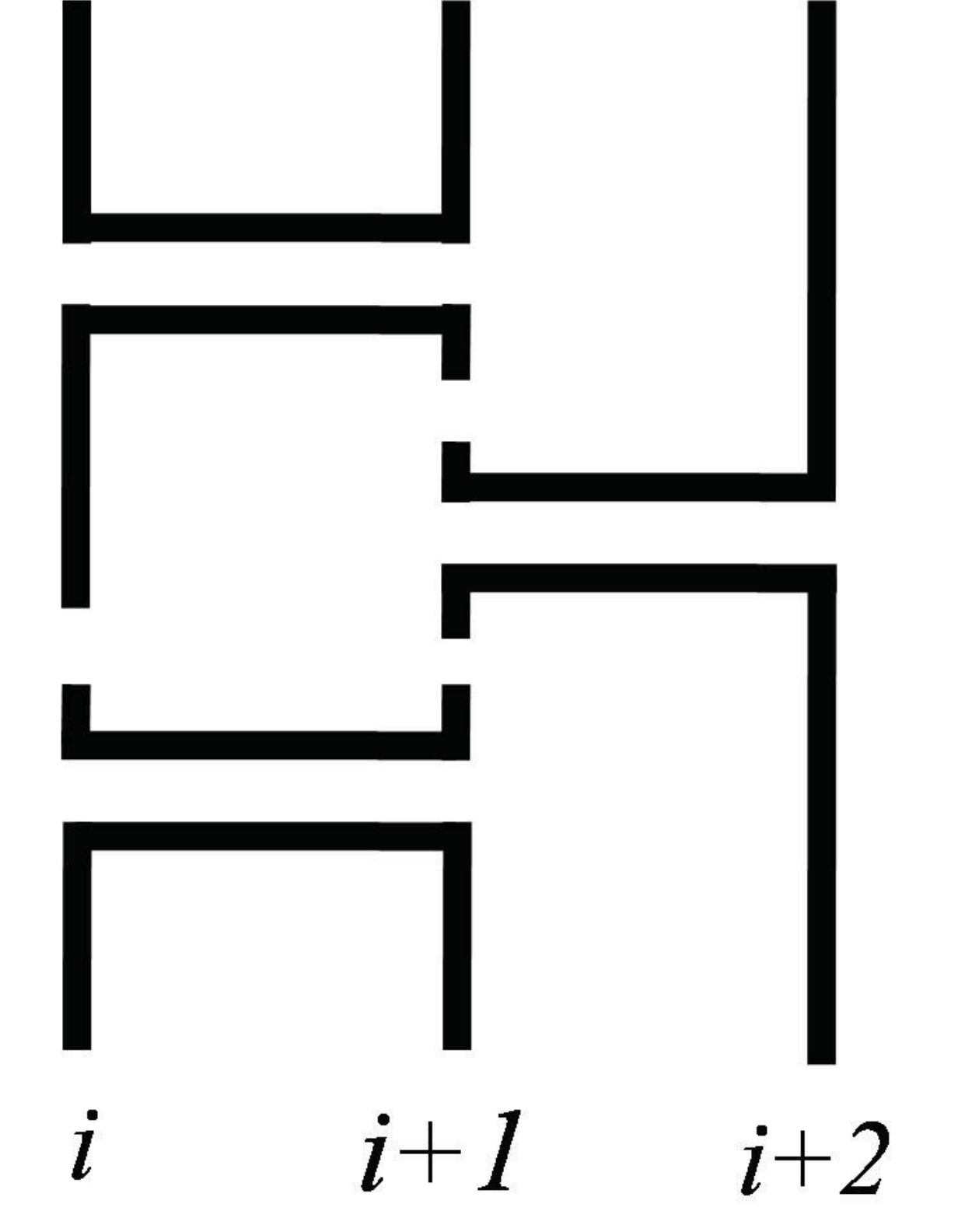}}}
\newcommand{\Unll}{\raisebox{-0.4\height}{\includegraphics[height=0.8cm]{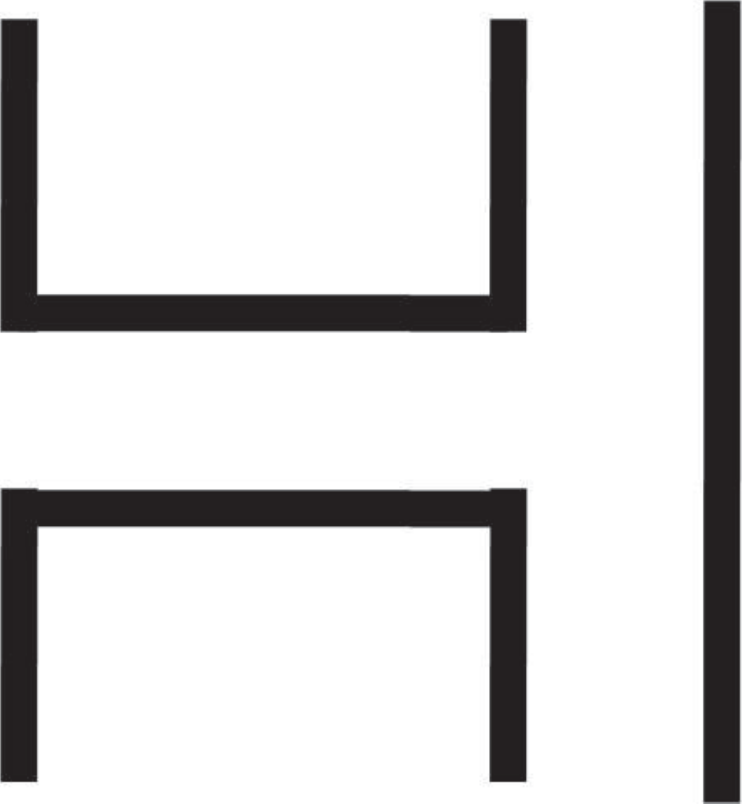}}}
\newcommand{\myloop}{\raisebox{-0.33\height}{\includegraphics[height=0.5cm]{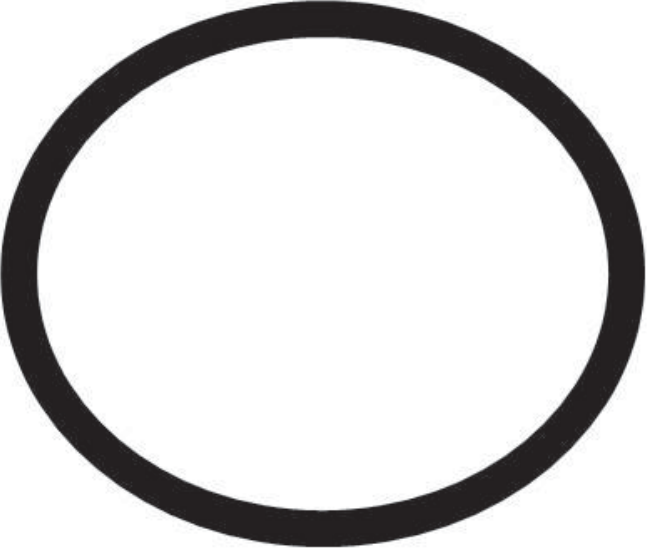}}}
\newcommand{\SicrossEi}{\raisebox{-0.4\height}{\includegraphics[height=1.5cm]{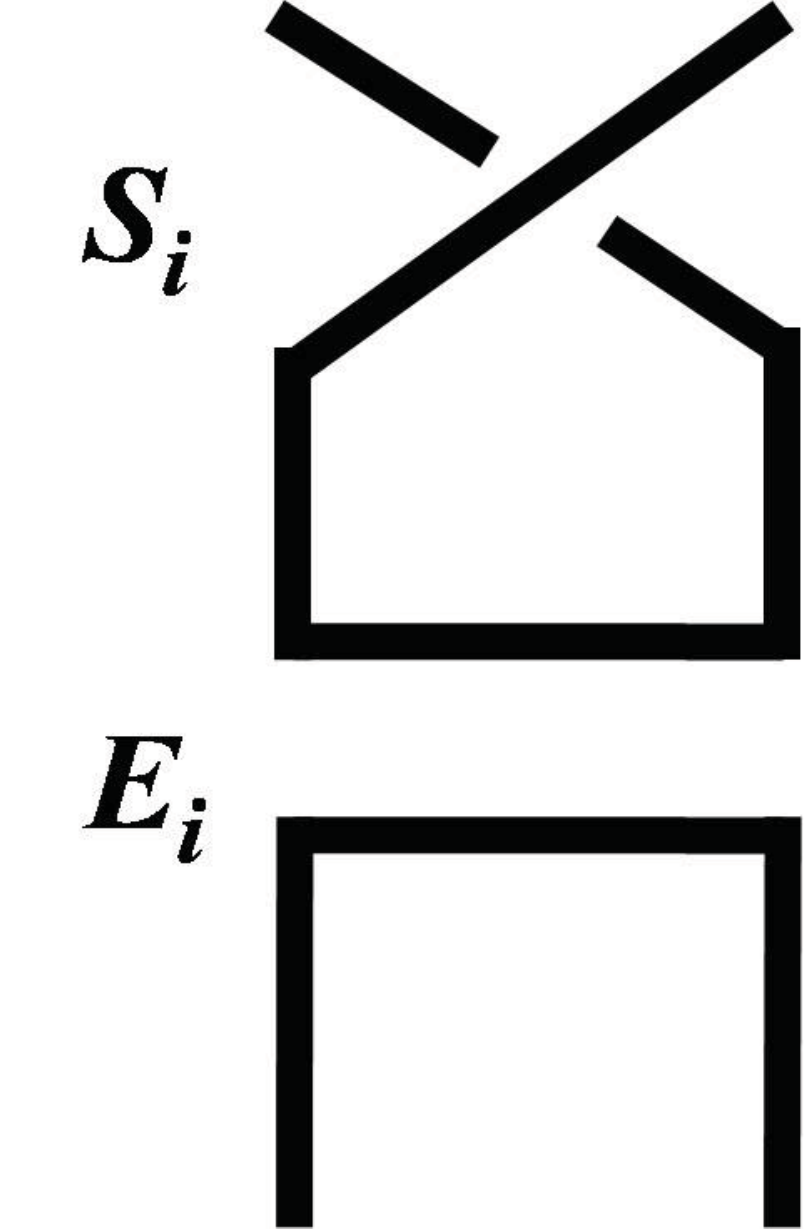}}}
\newcommand{\EiSicross}{\raisebox{-0.4\height}{\includegraphics[height=1.5cm]{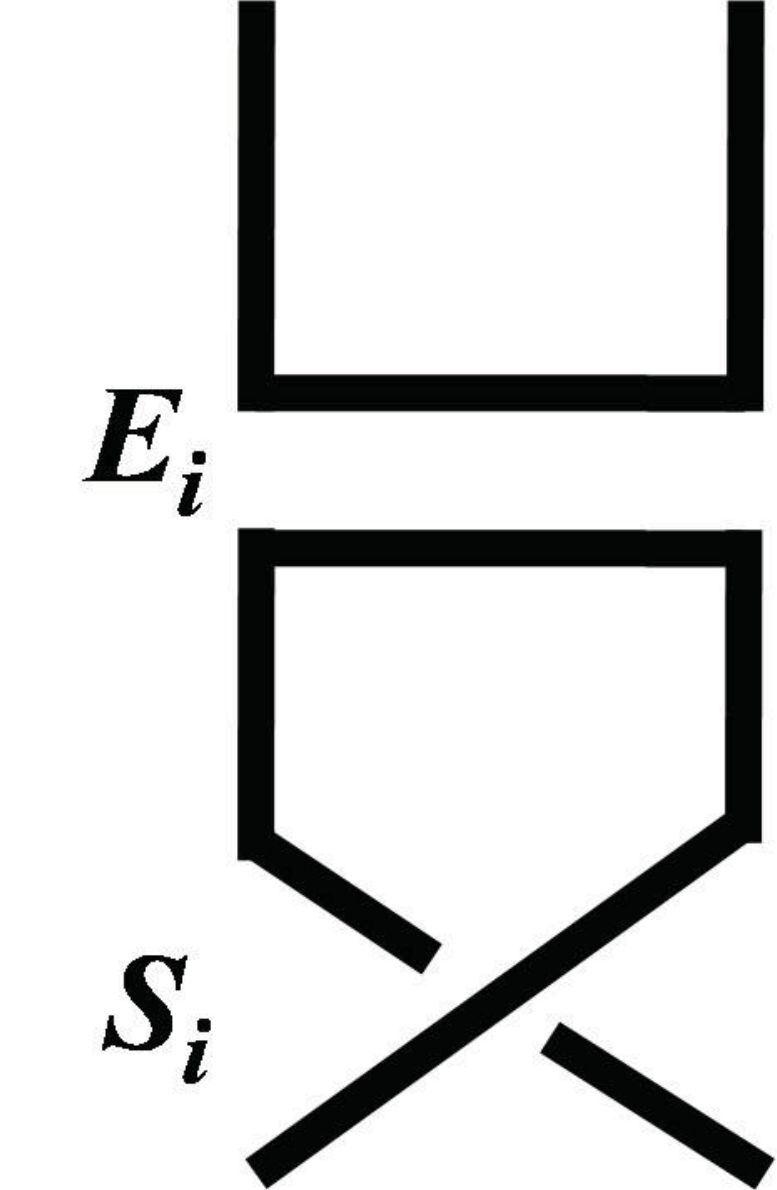}}}

\newcommand{\Ncross}{\raisebox{-0.4\height}{\includegraphics[height=0.8cm]{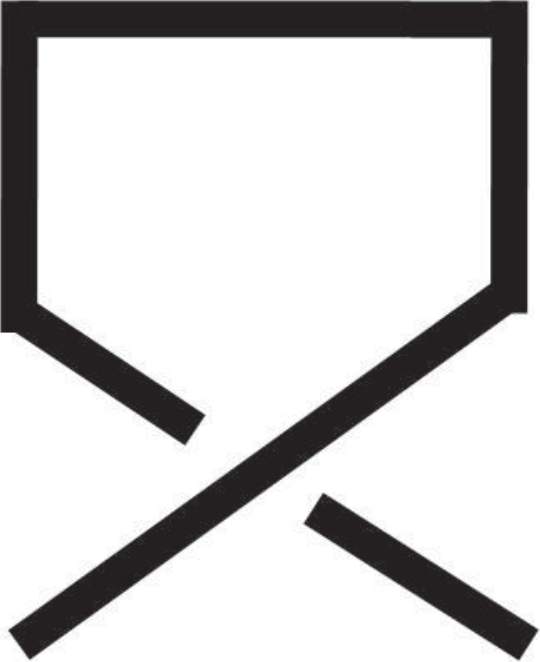}}}
\newcommand{\Ncup}{\raisebox{-0.4\height}{\includegraphics[height=0.5cm]{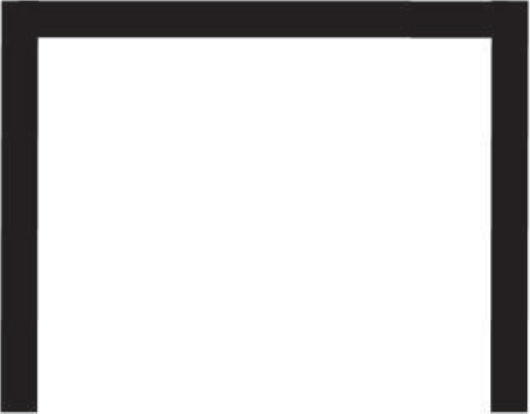}}}
\newcommand{\Ucros}{\raisebox{-0.4\height}{\includegraphics[height=0.8cm]{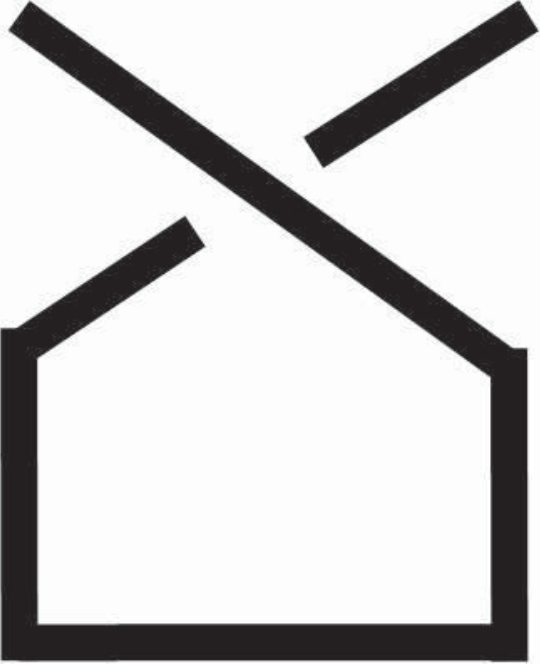}}}
\newcommand{\Ucup}{\raisebox{-0.4\height}{\includegraphics[height=0.5cm]{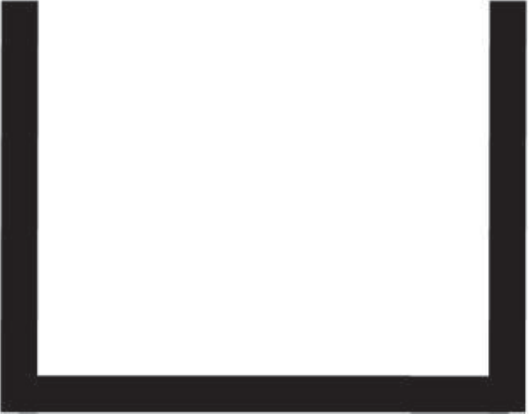}}}
\newcommand{\Ncros}{\raisebox{-0.4\height}{\includegraphics[height=0.8cm]{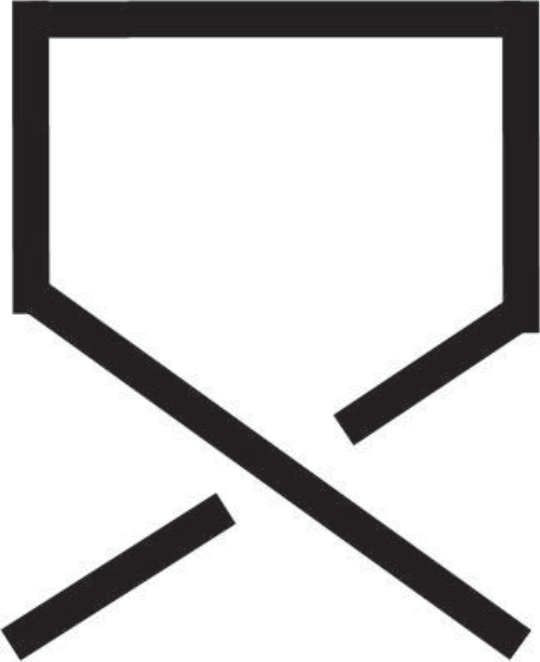}}}
\newcommand{\EiSiSj}{\raisebox{-0.4\height}{\includegraphics[height=1.5cm]{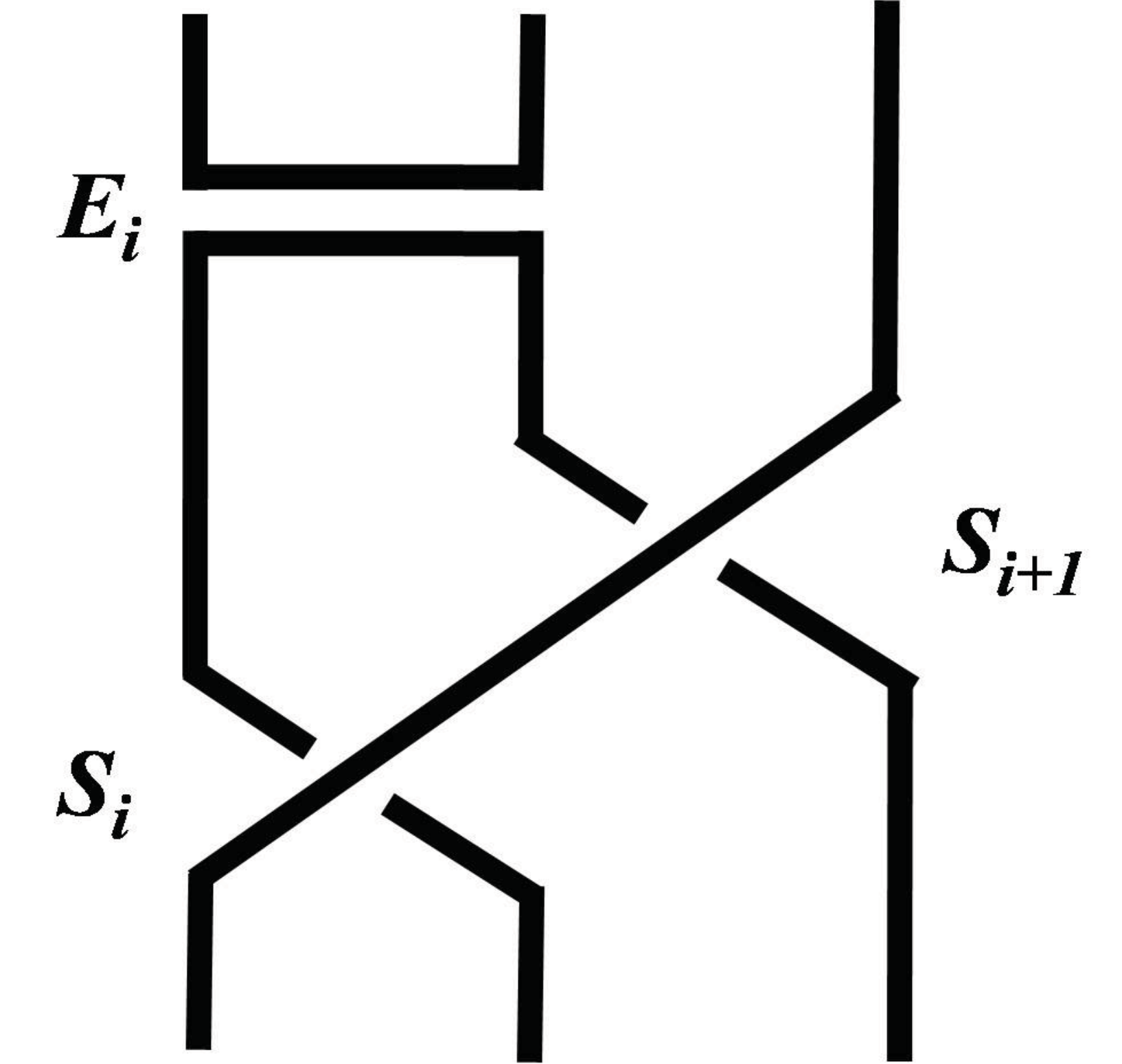}}}
\newcommand{\SiSjEj}{\raisebox{-0.4\height}{\includegraphics[height=1.5cm]{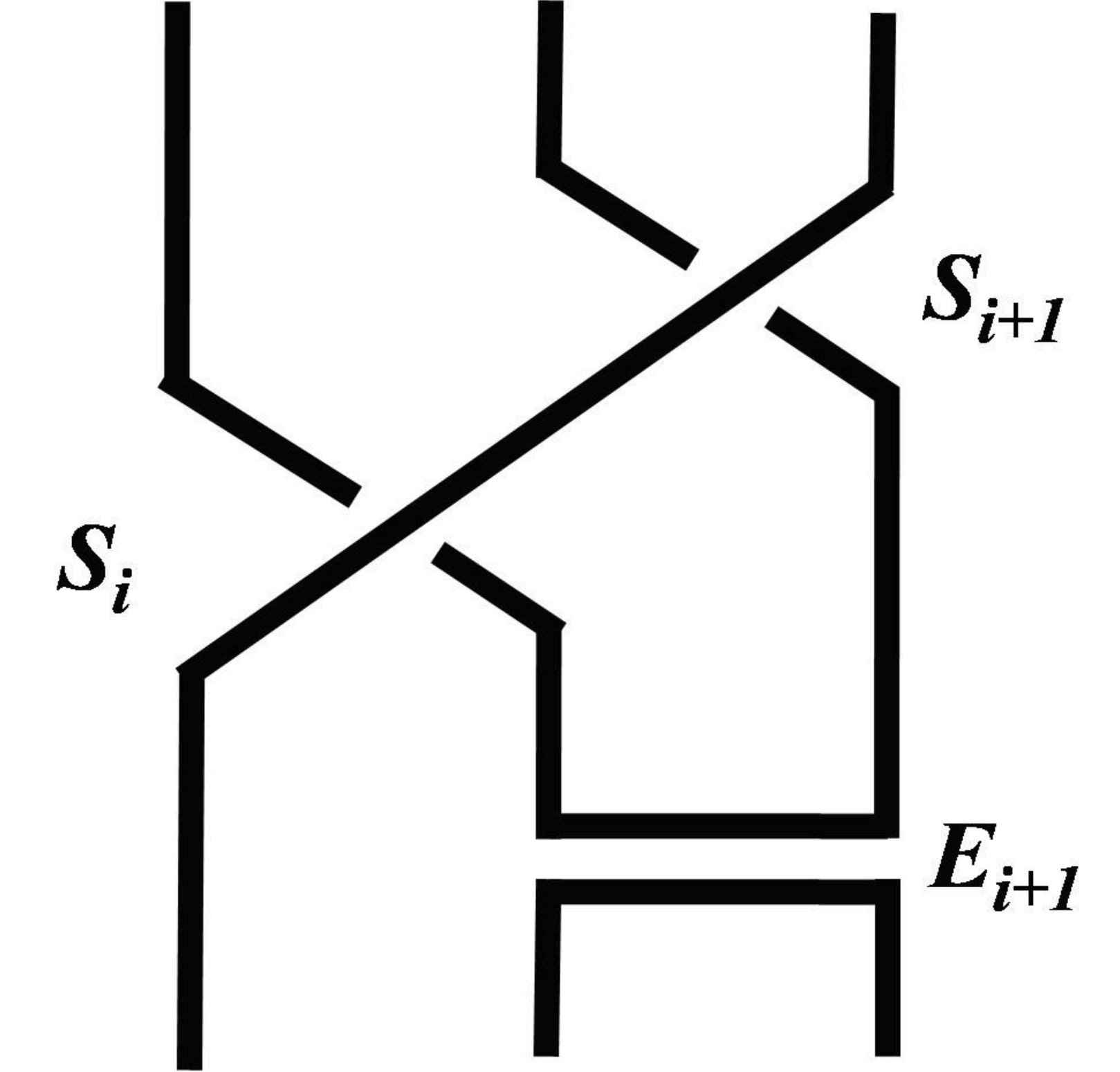}}}
\newcommand{\EiEj}{\raisebox{-0.4\height}{\includegraphics[height=1.2cm]{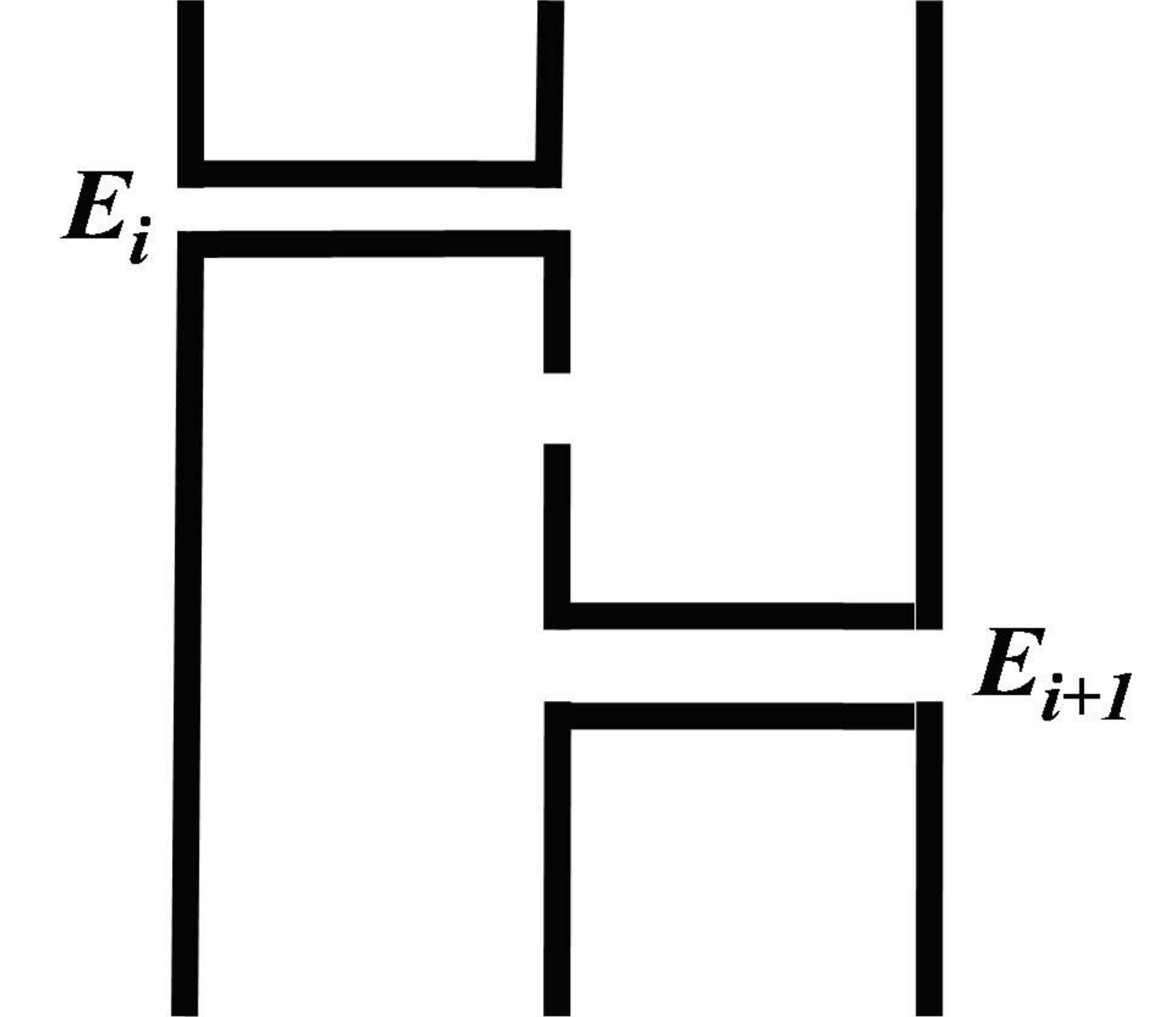}}}

\newcommand{\EiSjEj}{\raisebox{-0.4\height}{\includegraphics[height=1.5cm]{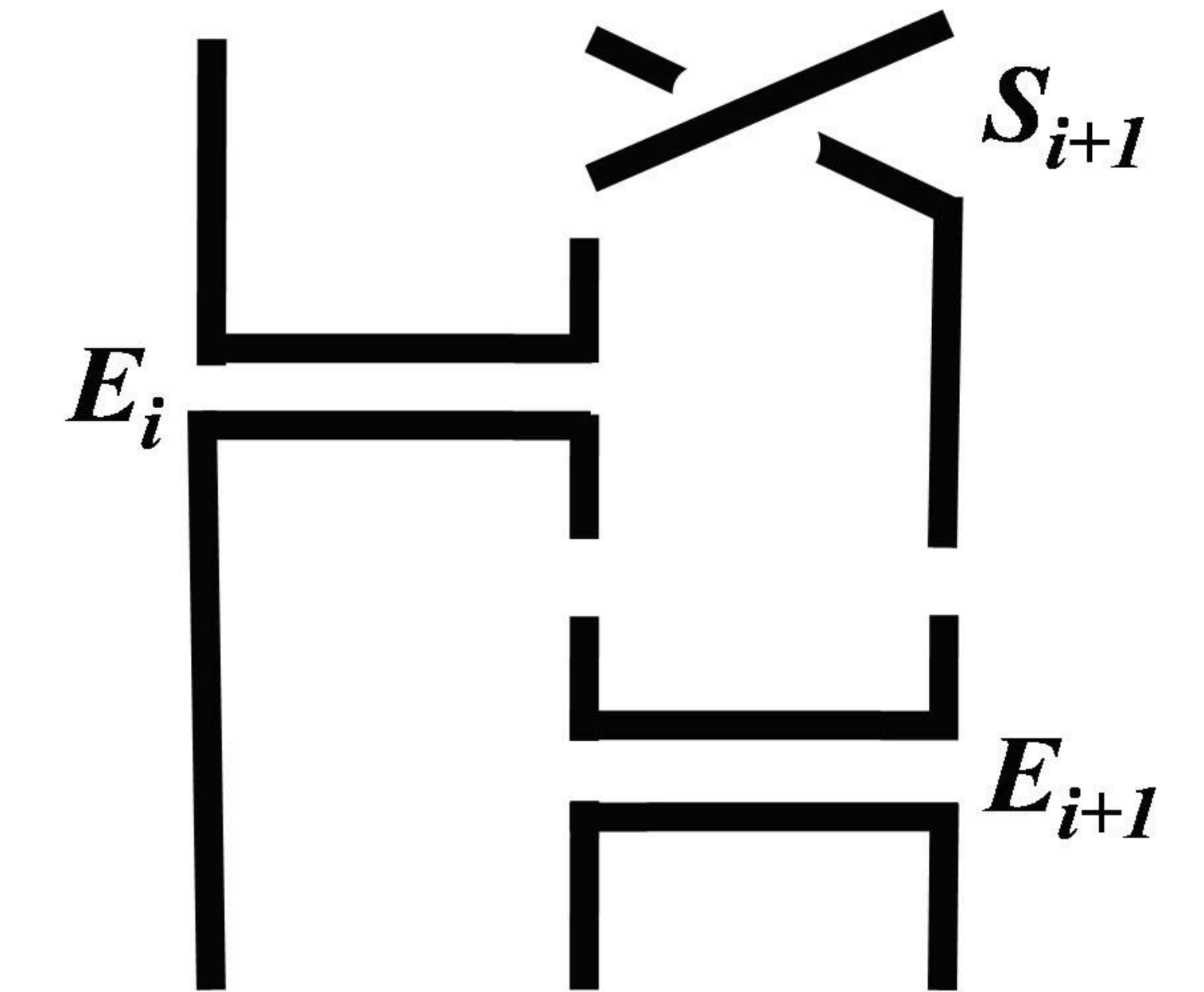}}}
\newcommand{\SihSjhEj}{\raisebox{-0.4\height}{\includegraphics[height=1.5cm]{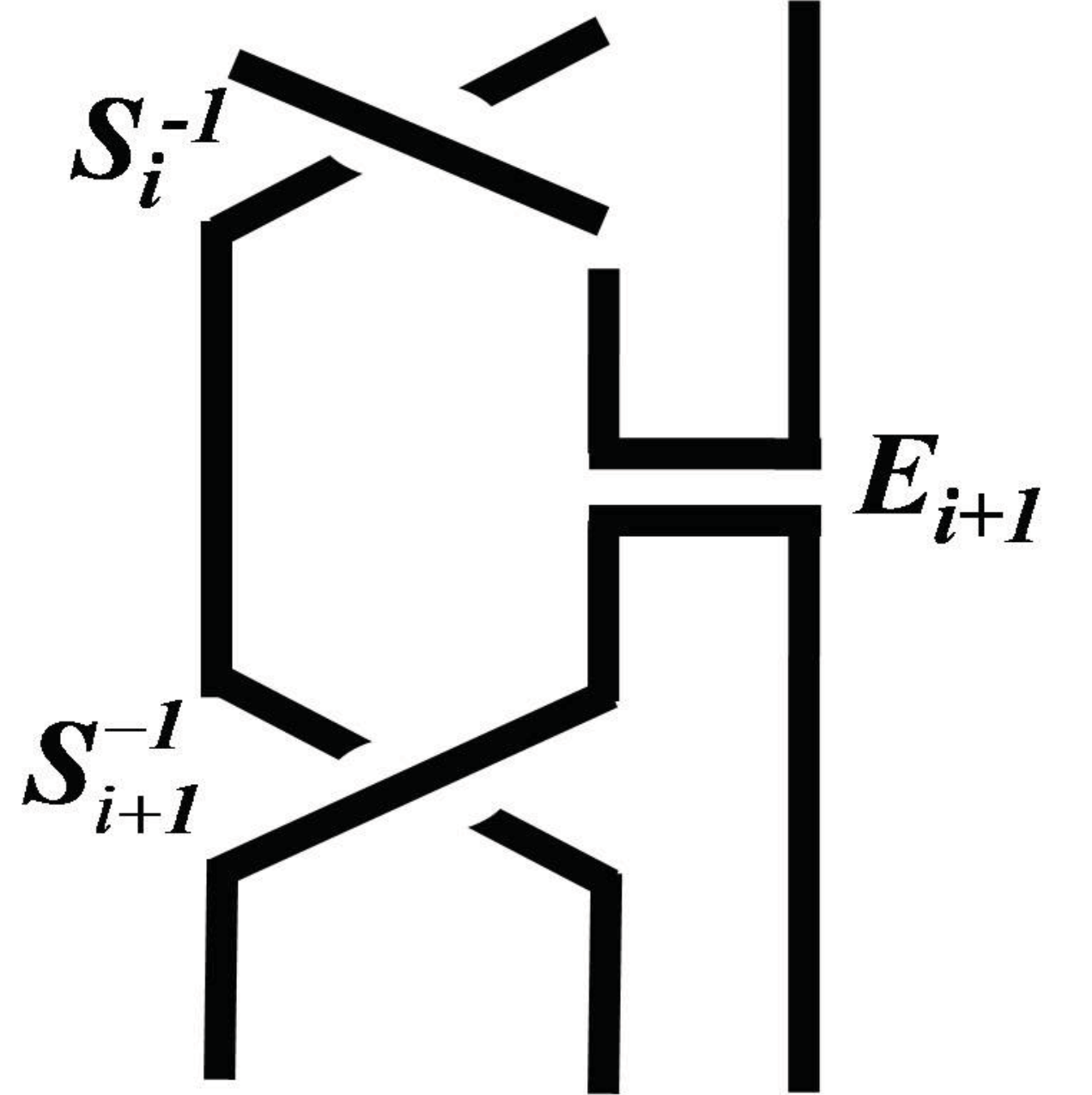}}}
\newcommand{\UcrossNn}{\raisebox{-0.4\height}{\includegraphics[height=1.5cm]{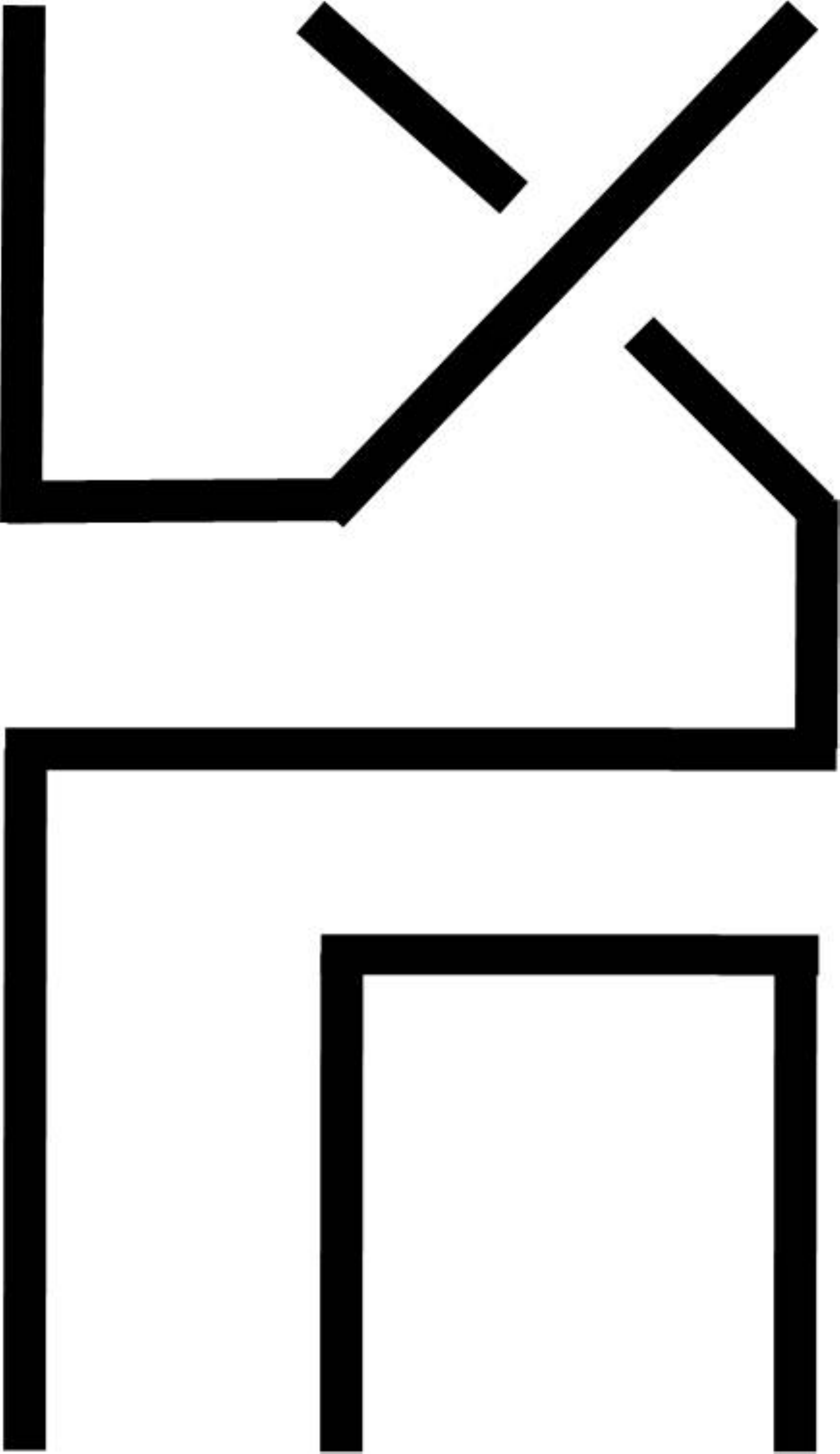}}}
\newcommand{\tcrossH}{\raisebox{-0.4\height}{\includegraphics[height=1.5cm]{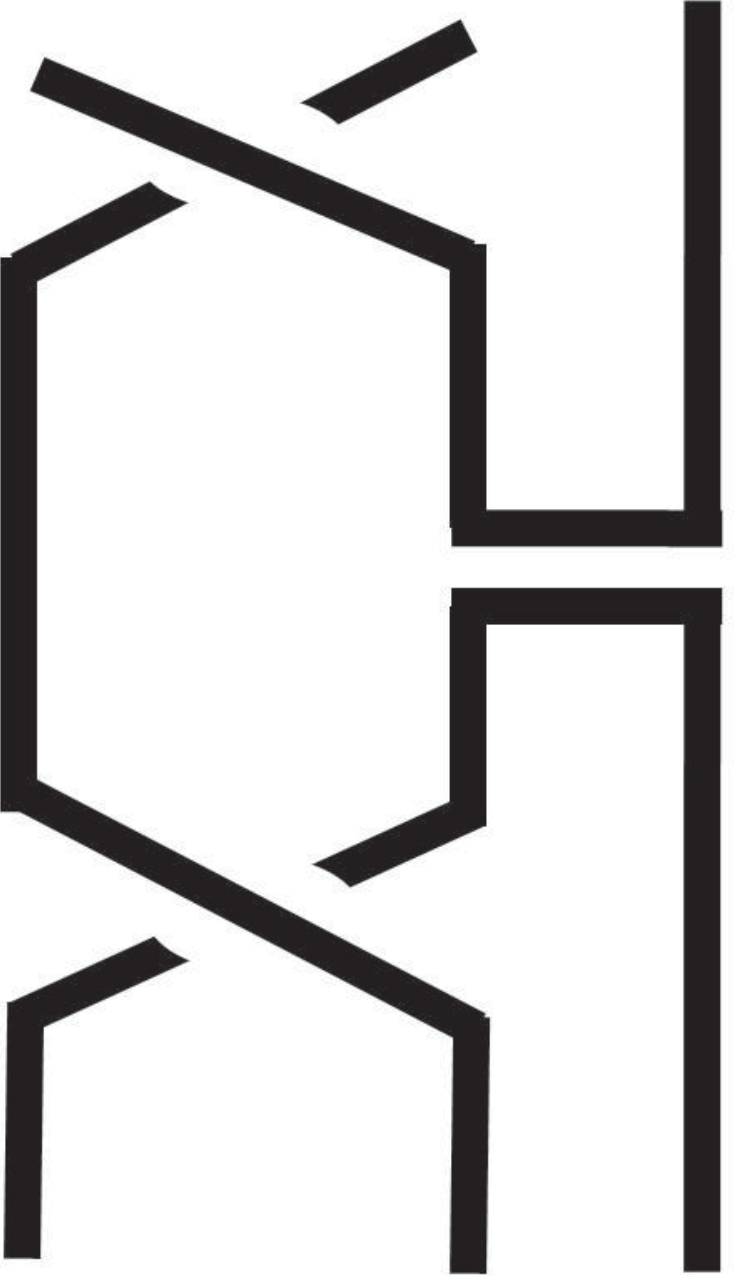}}}
\newcommand{\EiSjEjb}{\raisebox{-0.4\height}{\includegraphics[height=1.5cm]{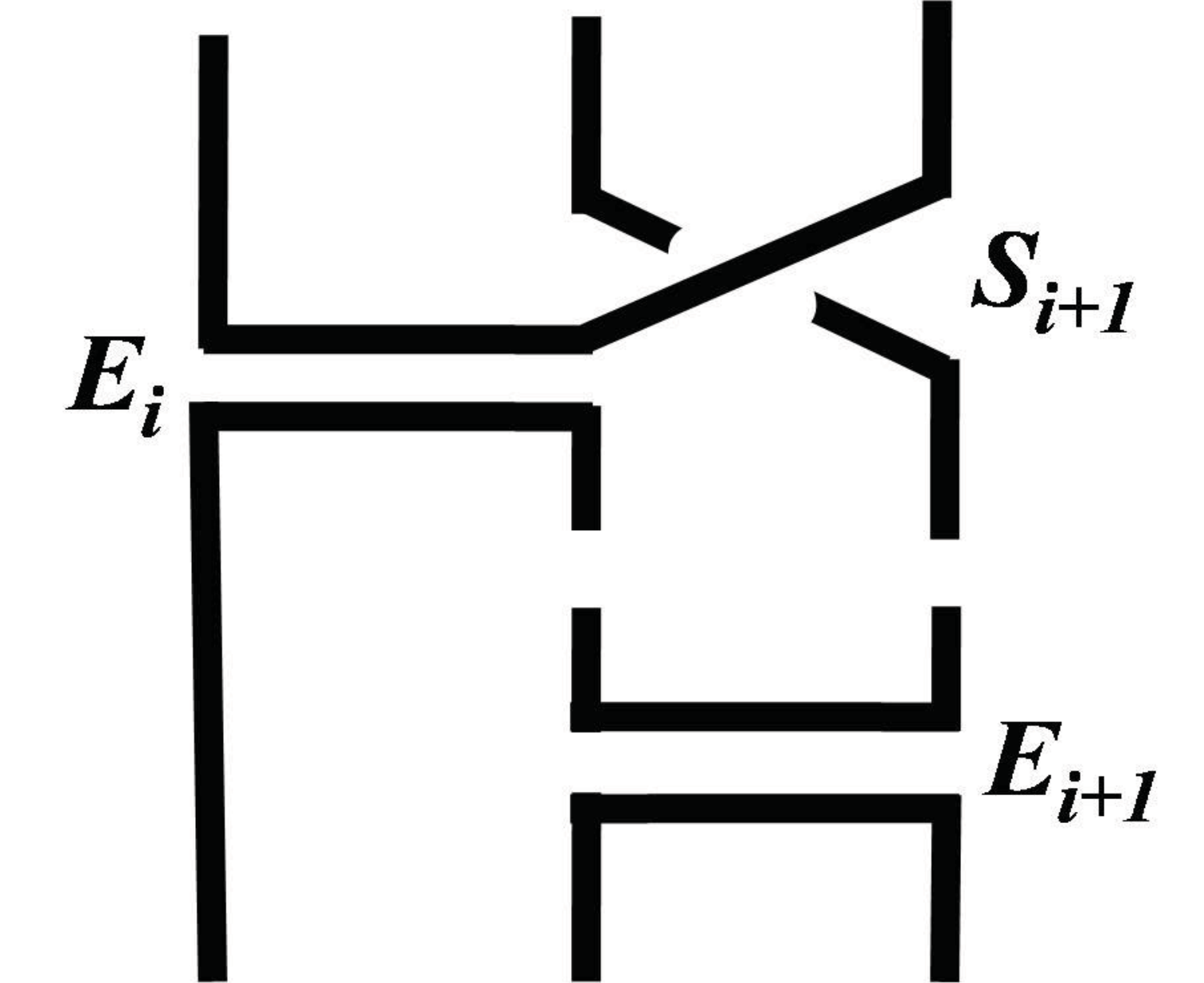}}}
\newcommand{\SihEjb}{\raisebox{-0.4\height}{\includegraphics[height=1.5cm]{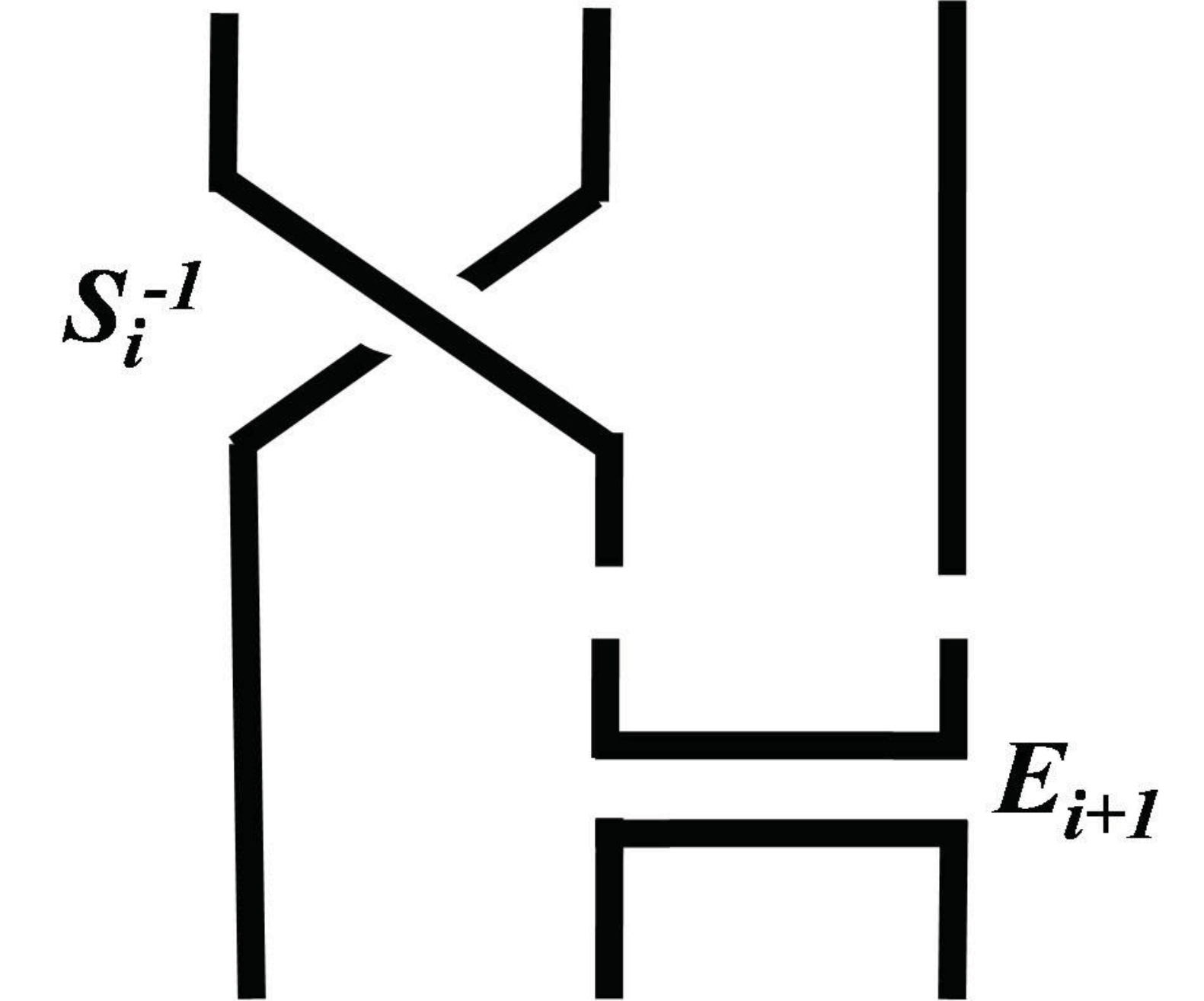}}}
\newcommand{\lUn}{\raisebox{-0.4\height}{\includegraphics[height=1.5cm]{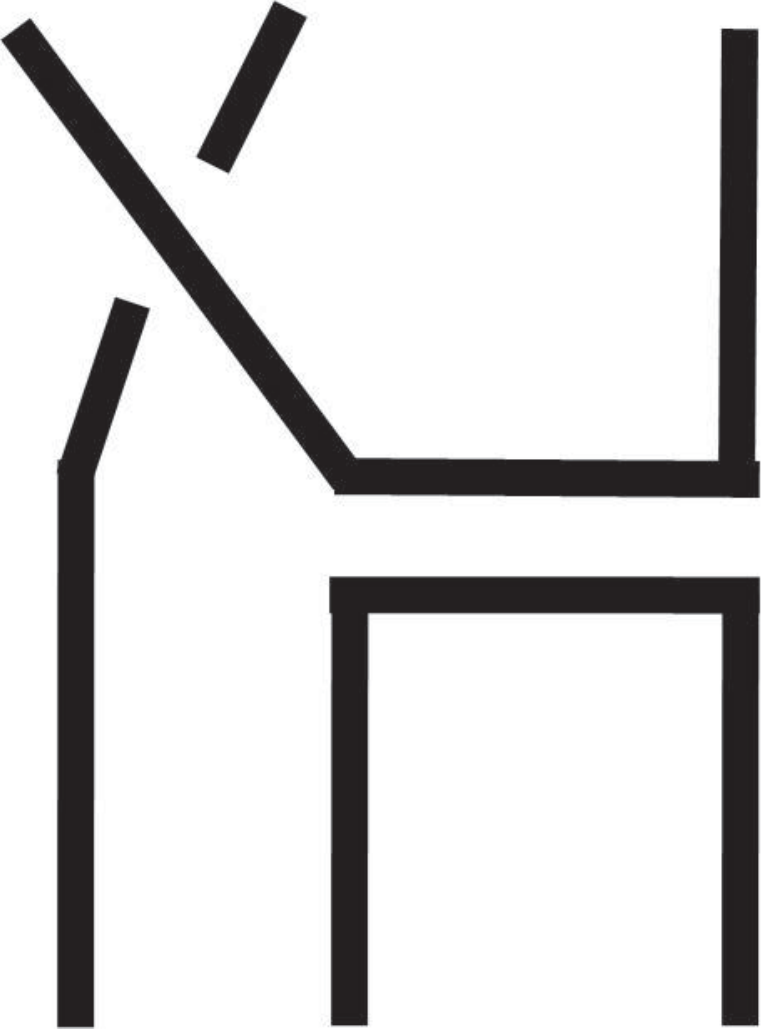}}}
\newcommand{\EiEjSj}{\raisebox{-0.4\height}{\includegraphics[height=1.5cm]{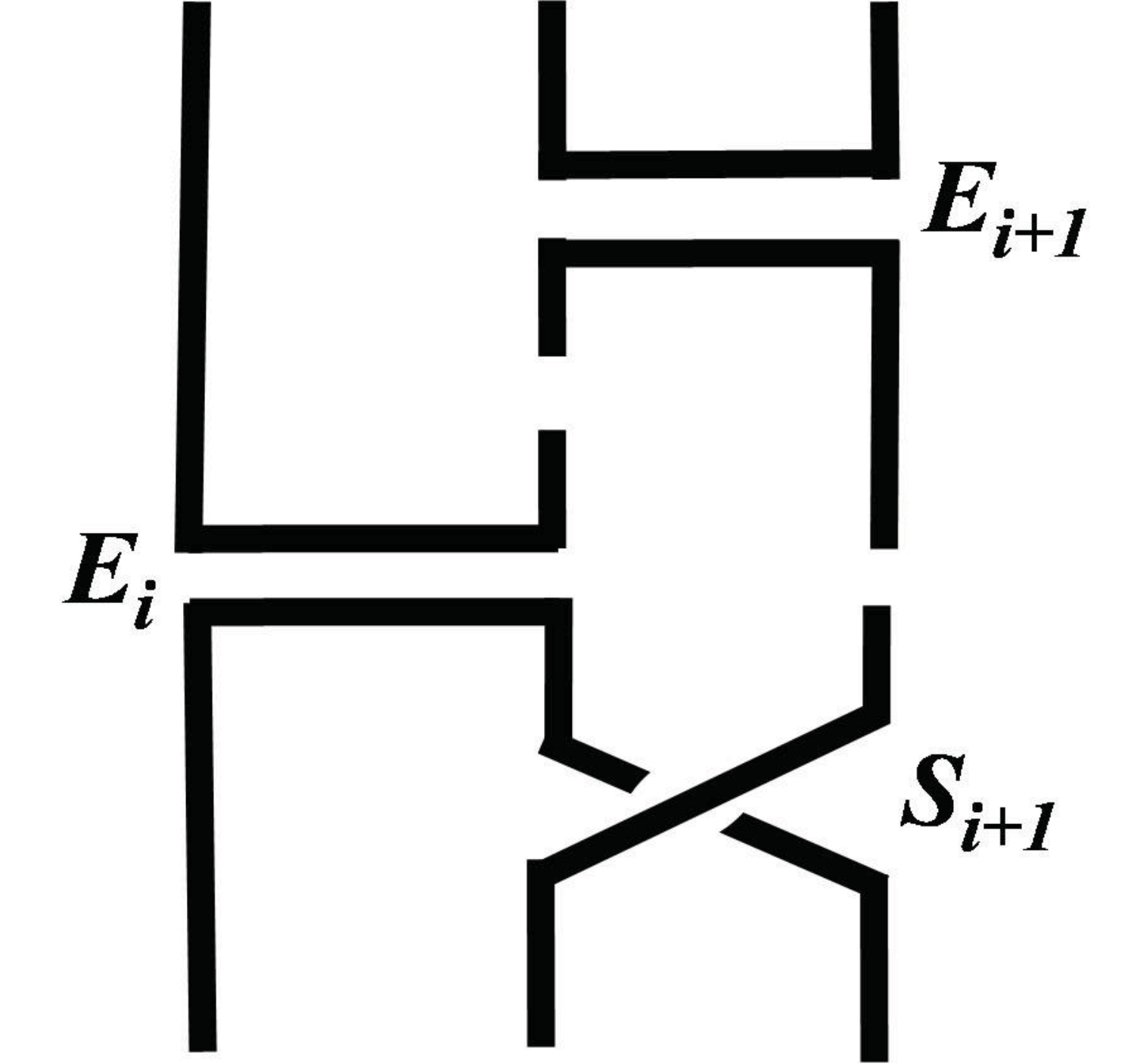}}}
\newcommand{\SihEj}{\raisebox{-0.4\height}{\includegraphics[height=1.2cm]{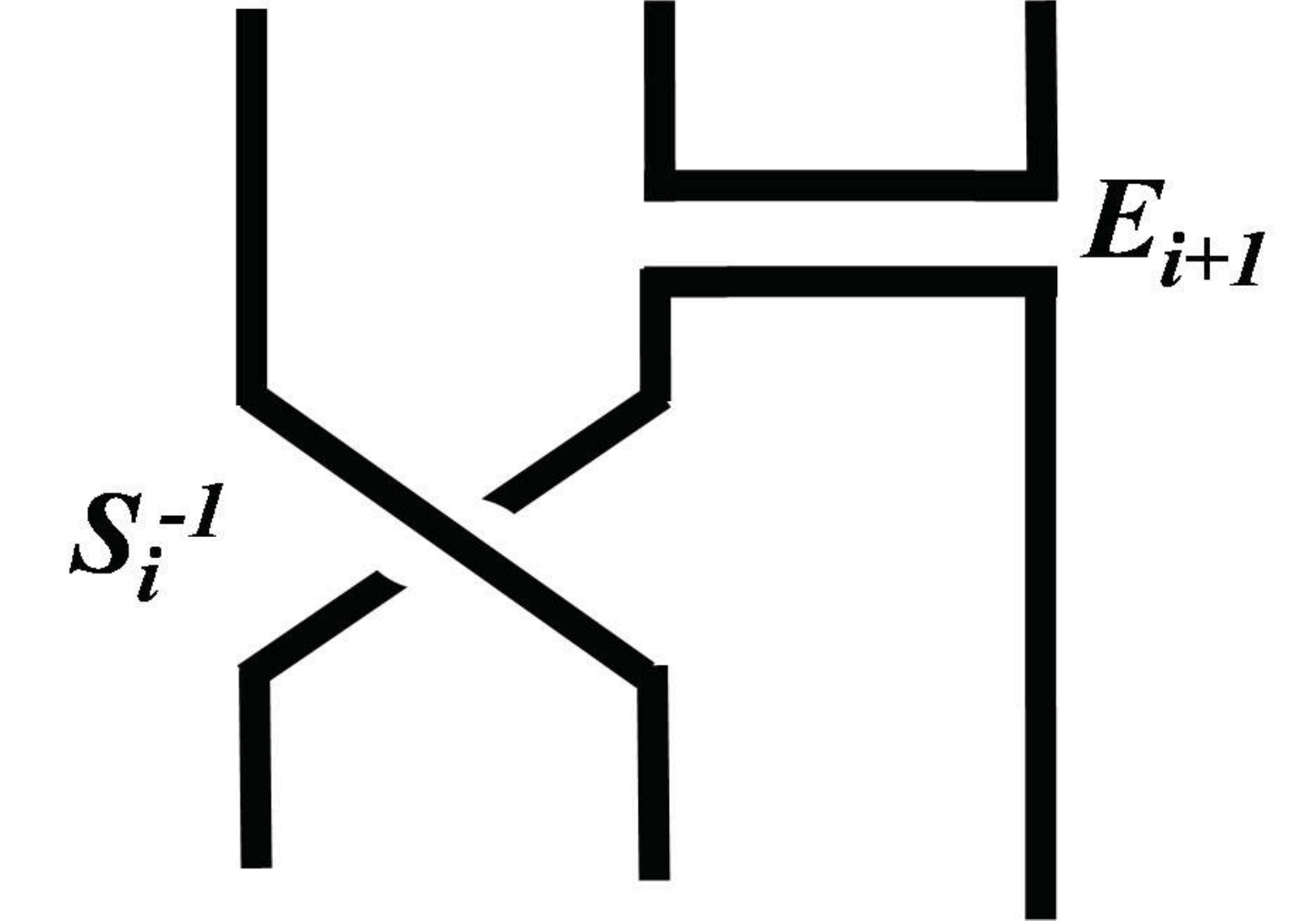}}}

\newcommand{\CcrossC}{\raisebox{-0.4\height}{\includegraphics[height=0.5cm]{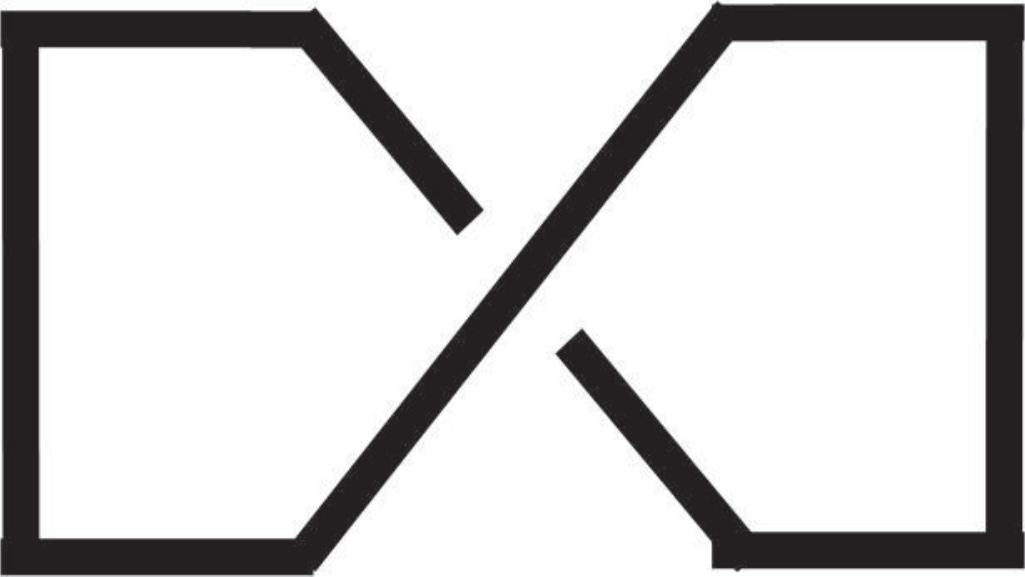}}}

\newcommand{\tsep}{\raisebox{-0.4\height}{\includegraphics[height=0.5cm]{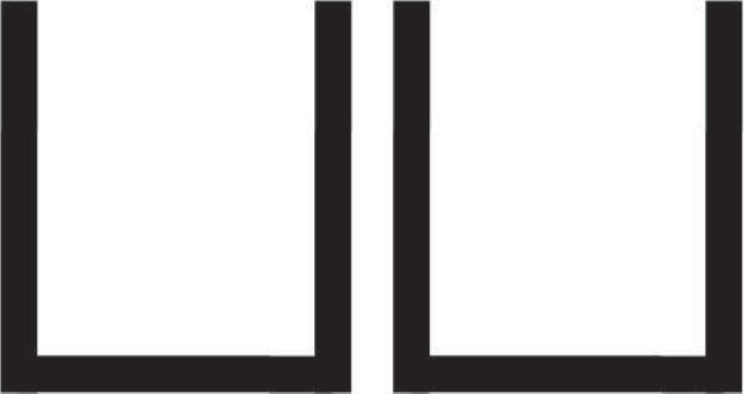}}}
\newcommand{\crossUu}{\raisebox{-0.4\height}{\includegraphics[height=0.8cm]{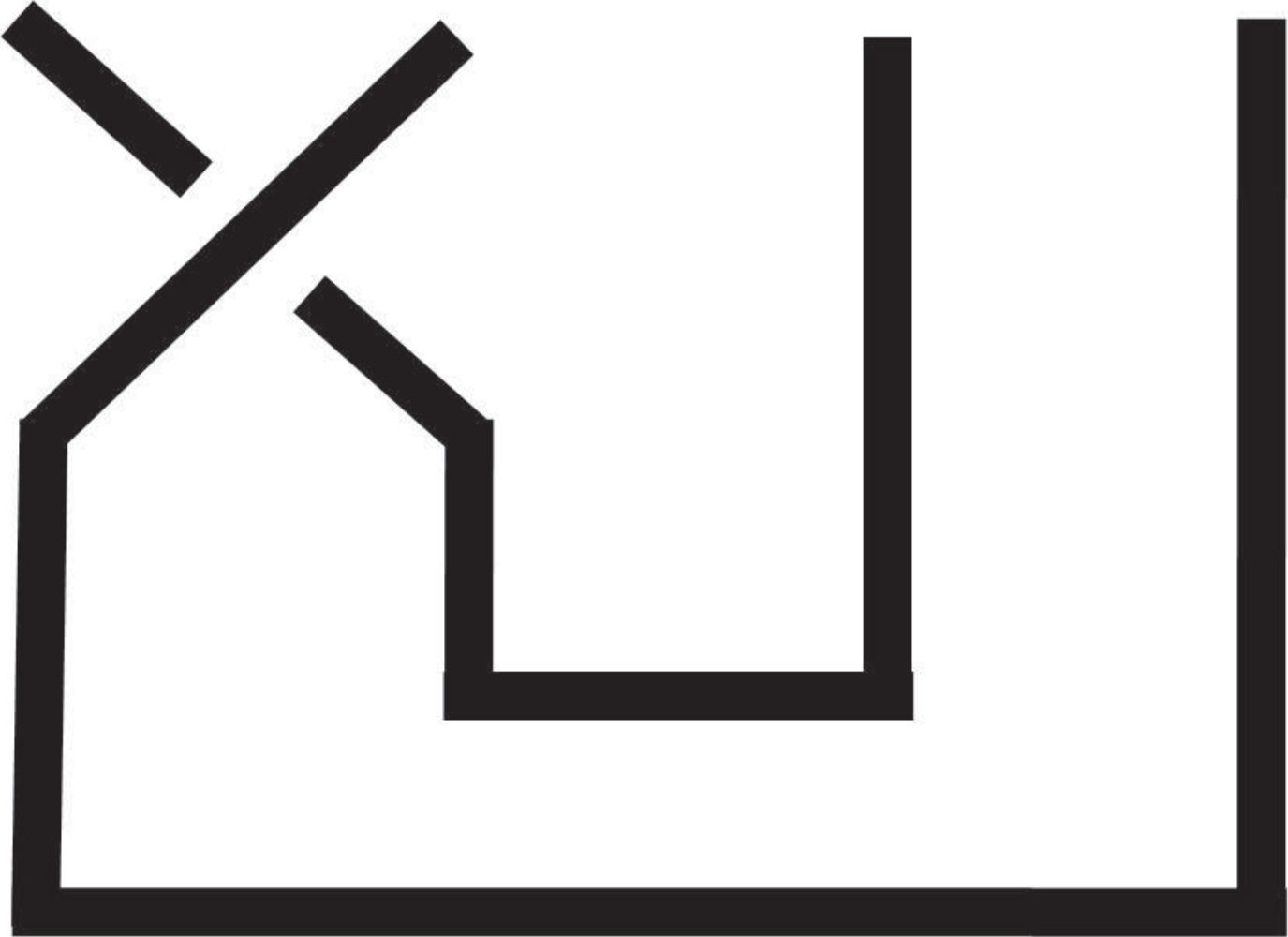}}}
\newcommand{\crosUu}{\raisebox{-0.4\height}{\includegraphics[height=0.8cm]{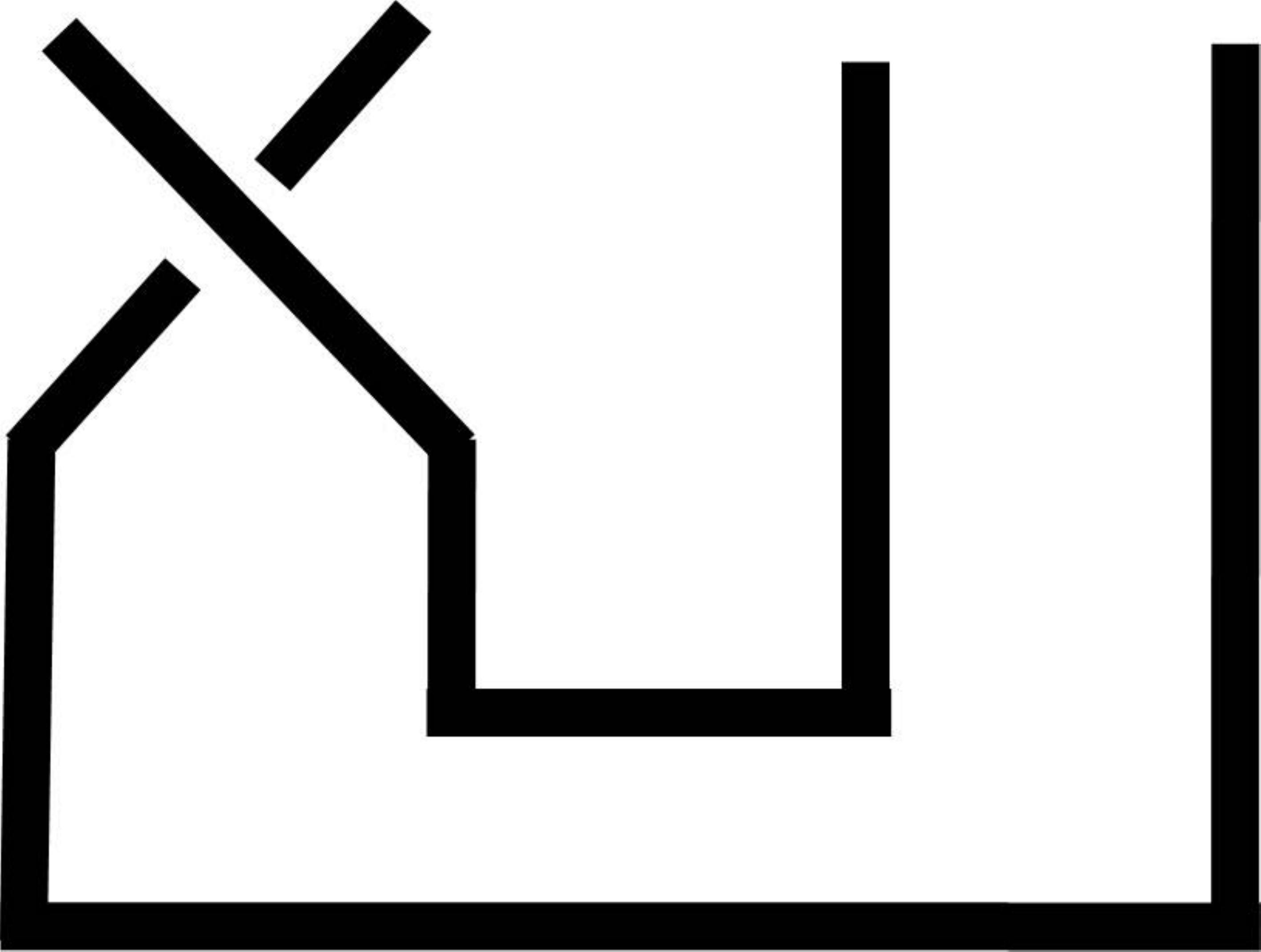}}}
\newcommand{\Uu}{\raisebox{-0.4\height}{\includegraphics[height=0.8cm]{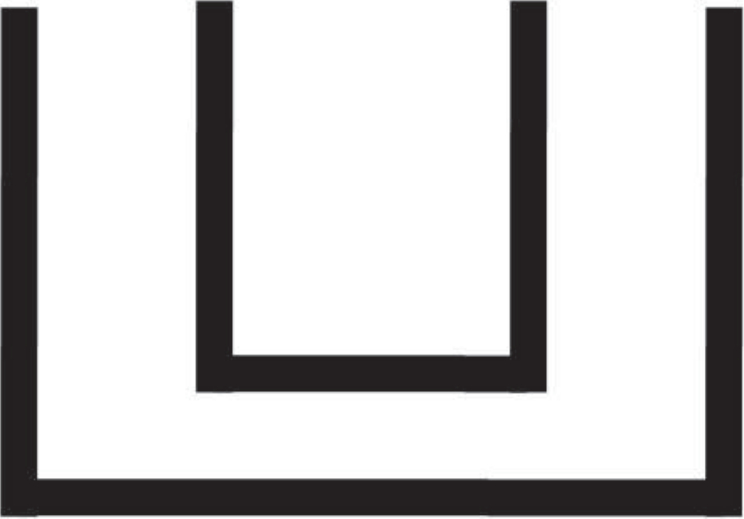}}}

\newcommand{\tcross}{\raisebox{-0.4\height}{\includegraphics[height=0.8cm]{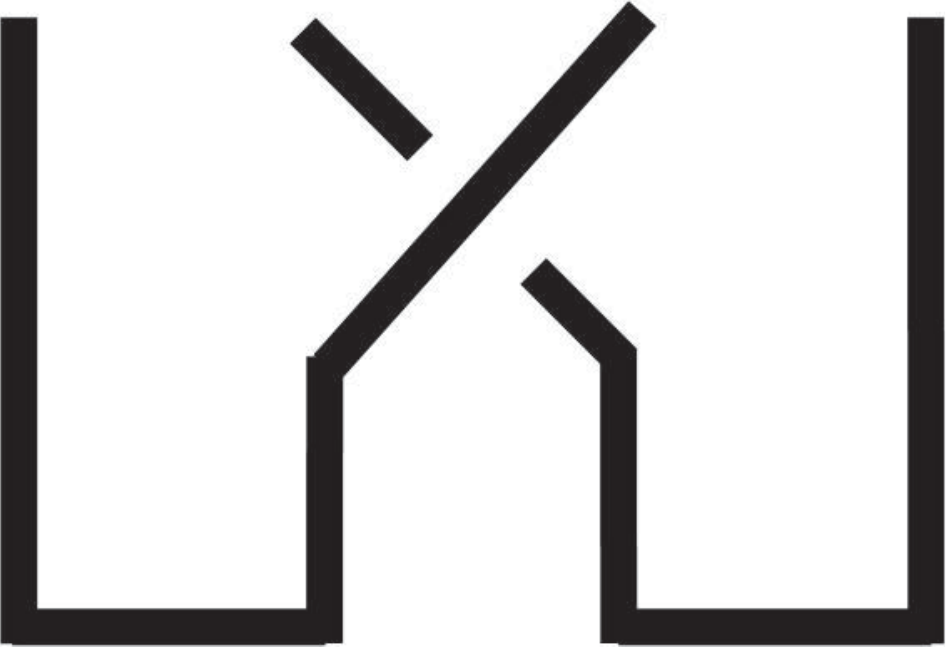}}}
\newcommand{\ntcros}{\raisebox{-0.4\height}{\includegraphics[height=0.8cm]{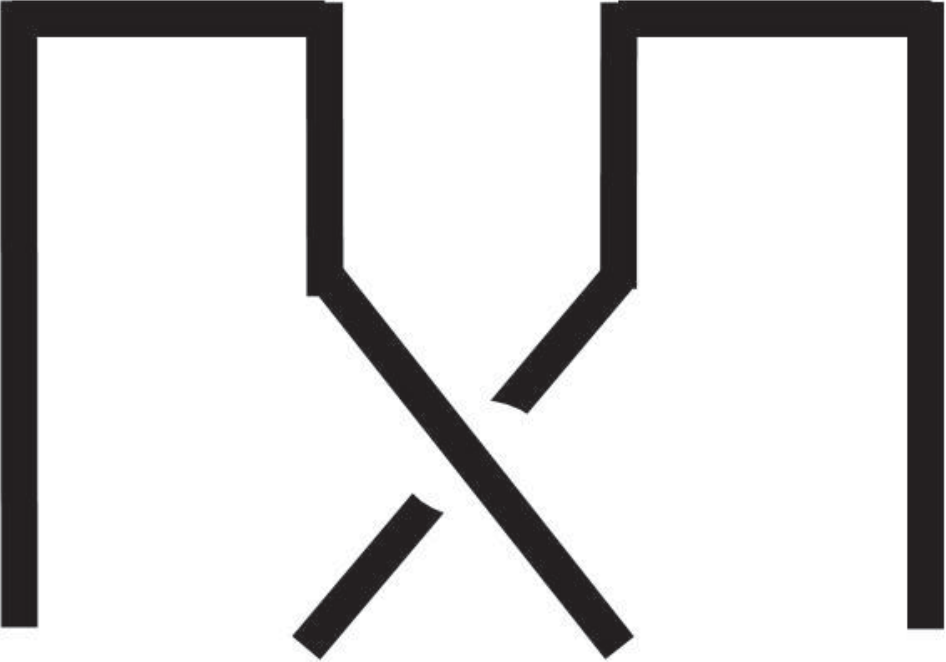}}}
\newcommand{\Nn}{\raisebox{-0.4\height}{\includegraphics[height=0.8cm]{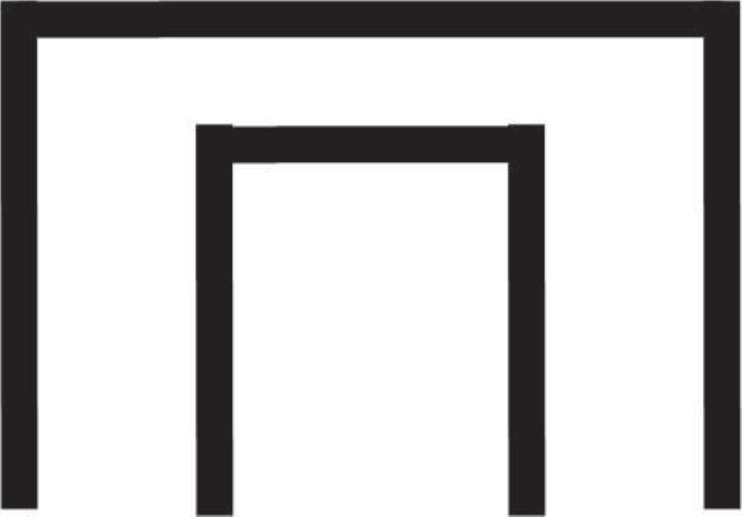}}}
\newcommand{\tncup}{\raisebox{-0.4\height}{\includegraphics[height=0.5cm]{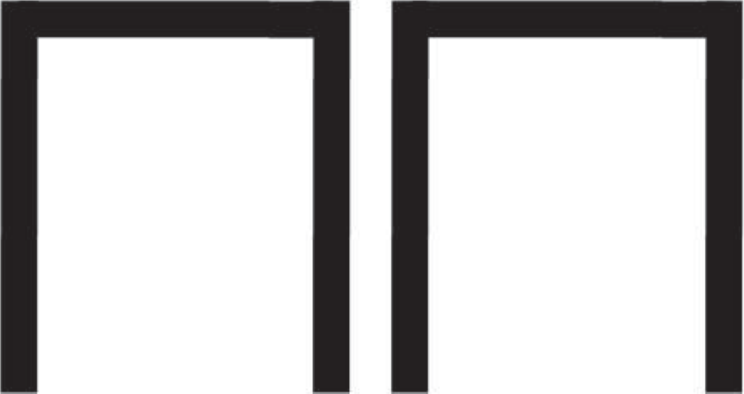}}}

\newcommand{\tloop}{\raisebox{-0.4\height}{\includegraphics[height=0.8cm]{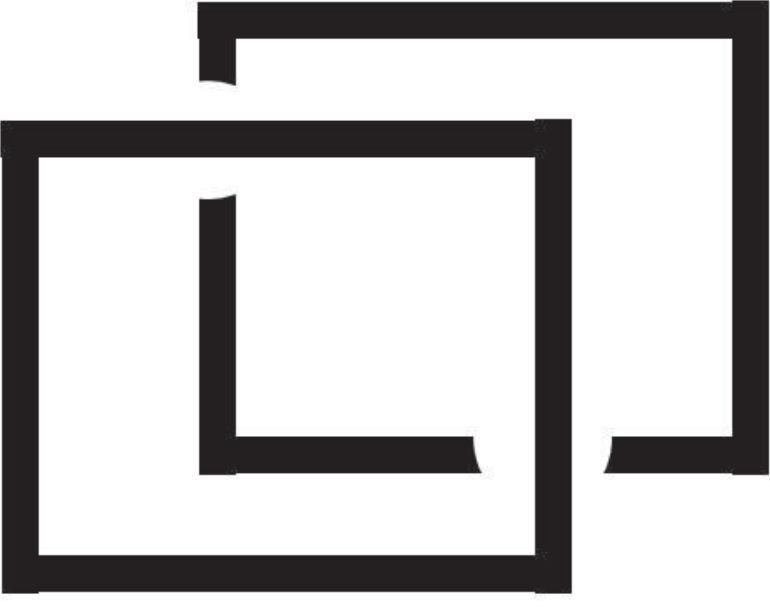}}}

\newcommand{\Unot}{\raisebox{-0.4\height}{\includegraphics[height=0.8cm]{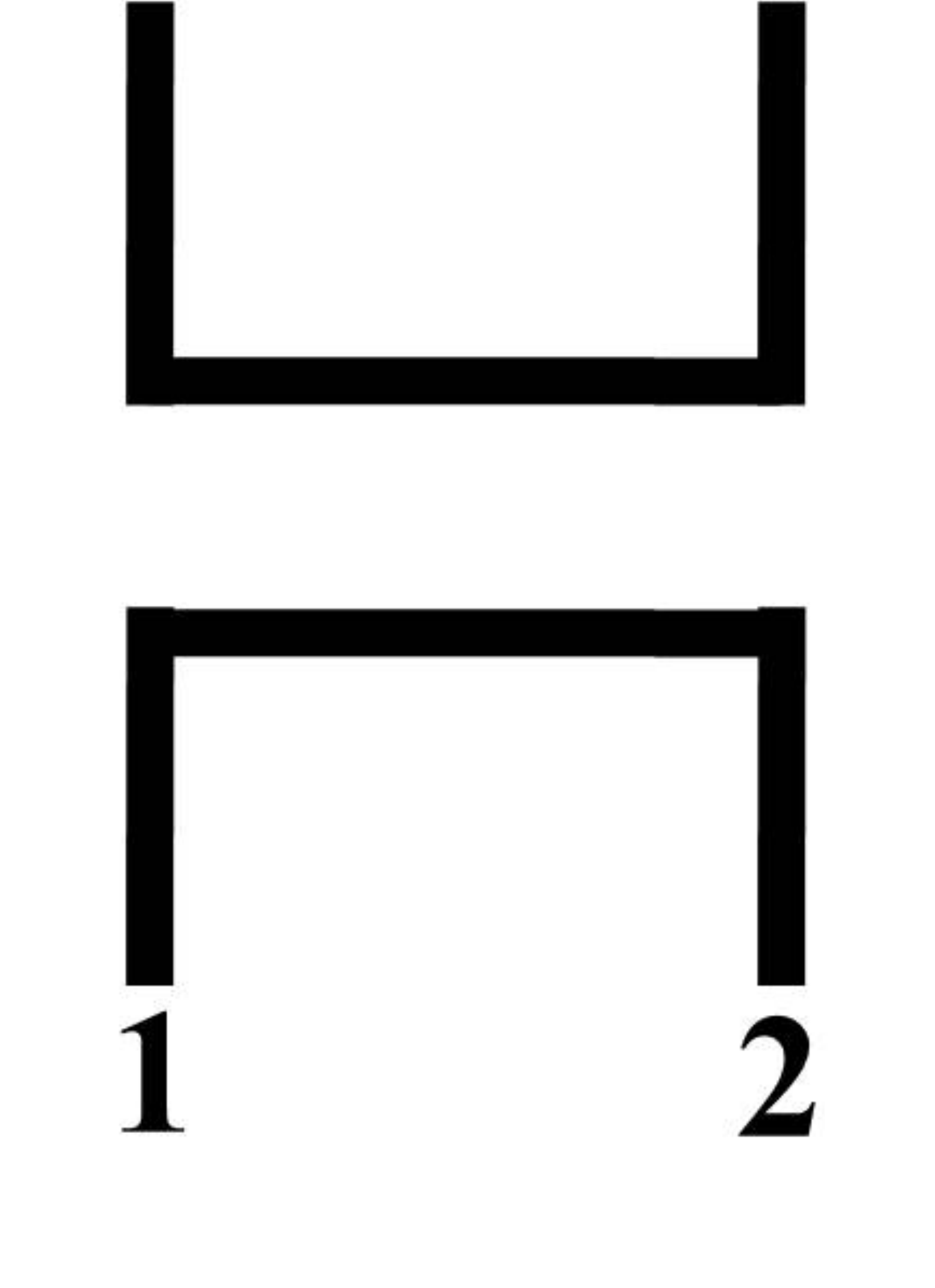}}}
\newcommand{\Utcross}{\raisebox{-0.4\height}{\includegraphics[height=0.8cm]{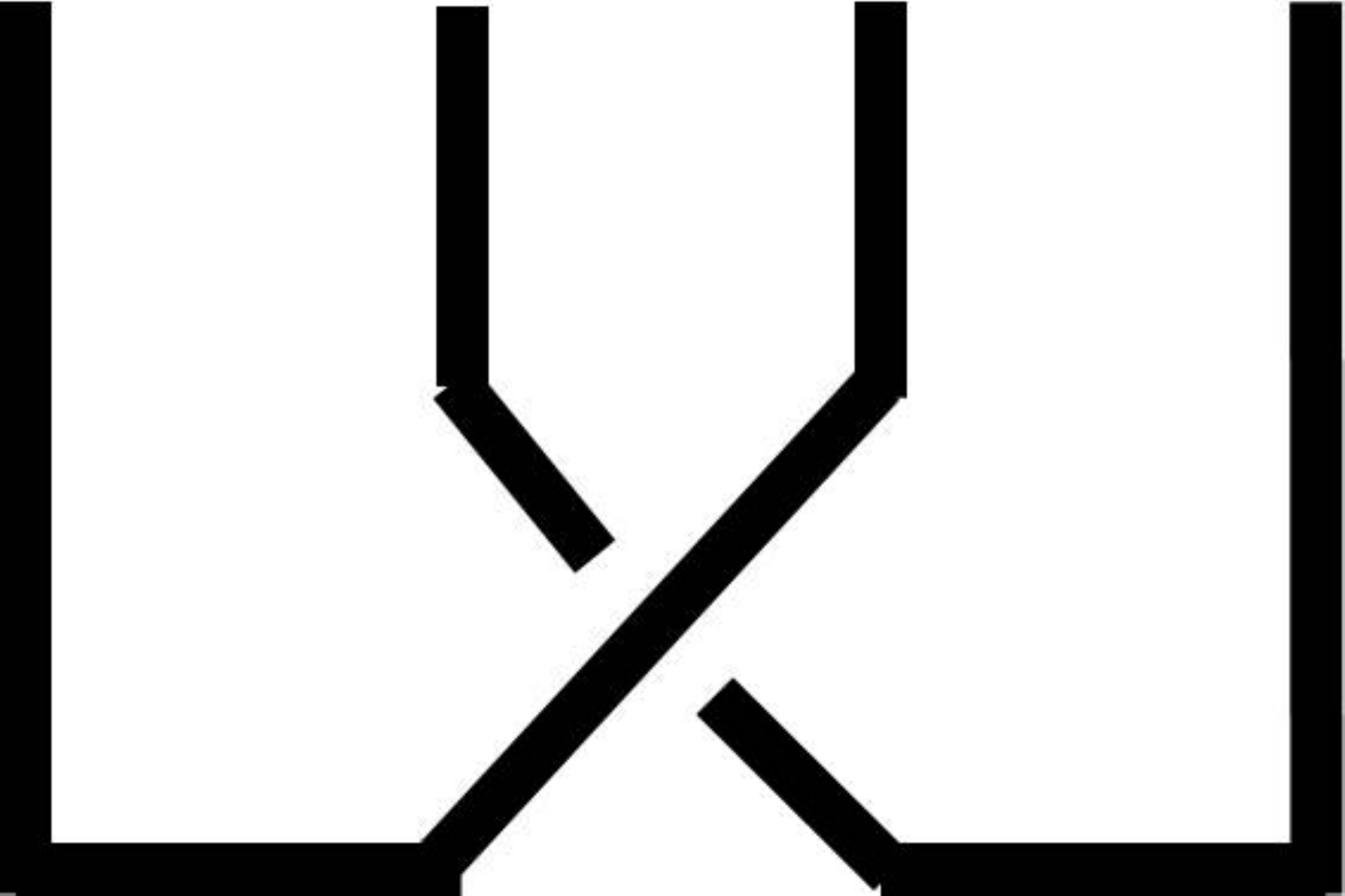}}}
\newcommand{\tsquare}{\raisebox{-0.4\height}{\includegraphics[height=0.5cm]{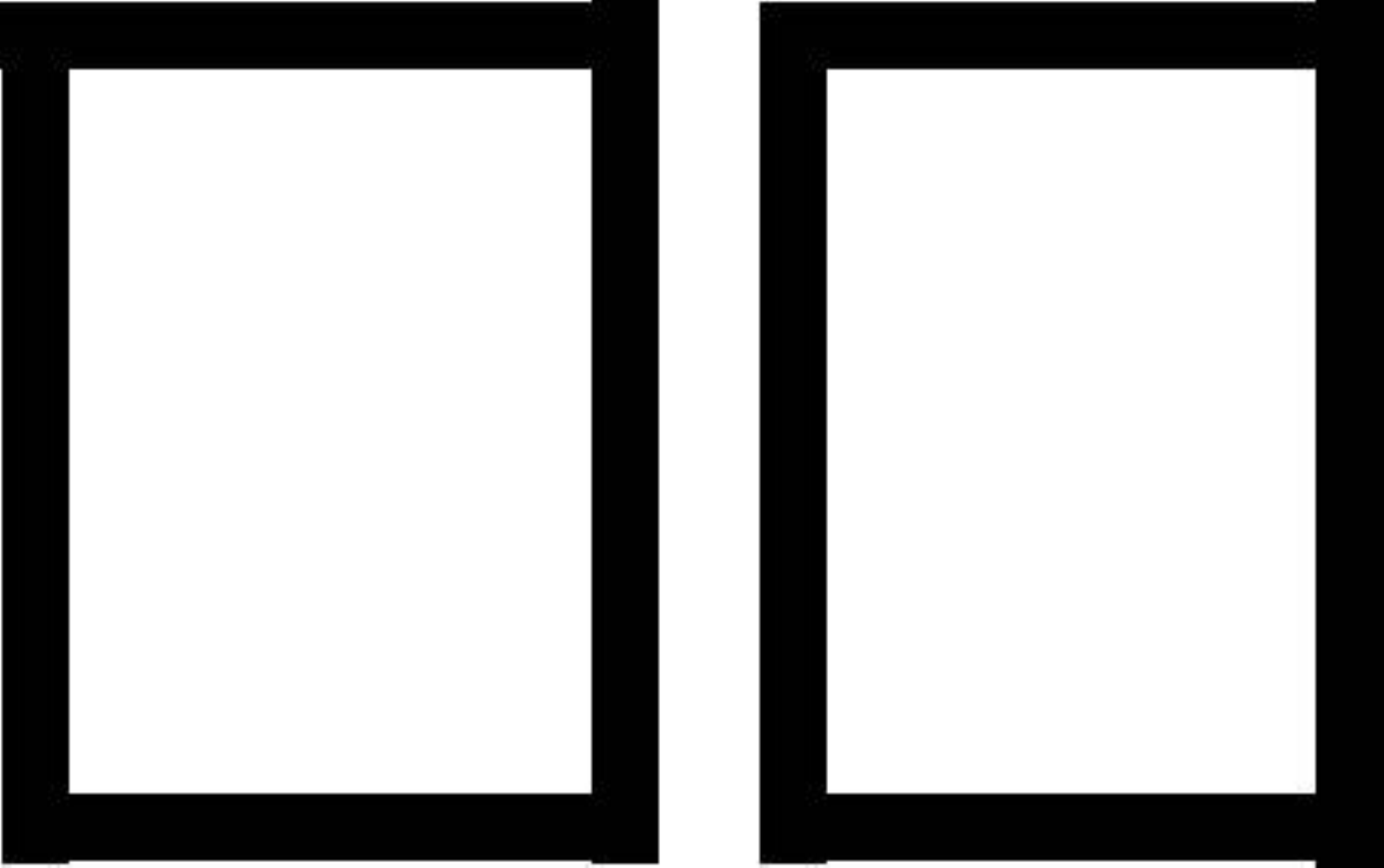}}}
\newcommand{\Nu}{\raisebox{-0.4\height}{\includegraphics[height=0.6cm]{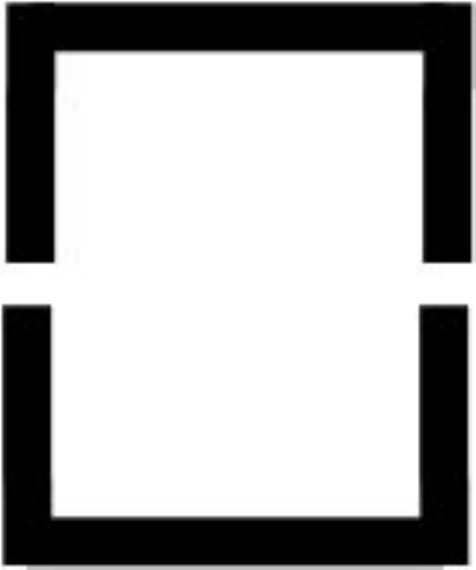}}}
\newcommand{\Crosss}{\raisebox{-0.4\height}{\includegraphics[height=0.8cm]{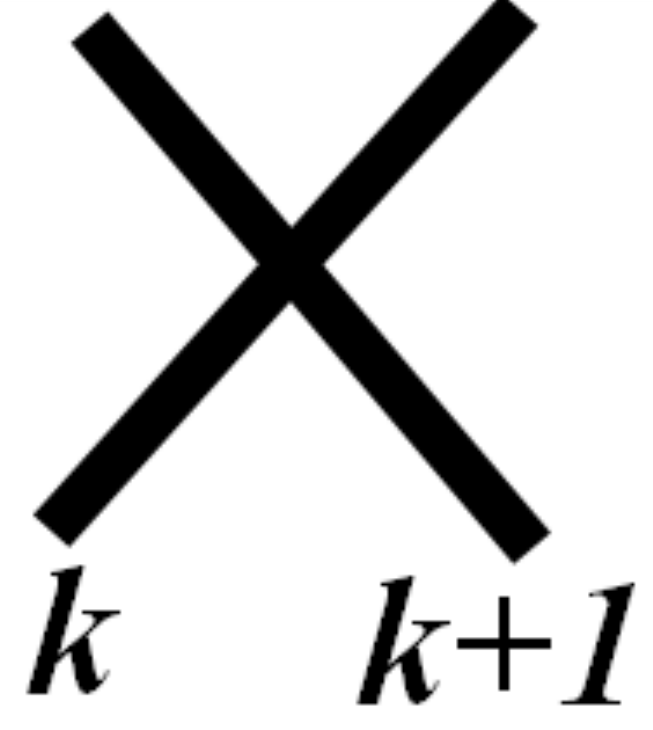}}}
\newcommand{\Unii}{\raisebox{-0.4\height}{\includegraphics[height=1cm]{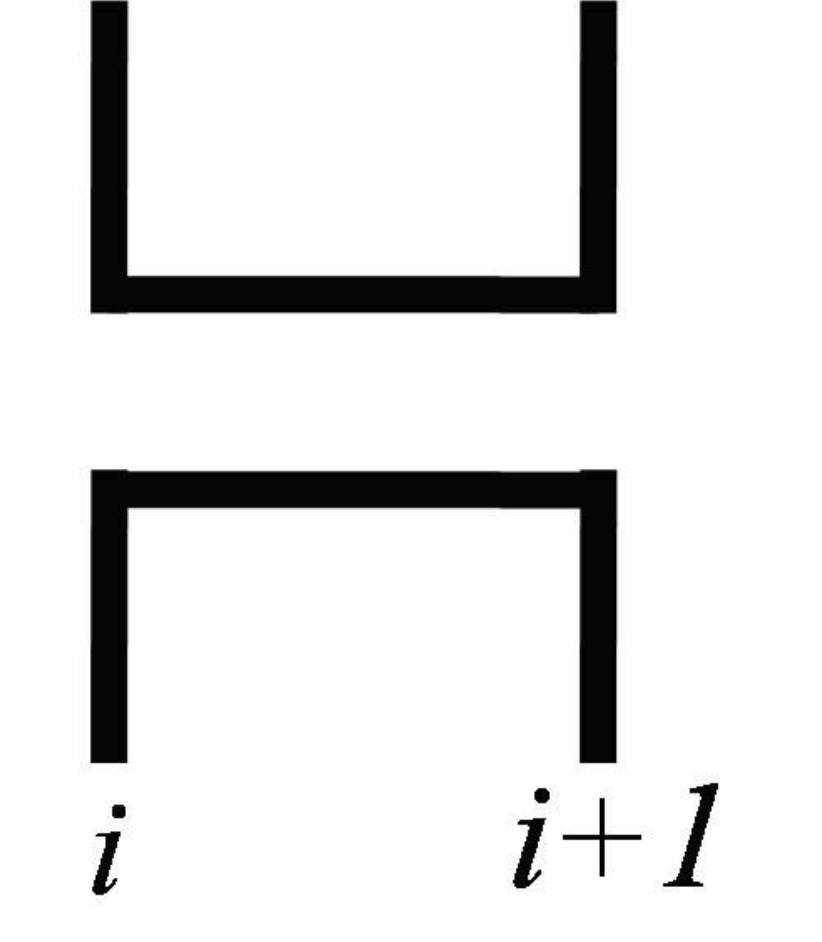}}}

\begin{document}

\title[Topological Basis Associated with BWMA.....] {Topological Basis Associated with BWMA, Extremes of $L_{1}$-norm in Quantum Information and Applications in Physics}

\author{Qing. Zhao$^{1}$, Ruo-Yang Zhang$^{2}$, Kang Xue$^{3}$ and Mo-Lin Ge$^{1,2}$}
\address{$^{1}$ College of Physics, Beijing Institute of Technology, Beijing, 100081, China}
\address{$^{2}$ Theoretical Physics Section, Chern Institute of Mathematics, Nankai University, Tianjin, 300071, China}
\address{$^{3}$ Dept. of Physics, Northeast Normal University, Changchun, 130024, China}

\ead{\mailto{geml@nankai.edu.cn}, {qzhaoyuping@bit.edu.cn}}


\begin{abstract}
  The topological basis associated with Birman-Wenzl-Murakami algebra (BWMA) is constructed and the three dimensional forms of braiding matrices S have been found for both $S^+=S$ and $S^+=S^{-1}$. A familiar spin-$1$ model related to braiding matrix associated with BWMA is discussed. The extreme points $(\theta=\pm\pi/2$ and $\pm\pi)$ of $L_{1}$-norm and von Neumann entropy are shown to be connected to each other. Through the general discussion and examples we then point out that the $L_{1}$-norm describes quantum entanglement.
\end{abstract}

\maketitle In the ref.\cite{1} two types of braiding matrices with
two distinct eigenvalues, i,e. those associated with Temperely-Lieb
algebra (TLA)\cite{2} and their corresponding solutions of
Yang-Baxter equation (YBE)\cite{3,4,5,6,7} had been discussed. Based on the
topological basis \cite{8,9,10} the $4$-dimensional braiding
operators were mapped onto 2-dimensional ones \cite{1,9}. We had shown
that the two types of braiding matrices are related to the extremes
of $L_{1}$-norm of Wigner's D-function\cite{1}. Especially the $2$-d
braiding matrix corresponding the Bell basis (the type-II) is
connected to the maximum of the $L_{1}$-norm, whereas the
permutation and extensions (the type-I) to the minimum. It hints
that the $L_{1}$-norm should relate to the quantum information. It
is natural to extend the discussions in the Ref.\cite{1} to the
solution of YBE with three distinct eigenvalues. Among them, the
most important ones belong to Birman-Wenzl-Murakami algebra
\cite{11,12,13}(BWMA). As is well-known that the forms of braiding
matrices with three distinct eigenvalues were given by references,
say in \cite{14,15}, and the connection with BWMA was shown in
Ref.\cite{15} for both standard and non-standard solutions. In
parallel to the ref.\cite{1}, in this paper we shall firstly set up
the topological base  $|e_{1}\rangle$, $|e_{2}\rangle$, and
$|e_{3}\rangle$ associated with BWMA, then map the $9$-d braiding
matrices to $3$-d forms. The physical application of BWMA is raised
through a familiar model which is different from the model
discussed in Ref. \cite{16}. We shall point out that in general the
extremes of $L_{1}$-norm of the D-functions ($\theta=\pi/2$ and $\pi$)
are related to those of von Neumann entropy. We also take the
spin-$1$ models as examples to favor the statement.

\section{Topological Basis for BWMA}
For self-contain the BWMA relations have been given in the Appendix A through the graphs for three states. Denoting the eigenvalues of a braiding matrix $S$ with three eigenvalues by $\lambda_1$, $\lambda_2$, and $\lambda_3$, where $S$ satisfies braid relation
\begin{equation}
  S_{12}S_{23}S_{12}=S_{23}S_{12}S_{23}\quad (S_1\equiv S_{12}=S\otimes I,\; S_2=S_{23}=I\otimes S)
\end{equation}
and without loss of generality by setting the eigenvalues of $S$ to be $\lambda_1$, $\lambda_2$ and $\lambda_3$ with
\begin{equation}
  \lambda_1\lambda_2=-1,\quad W=\lambda_1+\lambda_2,\quad \lambda_3=\sigma
\end{equation}
we have for $S$ with three distinct eigenvalues.
\begin{equation}
  S-S^{-1}=WI+\frac{1}{\sigma}(I+WS-S^2).\label{Eq1.3}
\end{equation}
Defining
\begin{equation}
  E=\frac{1}{\sigma W}(S^2-WS-I)
\end{equation}
(\ref{Eq1.3}) becomes
\begin{equation}
  S-S^{-1}=W(I-E).
\end{equation}
Where $S$ and $E$ occupy $k-th$ and $(k+1)-th$ sites and satisfy the relations shown in Appendix A, i. e. they form BWM algebra. Noting that a loop takes the value
\begin{equation}
  d=1+\frac{1}{W}(\sigma^{-1}-\sigma).\label{Eq1.6}
\end{equation}
Following the philosophy for T-L algebra to set up the topological basis $|e_1\rangle$ and $|e_2\rangle$ \cite{8,9}, we shall find the uni-orthogonal basis $|e_1\rangle$, $|e_2\rangle$ and $|e_3\rangle$ for $S$ and $E$ such that
\begin{equation}
  S_{12}|e_{\mu}\rangle=\lambda_{\mu}|e_{\mu}\rangle\quad(\mu=1,2,3) \label{Eq1.7}
\end{equation}
with
\begin{equation}
  S_{k,k+1}=\cross\quad\mathrm{and}\quad E_{k,k+1}=\Uniil,\quad d=\Nu=\myloop\quad  \label{Eq1.8}
\end{equation}
where the eigenvalues $\lambda_{\mu}$ may be complex. The graphic expressions \cite{17} of BWMA are shown in Appendix A. To satisfy all the relations for BWMA the base takes the forms:
\begin{eqnarray}
    |e_3\rangle&=&d^{-1}\tsep \label{Eq1.9}\\
    |e_i\rangle&=&f_i\left\{\tcross+\alpha_i\Uu+\beta_i\tsep\right\}\quad (i=1,2)\label{Eq1.10} .
\end{eqnarray}
In terms of the graphic calculations \cite{17} it can be proved that the (\ref{Eq1.7}) together with
\begin{equation}
  \langle e_3|e_i\rangle=0,\quad \langle e_i|e_j\rangle=\delta_{ij}\quad (i,j=1,2),
\end{equation}
lead to the constraints to the parameters $\alpha_i$, $\beta_i$ and normalization constant $f_i$:
\begin{equation}
 \alpha_i=\lambda_i\quad (\lambda_1\lambda_2=-1), \quad \alpha_i+\beta_i d=-\sigma^{-1}\quad (i=1,2)\label{Eq1.12}
\end{equation}
and
\begin{equation}
  f_i=\{d(\lambda_i^2+1)[-\lambda_i^{-1}d^{-1}(\sigma^{-1}+\lambda_i)+\lambda_i^{-1}\sigma+d]\}^{1/2} \label{Eq1.13}
\end{equation}
for $\lambda_{\mu}^{*}=\lambda_{\mu}(\mu=1,2,3)$, i.e. $S^{\dag}=S$ (hermitian), whereas
\begin{equation}
  f_i=\{(d-1)(\lambda_i+\lambda_i^{-1})(\sigma+\lambda_i d+\lambda_i^{-1})\}^{1/2}\label{Eq1.14}
\end{equation}
for $\lambda_{\mu}^{*}=\lambda_{\mu}^{-1}(\mu=1,2,3)$, i.e. $S^{\dag}=S^{-1}$ (unitary).
The (\ref{Eq1.12}) takes the same form for $S$ being hermitian or unitary. The only difference between hermitian and unitary consists in the different normalized constants $f_i$ and the parameters being complex for $S^+=S^{-1}$. The proof can be seen in the Appendix B.

\section{Three-Dimensional matrix forms of $E_{12}$, $E_{23}$, $S_{12}$ and $S_{23}$ for $S^{\dag}=S$}\label{Sec2}
In terms of the uni-orthogonally topological basis the direct calculation gives the $3$-D matrix forms of $E$ and $S$ acting on 1-st and 2-nd sites, 2-nd and 3-rd sites, respectively:
\begin{equation}
\eqalign{
(E_{A})_{\mu\nu}=\langle e_{\mu}|E_{12}|e_{\nu}\rangle \\
(E_{B})_{\mu\nu}=\langle e_{\mu}|E_{23}|e_{\nu}\rangle\quad(\mu,\nu=1,2,3)\\
A_{\mu\nu}=\langle e_{\mu}|S_{12}|e_{\nu}\rangle\\
B_{\mu\nu}=\langle e_{\mu}|S_{23}|e_{\nu}\rangle
}
\end{equation}
where A represents the braiding between the 1-st and 2-nd sites, whereas B for the braiding between the 2-nd and 3-rd sties.
The explicit 3-D matrix forms are shown to be$(\lambda_1\lambda_2=-1)$:
  \begin{equation}\label{Eq2.2}\fl
    E_{A}=\left[\begin{array}{ccc}
    0&&\\
    &0&\\
    &&d\\
    \end{array}\right], \qquad
    A=\left[\begin{array}{ccc}
    \lambda_1&&\\
    &\lambda_2&\\
    &&\lambda_3\\
    \end{array}\right]
    =\left[\begin{array}{ccc}
    \lambda_1&&\\
    &-\lambda_1^{-1}&\\
    &&\sigma\\
    \end{array}\right]
  \end{equation}
  \begin{equation}\fl
    E_{B}=d^{-1}(\lambda_1+\lambda_1^{-1})^{-1}\left[\!
    \begin{array}{lcr}
    (\lambda_1+\lambda_1^{-1})^{-1}f_1^{-2}&-(\lambda_1+\lambda_1^{-1})^{-1}(f_1f_2)^{-1}&f_1^{-1}\\
    -(\lambda_1+\lambda_1^{-1})^{-1}(f_1f_2)^{-1}&(\lambda_1+\lambda_1^{-1})^{-1}f_2^{-2}&-f_2^{-1}\\
    f_1^{-1}&-f_2^{-1}&\lambda_1+\lambda_1^{-1}\\
    \end{array}\!\right]
  \end{equation}
  \begin{equation}\label{Eq2.4}\fl
    B=d^{-1}(\lambda_1+\lambda_1^{-1})^{-1}\left[\begin{array}{lcr}
    \lambda_1^{-1}(\lambda_3(d-1)-\lambda_1^{-1})&(1-\lambda_1^{-1}\lambda_3-\lambda_1\lambda_3d)f_1^{-1}f_2&\lambda_1^{-1}f_1^{-1}\\
    (1-\lambda_1^{-1}\lambda_3-\lambda_1\lambda_3d)f_1^{-1}f_2& \lambda_1(\lambda_3(d-1)+\lambda_1)&\lambda_1f_2^{-1}\\
    \lambda_1^{-1}f_1^{-1}&\lambda_1f_2^{-1}&(\lambda_1+\lambda_1^{-1})\lambda_3^{-1}
    \end{array}\right]
  \end{equation}
  When $\lambda_1=q$,$\lambda_2=-q^{-1}$,$\lambda_3=q^{-2}$, i.e. the standard braiding matrix given in \cite {14,15}, we have
  \begin{equation}\fl
    E_{B}=\left[\begin{array}{ccc}
    d^{-1}(d^2-d-1)&-d^{-\frac{1}{2}}(d^2-d-1)^{\frac{1}{2}}&d^{-1}(d^2-d-1)^{\frac{1}{2}}\\
    -d^{-\frac{1}{2}}(d^2-d-1)^{\frac{1}{2}}&1&-d^{-\frac{1}{2}}\\
    d^{-1}(d^2-d-1)^{\frac{1}{2}}&-d^{-\frac{1}{2}}&d^{-1}\\
    \end{array}\right]
  \end{equation}
  \begin{equation}\fl
    B=\left[\begin{array}{ccc}
    q^{-4}d^{-1}(d-1)^{-1}&-q^{-2}d^{-\frac{1}{2}}(d-1)^{-1}(d^2-d-1)^{\frac{1}{2}}&d^{-1}q^{-1}(d^2-d-1)^{\frac{1}{2}}\\
    -q^{-2}d^{-\frac{1}{2}}(d-1)^{-1}(d^2-d-1)^{\frac{1}{2}}&(d-1)^{-1}(d-2)&d^{-\frac{1}{2}}q\\
    d^{-1}q^{-1}(d^2-d-1)^{\frac{1}{2}}&d^{-\frac{1}{2}}q&d^{-1}q^2
    \end{array}\right]\label{Eq2.6}
  \end{equation}
  The other relations can be sees in Appendix A.

\section{$3$-D Topological Basis of BWMA for  unitary S}\label{Sec3}
When $S$ is unitary with $q$ at root of unity, namely
  \begin{equation} \fl
    \lambda_{\mu}^{*}=\lambda_{\mu}^{-1}\quad (\mathrm{f.g.}\quad \lambda=q^m,\quad q \textrm{ at root of unity, where m may be $\pm$ inegers}),
  \end{equation}
the basis reads ($i=1,2$)
  \begin{equation}
    |e_i\rangle=f_i\left(\tcross+\alpha_i\Uu+\beta_i\tsep\right),
  \end{equation}
  \begin{equation}
    \langle e_i|=f_i^{*}\left(\ntcros+\alpha_i^{*}\Nn+\beta_i^{*}\tncup\right).
  \end{equation}

The matrix forms of $E$ and $S$ read
  \begin{equation} \fl
    E_B=d^{-1}\left[
    \begin{array}{ccc}
    |f_1|^{-2}(\lambda_1+\lambda_1^{-1})^{-2}&-(\lambda_1+\lambda_1^{-1})f_1^{-1}f_2^{*-1}&(\lambda_1+\lambda_1^{-1})^{-1}f_1^{-1}\\
    -(\lambda_1+\lambda_1^{-1})f_1^{*-1}f_2^{-1}& (\lambda_1+\lambda_1^{-1})^{-2}|f_2|^{-2}&-(\lambda_1+\lambda_1^{-1})^{-1}f_2^{-1}\\
    (\lambda_1+\lambda_1^{-1})f_1^{*-1}&-(\lambda_1+\lambda_1^{-1})^{-1}f_2^{*-1}&1\\
    \end{array}\right],
  \end{equation}
  \begin{equation}
    E_A=\left[\begin{array}{ccc}
    0&&\\
    &0&\\
    &&d\\
  \end{array}\right],
  \end{equation}
  \begin{equation} \fl
  \eqalign{
    S_{23}|e_1\rangle=&d^{-1}(\lambda_1+\lambda_1^{-1})^{-1}\cdot\\
    &\left\{\lambda_1^{-1}
    [\lambda_3(d-1)-\lambda_1^{-1}]|e_1\rangle-(1+\lambda_1\lambda_3+\lambda_1^{-1}\lambda_3d)f_1f_2^{-1}|e_2\rangle
    +\lambda_1^{-1}f_1^{*-1}|e_3\rangle\right\},\\
    S_{23}|e_2\rangle=&d^{-1}(\lambda_1+\lambda_1^{-1})\cdot\\
    &\left\{(1-\lambda_1^{-1}\lambda_3-\lambda_1\lambda_3d)f_1^{-1}f_2|e_1\rangle
    +\lambda_1[\lambda_3(d-1)+\lambda_1]|e_2\rangle
    -\lambda_1f_2^{*-1}|e_3\rangle\right\},\\
    S_{23}|e_3\rangle=&d^{-1}(\lambda_1+\lambda_1^{-1})^{-1}\cdot\\
    &\left\{\lambda_1^{-1}f_1^{-1}|e_1\rangle
    +\lambda_1f_2^{-1}|e_2\rangle
    +(\lambda_1+\lambda_1^{-1})\lambda_3^{-1}|e_3\rangle\right\}.
  }
  \end{equation}
  The 3-D matrix form is found:
  \begin{equation}\label{Eq3.7}
    A=\left[
    \begin{array}{ccc}
    \lambda_1&&\\
    &-\lambda_1^{-1}&\\
    &&\lambda_3\\
    \end{array}\right],
  \end{equation}
  \begin{equation}\label{Eq3.8}\fl
    B=\left[
    \begin{array}{ccc}
    \lambda_1^{-1}[\lambda_3(d-1)-\lambda_1^{-1}]&(1-\lambda_1^{-1}\lambda_3-\lambda_1\lambda_3d)f_1^{-1}f_2&\lambda_1^{-1}f_1^{-1}\\
    -(1+\lambda_1\lambda_3+\lambda_1^{-1}\lambda_3d)f_1f_2^{-1}&\lambda_1[\lambda_3(d-1)+\lambda_1]&\lambda_1f_2^{-1}\\
    \lambda_1^{-1}f_1^{*-1}&-\lambda_1f_2^{*-1}&(\lambda_1+\lambda_1^{-1})\lambda_3^{-1}\\
    \end{array}\right],
  \end{equation}
  with
  \begin{equation}
    \begin{array}{rcl}
    \because S-S^{-1}&=&W(I-E)=W(I-E^{+})\\
       S^{\dag}-(S^{-1})^{\dag}&=&W^*(I-E^{\dag})\\
    S^{-1}-S&=&-W(I-E^{\dag})\\
    S-S^{-1}&=&W(I-E^{\dag})\\
    \end{array}
  \end{equation}
   i.e. $E$ must be real.
  \begin{equation}  \fl
    W=\lambda_1+\lambda_2=\lambda_1-\lambda_1^{-1}\qquad W^*=\lambda_1^*-\lambda_1^{*-1}=-\lambda_1+\lambda_1^{-1}=-W.
  \end{equation}
  Following Ref. \cite{15} for $q$ at root of unity it allows 'non-standard' braiding matrices, say, for $B_n$ some of $\lambda_k=q$, and for the others $\lambda_i=-q^{-1}$, then the general form of $\sigma$ is
  \begin{equation}
    \sigma=\lambda_3=q^{m} \quad(|m|\le 2n).
  \end{equation}
  It leads to
  \begin{equation}
  \begin{array}{rcl}
  d&=&1-\frac{1}{\lambda_1+\lambda_2}(\sigma-\sigma^{-1})=1-\frac{q^{m}-q^{-m}}{q-q^{-1}}\\
   &=&1-\frac{\sin{m\alpha}}{\sin{\alpha}}  \quad(\frac{\pi}{N}=\alpha, q=e^{i\frac{\pi}{N}})\\
   &=& 1-[m]_q\\
  \end{array}
  \end{equation}
 where $|m|$ is the difference between the positive power number and negative ones in the power of $q$ in the third eigenvalues of $S$ for the fundamental representations of $B_n$, $D_n$ and $C_n$.

\section{Spin-1 model associated with BWM algebra}
As was pointed out in \cite{14,15} that for $B_n$ algebra the corresponding braiding matrix $S$ has three distinct eigenvalues $\lambda_1=q$,$\lambda_2 =-q^{-1}$ whereas as the third one is given by
 \begin{equation}
 \lambda_{3}=\sigma=\prod_{k=1} u_{k}^{-2}
 \end{equation}\label{Eq5.1}
 where $u_{k}$ can be either $q$ (standard solution) or $-q^{-1}$ for nonstandard solution, so in general, we are able to take
 \begin{equation}
 \sigma=\lambda_{3}=q^{m}
 \end{equation}
 where $m$ can be arbitrary integers. To satisfy the spectral parameter dependent Yang-Baxter equation, the corresponding $\check{R}(x)$-matrix takes the form \cite{15}
\begin{equation} \fl
  \check{R}_{\alpha}(x)=(x-1)(x-x_{\alpha})S+Wx(x-1)E-Wx(x-x_{\alpha})I, \quad \alpha=a,b
 \end{equation}
where $W=q-q^{-1}$, $x_a=-q\sigma^{-1}=-q^{(1-m)}$, $x_b=q^{-1}\sigma^{-1}=q^{-(m+1)}$.\\
In order to obtain the rational limit of the $\check{R}(x)$ the type-I solution is given by
\begin{equation}
  \check{R}_b(x)=\mathrm{const}\,\left\{I-\left(\frac{x-1}{Wx}\right)S-\frac{x-1}{x-q^{-(m+1)}}E\right\}
\end{equation}
Under the rational limit and $q\rightarrow1$ we set $h\rightarrow0$ for
\begin{equation}
  x=e^{hu} \quad\mathrm{and}\quad q=e^{-\frac{h\gamma}{2}}
\end{equation}
it leads to
\begin{equation}
  \check{R}_b(x)|_{h\rightarrow0}=\mathrm{const}\cdot\left\{I+\frac{u}{\gamma}S-\frac{u}{u-\gamma(\frac{m+1}{2})}E\right\}|_{q=1}
\end{equation}
that under the rescaling $\frac{u}{\gamma}\rightarrow u$ becomes
\begin{equation}
  \check{R}_b(u)|_{h\rightarrow0}\equiv\check{R}(u)=\mathrm{const}\cdot\left\{I+uT-\left(\frac{u}{u-\beta}\right)M\right\}
\end{equation}
where $T=S|_{q=1}$, $M=E|_{q=1}$, and $\beta=\frac{m+1}{2}$.\\
Following the standard way of Baxter \cite{4} the Hamiltonian can be given by
\begin{equation}
  H_{k,k+1}\propto\frac{\partial\check{R}_{k,k+1}(u)}{\partial u}|_{u=0}=T_{k,k+1}+\frac{1}{\beta}M_{k,k+1}
\end{equation}
Here the $k-th$ site has been indicated explicitly because $T$ and $M$ occupy the $k$-th and $(k+1)$-th sites. The $M_{k,k+1}$ is a new term added to the permutation-like operator $T_{k,k+1}$ due to BWMA.\\
In particular when $m=-3$, i.e. $\sigma=q^{-3}$ we have
\begin{equation}
  H_{k,k+1}=T_{k,k+1}-M_{k,k+1}
\end{equation}
It is worthy noting that $m=-3$ means the solution of YBE being "nonstandard"\cite{15}

The M works in the block for ${S_{z}(k)+S_{z}(k+1)=0}$,
 where $S_{z}(k)$ means the third component of spin-$1$ at $k$-th site, or in terms of the basis for spin-$1$:
\begin{equation} \fl
\eqalign{
  T_{k,l}=&\left|1,1\right\rangle_{k.l\
  k,l}\left\langle1,1\right|+\left|-1,-1\right\rangle_{k.l\ k,l}\left\langle-1,-1\right|+\left|0,0\right\rangle_{k.l\ k,l}\left\langle0,0\right| \\
        &+\left|1,0\right\rangle_{k,l\ k,l}\left\langle0,1\right|+\left|0,1\right\rangle_{k.l\ k,l}\left\langle1,0\right|+\left|0,-1\right\rangle_{k.l\ k,l}\left\langle-1,0\right| \\
        &+\left|-1,0\right\rangle_{k.l\ k,l}\left\langle0,-1\right|+\left|1,-1\right\rangle_{k.l\ k,l}\left\langle-1,1\right|+\left|-1,1\right\rangle_{k.l\ k,l}\left\langle1,-1\right|    \label{Eq5.11}
}
\end{equation}
where the $|S_z,S_z\rangle_{k,l}$ represents the state which occupies $k$-th and $l$-th sites. In general, $l$ is not necessary to be $k+1$.
\begin{equation} \fl
\eqalign{
  M_{k,l}=&\left|0,0\right\rangle_{k.l\ k,l}\left\langle0,0\right|+\big(\left|1,-1\right\rangle_{k,l}+\left|-1,1\right\rangle_{k,l}\big)\left(_{k,l}\left\langle1,-1\right|+_{k,l}\left\langle-1,1\right|\right)
  \nonumber\\
  &+e^{i\varphi}\left|0,0\right\rangle_{k,l}\left(_{k,l}\left\langle1,-1\right|+_{k,l}\left\langle-1,1\right|\right)
  \nonumber\\
  &+e^{-i\varphi}\left(\left|1,-1\right\rangle_{k,l}+\left|-1,1\right\rangle_{k,l}\right)_{k,l}\left\langle0,0\right|    \label{Eq5.12}
  }
\end{equation}
where $\varphi$ is real and $T^2_{k,l}=I$, $M^2_{k,l}=3M_{k,l}$ (i.e. $d=3$, here the braiding matrix is not unitary), then $T_{k,k+1}$ and $M_{k,k+1}$ satisfy BWM algebra.
It is easy to check that for spin-$1$ and $l=k+1$, in terms of \\
$S_{+}=\sqrt{2}\left[
\begin{array}{ccc}
0&1&0\\
0&0&1\\
0&0&0
\end{array}
\right]$,
$S_{-}=\sqrt{2}\left[
\begin{array}{ccc}
0&0&0\\
1&0&0\\
0&1&0
\end{array}
\right]$
and
$S_{3}=\sqrt{2}\left[
\begin{array}{ccc}
1&0&0\\
0&0&0\\
0&0&-1
\end{array}
\right]$,\\
 we have folowing Baxter \cite{4},
\begin{equation}\label{Eq5.13}
  \frac{\partial\check{R}_{k,k+1}(u)}{\partial u}|_{u=0}=H_{k,k+1}=T_{k,k+1}-M_{k,k+1}=\overrightarrow{S}_k\cdot\overrightarrow{S}_{k+1}
\end{equation}
for $\varphi=\pi$. Therefore, for spin-$1$
\begin{equation}\label{Eq5.14}
  H=\sum_{k=1}^{N}H_{k,k+1}=\sum_{k=1}^{N}\left(T_{k,k+1}-M_{k,k+1}\right)=\sum_{k=1}^{N}=\overrightarrow{S}_k.\overrightarrow{S}_{k+1}
\end{equation}
 It is interesting to note that the Hamiltonian (\ref{Eq5.14}) is known well for long time. Especially, it is not permutation operator, but plays role, say, in the Haldane conjuncture \cite{17}. Here we have obtained $(\overrightarrow{S}_k.\overrightarrow{S}_{k+1})$ for spin-1 whose Hamiltonian is associated with BWM algebra.
Furthermore, as a demonstration example we show how to solve the model with $N=4$ in terms of the topological basis given by (\ref{Eq1.9}), (\ref{Eq1.10}).

Graphically the Hamiltonian can be expressed by the operators for $N=4$
\begin{equation}\label{Eq5.15}
  \hat{\mathscr{H}}=J\sum_{k=1}^{4}\left(\Crosss-\Uniil\right)=J\sum_{k=1}^{4}\left(\hat{T}_{k,k+1}-\hat{M}_{k,k+1}\right)
\end{equation}
Its 9-d representation is given by acting the operator $\mathscr{H}$ on 9-d basis. Whereas acting (\ref{Eq5.15}) on the 3-d basis (\ref{Eq1.9}),(\ref{Eq1.10}) for $m=-3$, we find
\begin{eqnarray}
  \mathscr{H}\left|e_1\right\rangle&=&\frac{J}{3}\left(-3\left|e_1\right\rangle+\sqrt{15}\left|e_2\right\rangle\right)\nonumber\\
  \mathscr{H}\left|e_2\right\rangle&=&\frac{J}{3}\left(\sqrt{15}\left|e_1\right\rangle-9\left|e_2\right\rangle+4\sqrt{3}\left|e_3\right\rangle\right)\\
  \mathscr{H}\left|e_3\right\rangle&=&\frac{J}{3}\left(4\sqrt{3}\left|e_2\right\rangle-12\left|e_3\right\rangle\right)\nonumber
\end{eqnarray}
The $\mathscr{H}$ can be diagonalized in terms of the eigenstates $\left|g_{\mu}\right\rangle$: ($\mu=1,2,3$)
\begin{equation} \label{Eq5.17}
  \mathscr{H}\left|g_{\mu}\right\rangle=E_{\mu}\left|g_{\mu}\right\rangle
\end{equation}
where
\begin{equation*}
  E_1=-6J, \quad E_2=-2J, \quad E_3=0\nonumber\\
\end{equation*}
and
\begin{eqnarray}
  \left|g_1\right\rangle&=&\frac{1}{\sqrt{6}}\left(\left|e_1\right\rangle-\sqrt{15}\left|e_2\right\rangle+2\sqrt{15}\left|e_3\right\rangle\right)\nonumber\\
  \left|g_2\right\rangle&=&\frac{1}{2\sqrt{3}}\left(\sqrt{5}\left|e_1\right\rangle-\sqrt{3}\left|e_2\right\rangle-2\left|e_3\right\rangle\right)\nonumber\\
  \left|g_3\right\rangle&=&\frac{1}{3}\left(\sqrt{5}\left|e_1\right\rangle+\sqrt{3}\left|e_2\right\rangle+\left|e_3\right\rangle\right)  \label{Eq5.18}
\end{eqnarray}
How to extend the approach to any $N$ by using the topological basis more than four sites to solve the eigenvalues problem with the help of topological basis is far beyond the current discussion. Here we only discuss a four spin$-1$ model which may be a hint to look for how to solve the $N$-site chain problem based on the topological basis.\\

\section{Four Spin Model}
The relations (\ref{Eq5.11}) and (\ref{Eq5.12}) are defined for any $k$ and $l$. To obtain the Hamiltonian (\ref{Eq5.13}), the nearest neighborhood has been imposed through putting $l=k+1$. However for any $i$ and $j$, the operator $S_{ij}$ can be recast to
\begin{eqnarray}
  S_{k,l}=&\left|1,1\right\rangle_{kl\
  kl}\left\langle1,1\right|+\left|-1,-1\right\rangle_{kl\ kl}\left\langle-1,-1\right|+\left|0,0\right\rangle_{kl\ kl}\left\langle0,0\right|\nonumber\\
        &+\left|1,0\right\rangle_{kl\ kl}\left\langle0,1\right|+\left|0,1\right\rangle_{kl\ kl}\left\langle1,0\right|+\left|0,-1\right\rangle_{kl\ kl}\left\langle-1,0\right|\nonumber\\
        &+\left|-1,0\right\rangle_{kl\ kl}\left\langle0,-1\right|.  \label{Eq6.1}
\end{eqnarray}

It can be checked in terms of (\ref{Eq5.11}), (\ref{Eq5.12})and (\ref{Eq6.1}), it holds for any $i$ and $k$:
\begin{equation}\label{Eq6.2}
  \overrightarrow{S}_{i}\cdot\overrightarrow{S}_{k}=T_{i,k}-M_{i,k}
\end{equation}
where at any $i$-th site,
\begin{eqnarray}
 S_{+}=\sqrt{2}\left(\left|1\right\rangle\left\langle0\right|+\left|0\right\rangle\left\langle-1\right|\right), \nonumber\\ S_{-}=\sqrt{2}\left(\left|0\right\rangle\left\langle1\right|+\left|-1\right\rangle\left\langle0\right|\right), \nonumber\\
 S_{3}=\left|1\right\rangle\left\langle1\right|-\left|-1\right\rangle\left\langle-1\right|.
\end{eqnarray}\label{Eq6.3}
Noting that (\ref{Eq6.2}) is valid for any {i} and {k}. Because ${S_{i}}^2=S_{i}(S_{i}+1)=2$ and total spin
\begin{equation}
{\overrightarrow{S}}^2=\left(\sum_{i=1}^{N}\overrightarrow{S}_{i}\right)^2=\sum_{i=1}^{N}{\overrightarrow{S}_{i}}^2+\sum_{i\ne j}^{N}(\overrightarrow{S}_i.\overrightarrow{S}_{j})=2N+2\sum_{i<j}^{N}(\overrightarrow{S}_i.\overrightarrow{S}_{j}).
\end{equation}
for $N=4$ we have
\begin{equation}
\overrightarrow{S}^2=2N+2\sum_{i<j}^{4}(T_{i,j}-M_{i,j}).
\end{equation}\label{Eq6.4}
There appear the additional terms other than $(T_{i,i+1}-M_{i,i+1})$, i.e. the term $(T_{13}-M_{13})$ and $(T_{24}-M_{24})$. Taking into account of
\begin{equation}
(T_{1,3}-M_{1,3})\left|g_{k}\right\rangle=(T_{2,4}-M_{2,4})\left|g_{k}\right\rangle,
\end{equation}
we find
\begin{equation}\label{Eq6.5}
\sum_{i=1}^{2}(T_{i,i+2}-M_{i,i+2})\left|g_{k}\right\rangle=\mu_{k}\left|g_{k}\right\rangle
\end{equation}
where $\mu_{1}=1$,$\mu_{2}=-1$, $\mu_{3}=-3$, $(k=1,2,3)$
and
\begin{equation}  \fl
\eqalign{
\overrightarrow{S}^{2}\left|g_{k}\right\rangle &=\{\sum_{i=1}^{4}\overrightarrow{S}_i^{2}+2\sum_{i=1}^{4}(T_{i,i+1}-M_{i,i+1})+2\sum_{i=1}^{2}(T_{i,i+2}-M_{i,i+2})\}\left|g_{k}\right\rangle \nonumber\\
&=2(
4+E_{k}J^{-1}+2\mu_{k})\left|g_{k}\right\rangle  \label{Eq6.6}
}
\end{equation}
Substituting (\ref{Eq5.17}) and (\ref{Eq6.5}) into (\ref{Eq6.6}),we obtain for $N=4$
 \begin{equation}
 \overrightarrow{S}^{2}\left|g_{k}\right\rangle=0\left|g_{k}\right\rangle
 \end{equation}\label{Eq6.7}

It is interesting to note that the topological eigenstates $\left|g_{k}\right\rangle (k=1,2,3)$ are spin singlet. From the point of view of Lie algebra, the direct product of four spin 1 can be decomposed to 5 subspaces; however, only the singlet with multiplicities three is the eigenstates of $\mathscr{H}$ for $N=4$.\\

\section{$D_{mm'}^{J}(\theta,\varphi)$ function as the N-dimensional solutions of YBE}
The Yang-Baxterization (parametrization) of $N^2\times N^2$ braiding matrices can be made in the standard way, say, following Jimbo, Jones, and others $\cdots$ \cite{14,19,20}. For $N\times N$ YBE there is another way to introduce spectral parameter to a given braiding matrix. The basic idea comes from the Wigner D-function \cite{21}.
\begin{equation} \fl
  D(n)=e^{i\theta\overrightarrow{m}\cdot\overrightarrow{J}}=e^{\xi J_{+}-\xi^{*}J_{-}}, \quad \xi=-\frac{\theta}{2}e^{-i\varphi},\quad J_{\pm}=J_{x}\pm iJ_{y}, \quad J_0=J_3
\end{equation}
\begin{equation} \fl
  \overrightarrow{m}=(\sin\varphi,-\cos\varphi,0),\quad \overrightarrow{n}=(\sin\theta\cos\varphi,\sin\theta\sin\varphi,\cos\theta),\quad n=-\tan\frac{\theta}{2}e^{-i\varphi}
\end{equation}
It can be proved that
\begin{equation}
  D^{J}(\theta_1,0)D^{J}(\theta_2,\varphi)D^{J}(\theta_3,0)=D^{J}(\theta_3,\varphi)D^{J}(\theta_2,0)D^{J}(\theta_1,\varphi)
\end{equation}
provided it holds
\begin{equation}
  \cos\varphi=\frac{1}{2}\left[\frac{\tan\frac{\theta_1}{2}+\tan\frac{\theta_3}{2}-\tan\frac{\theta_2}{2}} {\tan\frac{\theta_1}{2}\tan\frac{\theta_2}{2}\tan\frac{\theta_3}{2}}-1\right] \label{Eq:6.4}
\end{equation}
When $\theta_1=\theta_2=\theta_3=\theta$, (\ref{Eq:6.4}) reduces to \cite{22}
\begin{equation}\label{Eq63}
  \cos\varphi=\frac{\cos\theta}{1-\cos\theta}
\end{equation}
which is the condition that (\ref{Eq:6.4}) reduces to braid relation.
Denoting by
\begin{equation}
  A(\theta)=D^{J}(\theta,0), \quad B(\theta,\varphi)=D^{J}(\theta,\varphi)
\end{equation}
the ($N\times N$) YBE reads \cite{23}
\begin{equation}
  A(\theta_1)B(\theta_2,\varphi)A(\theta_3)=B(\theta_3,\varphi)A(\theta_2)B(\theta_1,\varphi)\\
 \end{equation}
or
\begin{equation}
A(x)B(xy)A(y)=B(y)A(xy)B(x), cos\theta=\frac{(1-x)}{\sqrt{2(1+x^{2})}}
\end{equation}
is satisfied under (\ref{Eq:6.4}) .

The examples for $N=2$ had been given in the ref.[1]. Here it holds for any ($2J+1$)-dimensional representations.

Let us condiser spin-1 example. The $D^{1}(\theta,\varphi)$ takes the form \cite{21}
\begin{equation}\label{Eq:6.7}
  D^{1}_{mm'}(\theta,\varphi)=\left(\begin{array}{ccc}
                                       \frac{1+\cos\theta}{2} & -\frac{\sin\theta}{\sqrt{2}}e^{-i\varphi} & \frac{1-\cos\theta}{2}e^{-2i\varphi} \\
                                       \frac{\sin\theta}{2}e^{i\varphi} & \cos\theta & -\frac{\sin\theta}{\sqrt{2}} \\
                                       \frac{1-\cos\theta}{2}e^{2i\varphi} & \frac{\sin\theta}{\sqrt{2}}e^{i\varphi} & \frac{1+\cos\theta}{2} \\
                                     \end{array}\right)
\end{equation}
The 3-d braiding relation is given by
\begin{equation}
  A(\theta)B(\theta,\varphi)A(\theta)=B(\theta,\varphi)A(\theta)B(\theta,\varphi)
\end{equation}
where
\begin{equation}
  \cos\varphi=\frac{\cos\theta}{1-\cos\theta}
\end{equation}
\begin{equation}\label{Eq:6.10}
  A=D^{1}(\theta,\varphi=0),\quad B=D^{1}(\theta,\varphi)
\end{equation}
For the type-I, i.e. for $\varphi=2\pi/3$, $\theta=\pm\pi$ in (\ref{Eq:6.7}), after the unitary transformation, (\ref{Eq:6.7}) becomes
\begin{equation}\label{Eq:6.11}  \fl
\eqalign{
  A_{\rm{I}}=V^{\dag}A(\theta=\pi)V=-\left(\begin{array}{ccc}
                                      1 & 0 & 0 \\
                                      0 & -1 & 0 \\
                                      0 & 0 & 1 \\
                                    \end{array}\right),\quad
  V=\left(\begin{array}{ccc}
    \frac{1}{2} & \frac{1}{\sqrt{2}} & \frac{1}{2} \\
    \frac{i}{\sqrt{2}} & 0 & -\frac{i}{\sqrt{2}} \\
    -\frac{1}{2} & \frac{1}{\sqrt{2}} & -\frac{1}{2} \\
  \end{array}\right)                                   \\
 B_{\rm{I}}=V^{\dag}B(\theta=-\pi,\varphi=2\pi/3)V=-\frac{1}{4}\left(\begin{array}{ccc}
  1 & i\sqrt{6} & -3 \\
  -i\sqrt{6} & 2 & -i\sqrt{6} \\
  -3 & i\sqrt{6} & 1
\end{array}\right)
}
\end{equation}
On the other hand, for $S^+=S$ on substituting
\begin{equation}
  \lambda_1=q,\quad \lambda_2=-q^{-1}, \quad \lambda_3=q^s\quad (q\rightarrow1)
\end{equation}
into (\ref{Eq2.2}) and (\ref{Eq2.4}) the ($3\times 3$) matrices $A$ and $B$ given by the topological basis and on account of
\begin{equation}\fl
  d=1-\left.\frac{q^s-q^{-s}}{q-q^{-1}}\right|_{q\rightarrow1}=1-s,\quad f_1=[2S(S-3)]^{\frac{1}{2}},\quad f_2=[2S(S-1)]^{\frac{1}{2}}
\end{equation}
we obtain for $s=-3$
\begin{equation} \fl
  A_{\rm{I}}'=-\left(\begin{array}{ccc}
    q & & \\
    & -q^{-1} & \\
    & & q^s\\
  \end{array}\right)_{q\rightarrow1},\quad
  B_{\rm{I}}'=-\frac{1}{4}\left(\begin{array}{ccc}
    1 & -\sqrt{6} & 3\\
    -\sqrt{6} & 2 & \sqrt{6} \\
    3 & \sqrt{6} & 1 \\
  \end{array}\right)
\end{equation}
that under similar transformation becomes
\begin{equation}\label{Eq:6.15} \fl
  T^{\dag}A_{\rm{I}}'T=A_{\rm{I}}',
  \quad T^{\dag}B_{\rm{I}}'T=-\frac{1}{4}\left(\begin{array}{ccc}
    1 & i\sqrt{6} & -3\\
    -i\sqrt{6} & 2 & -i\sqrt{6} \\
    -3 & i\sqrt{6} & 1\\
  \end{array}\right),
  \quad T^{\dag}=\left(\begin{array}{ccc}
    1 & 0 & 0\\
    0 & i & 0\\
    0 & 0 & -1\\
  \end{array}\right)
\end{equation}
(\ref{Eq:6.15}) is identified with (\ref{Eq:6.11}).
Namely, as was pointed out for $S^{\dag}=S$ that the type-I ($3\times 3$) braiding matrices based on the topological basis are the same as those given by $D^{1}(\theta,\varphi)$ with $\varphi=\frac{2\pi}{3}, \theta=\pm\pi$.

For the type-II, i.e. $\varphi=\pm\pi/2$, $\theta=\pm\pi/2$ with the same transformation $V$ as given by (\ref{Eq:6.11}) the $D^{1}$-function gives
\begin{equation}\label{Eq6.16}
  A=\left(\begin{array}{ccc}
    -i & 0 & 0\\
    0  & 1 & 0\\
    0  & 0 & i\\
  \end{array}\right),
  \quad B=\frac{1}{2}\left(\begin{array}{ccc}
    1 & -\sqrt{2} & 1\\
    \sqrt{2} & 0 & -\sqrt{2} \\
    1 & \sqrt{2} & 1\\
  \end{array}\right)
\end{equation}
On the other hand, for $S^{\dag}=S^{-1}$ the ($3\times 3$) braiding matrices (\ref{Eq3.7}) and (\ref{Eq3.8}) based on the topological basis and on account of
\begin{equation}
  \lambda_1=e^{i\pi/4},\quad \lambda_2=-\lambda_1^{-1}=-e^{-i\pi/4},\quad \lambda_3=\sigma=-e^{i\pi/4}
\end{equation}
are equal to $(f_1=\frac{1}{\sqrt{2}}, f_2=\frac{1}{2})$
\begin{equation}
  A_{\rm{II}}=\left(\begin{array}{ccc}
    -i & 0 & 0 \\
     0 & 1 & 0 \\
     0 & 0 & i \\
  \end{array}\right),
  B_{\rm{II}}=-\left(\begin{array}{ccc}
    -\frac{1}{2} & \frac{i}{\sqrt{2}} & \frac{1}{2} \\
    \frac{i}{\sqrt{2}} & 0 & \frac{i}{\sqrt{2}} \\
    \frac{1}{2} & \frac{i}{\sqrt{2}} & -\frac{1}{2} \\
  \end{array}\right)
\end{equation}
that after the transformation becomes
\begin{equation}\label{Eq6.18}
  T^{\dag}A_{\rm{II}}T=\left(\begin{array}{ccc}
    -i & 0 & 0 \\
     0 & 1 & 0 \\
     0 & 0 & i \\
  \end{array}\right), \quad
  T^{\dag}B_{\rm{II}}T=\left(\begin{array}{ccc}
    \frac{1}{2} & -\frac{1}{\sqrt{2}} & \frac{1}{2} \\
    \frac{1}{\sqrt{2}} & 0 & -\frac{1}{\sqrt{2}} \\
    \frac{1}{2} & \frac{1}{\sqrt{2}} & \frac{1}{2} \\
  \end{array}\right)
\end{equation}
(\ref{Eq6.18}) are the same as those given by $D^{1}$-function, subjecting to the unitary transformation V for the type-II as shown in (\ref{Eq6.16}).

In short conclusion it turns out that the Yang-Baxterization for ($N\times N$) YBE takes the different way from the ($N^2\times N^2$) representations. The resultant solutions of YBE are simply the Wigner's $D^{J}$-functions. When $\theta_1=\theta_2=\theta_3=\theta$ it reduces to the braiding matrix. Two examples for spin-$\frac{1}{2}$ and spin-$1$ have been checked in Ref.[1] and in this section, respectively. However, the explicit correspondences between $(N\times N)$ and ($N^2\times N^2$) braiding matrices for any $J$ for type-II are still a challenge problem.

It is emphasized that the Yang-Baxterization for ($N\times N$) solutions of YBE through the D-function is a quite new parametrization way based on the topological basis in different from Refs.\cite{14}.


\section{Relationship between Extremes of $L_{1}$-norm of D-function and von Neumann Entropy}
As was pointed out in Ref[1] that the extreme of $D^{j}$-functions may take multiple values for the different $j$ with $2j+1$ components. However, for any half-integer $j$ (also for $j$ integer, but $J_z=0$ should be excluded, obviously) there exist the common maximum $\theta=\pm\pi/2$ and minimum $\theta=\pm\pi$.
The Bell basis can be regarded as a linear combination of the natural basis $|\psi_0\rangle=(|\uparrow\uparrow\rangle, |\downarrow\uparrow\rangle, |\uparrow\downarrow\rangle, |\downarrow\downarrow\rangle)^T$, i.e. subject to the rotational transformation with $\theta=\pi/2$ \cite{24,25}. In general

\begin{equation}
|\Psi(\theta,\varphi)\rangle=D^j(\theta,\varphi)|\Psi_0\rangle
\end{equation}
where $|\Psi_0\rangle$ is an $(2j+1)$ vector serving as natural basis. For $j=1/2$, $|\Psi_0\rangle=|\Psi\rangle$ and $|\Psi(\pm\pi/2, \pm\pi/2)\rangle$ stands for the Bell basis \cite{24,25} which possesses the maximum of entanglement. For any $\theta$ other than $\pm\pi/2$ (the maximum of $L_1$-norm of $D^{1/2}(\theta)$), it decreases the entanglement. We naturally think the extreme of D-function should indicate the entangling degree. One of the descriptions of entangling degree is von Neumann entropy. We should show that the extreme points of the entropy and $L_1$-norm of D-function shares the common $\theta$-values.
The von Neumann entropy is defined by
\begin{equation}
S(\rho)=-Tr(\rho \textrm{log}_2\rho)
\end{equation}
where $\rho$ is the reduced density matrix of a quantum state. Following  the YBE in $(2J+1)$ dimensions the Schmidt decomposition of a entangled state in $(2J+1)$ dimensions, (i.e. acting on topological basis) $|\Psi(\theta,\varphi)\rangle$ can be written in the form for a fixed m:
\begin{equation}
|\Psi\rangle_{ab}=\sum_{m'}{D^{j}_{m'm}(\theta,\varphi)|a_{m'}\rangle|b_{m'}\rangle}
\end{equation}
where m and m' take over $N=2j+1$, and $|a\rangle$, $|b\rangle$ are "natural states".
Since the reduced density operator of the subsystem a is
\begin{equation}
\rho^{m}_{a}=\sum_{m'}|D^j_{m'm}(\theta,\varphi)|^2|a_{m'}\rangle \langle a_{m'}|
=\sum_{m'}|d^j_{m'm}(\theta)|^2|a_{m'}\rangle \langle a_{m'}|
\end{equation}
and for a fixed m
\begin{equation}\label{Eq6.19}
S(\rho_a)\leq2 \textrm{log}_2(\sum_{m'}|d^j_{m'm}(\theta)|)
\end{equation}
the extremes of $S(\rho_a)$ are shown in the following examples in comparison to the $L_1$-norm, i.e. $f_m=\sum_{m'}|d^j_{m'm}(\theta)|$.

Example:
When $\theta=\pm\pi$, the bipartite state is a direct product sate which is separable, therefore $S(\rho_a)$ and $f$ arrive at their minimum value simultaneously and the equality of (\ref{Eq6.19}) holds. When $\theta=\pm\pi/2$, $|d^{1/2}_{-1/2\,1/2}|=|d^{1/2}_{1/2\,1/2}|=1/\sqrt{2}$, the bipartite state reaches the maximum of entanglement. We have $S(\rho_a)=2\log_2f=\log_22=1$ shown in Fig.1 and Fig.2.

\begin{figure}[!htb]
\centering
\includegraphics[width=0.5\columnwidth,clip]{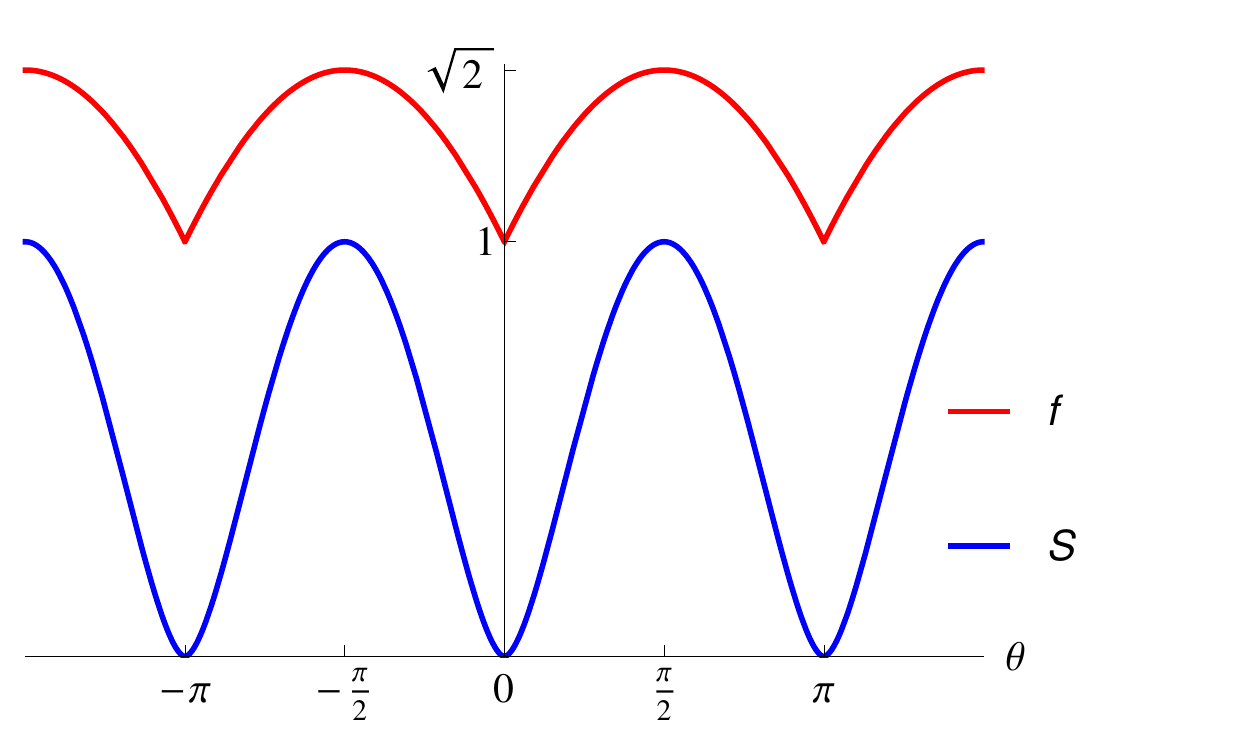}
\caption{}
\end{figure}

\begin{figure}[!htb]
\centering
\includegraphics[width=0.5\columnwidth,clip]{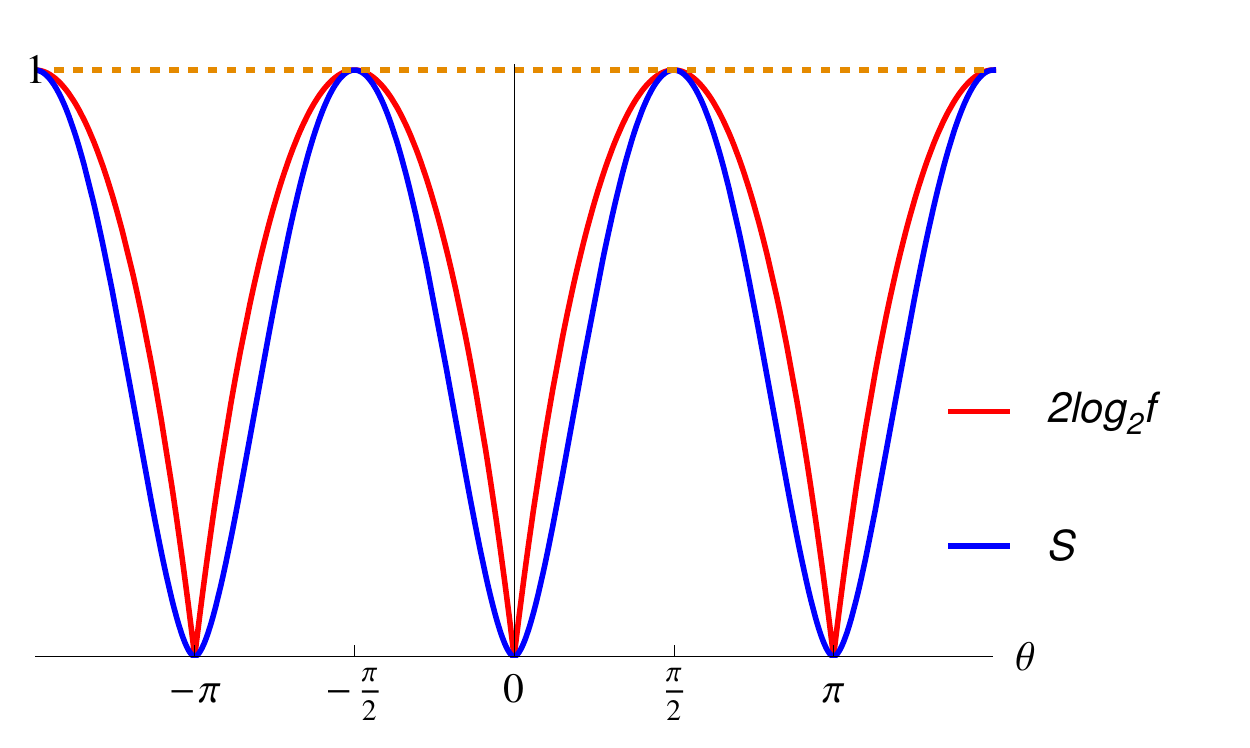}
\caption{}
\end{figure}

Example 2: {$j=1,\ m=\pm1$}


When $\theta=\pm\pi$, the state is a separable, $S(\rho_a)$ and $f$ arrive at their minimum value simultaneously and the equality of (\ref{Eq6.19}) holds. When $\theta=\pm\pi/2$, $S(\rho_a)$ and $f$ reach their maximum value simultaneously. Of course, the maximum value of $\sum_m'|d^j_{m'm}|$ is not the same as of $S(\rho_a)$, since $S(\rho_a)<2\log_2f<\log_23$, but both of them occur at $\theta=\pm\pi/2$, see Fig.3 and Fig.4.
\begin{figure}[!htb]
\centering
\includegraphics[width=0.5\columnwidth,clip]{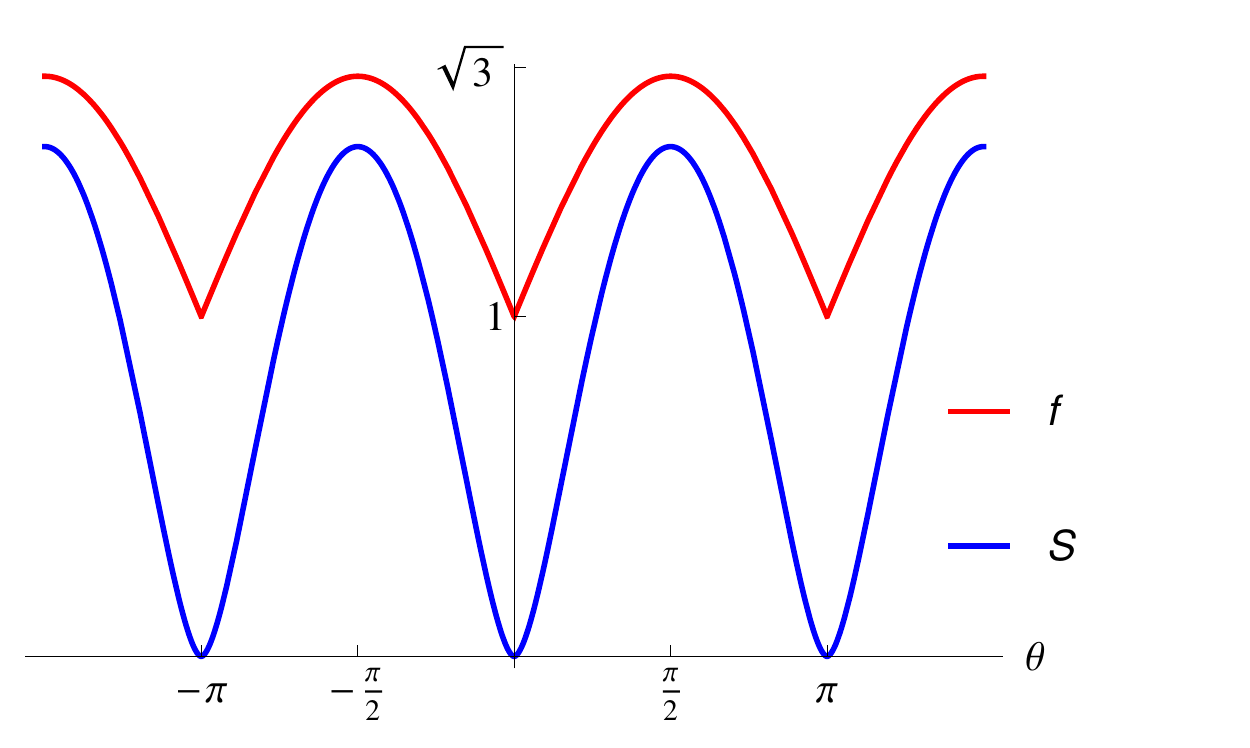}
\caption{}
\end{figure}

\begin{figure}[!htb]
\centering
\includegraphics[width=0.5\columnwidth,clip]{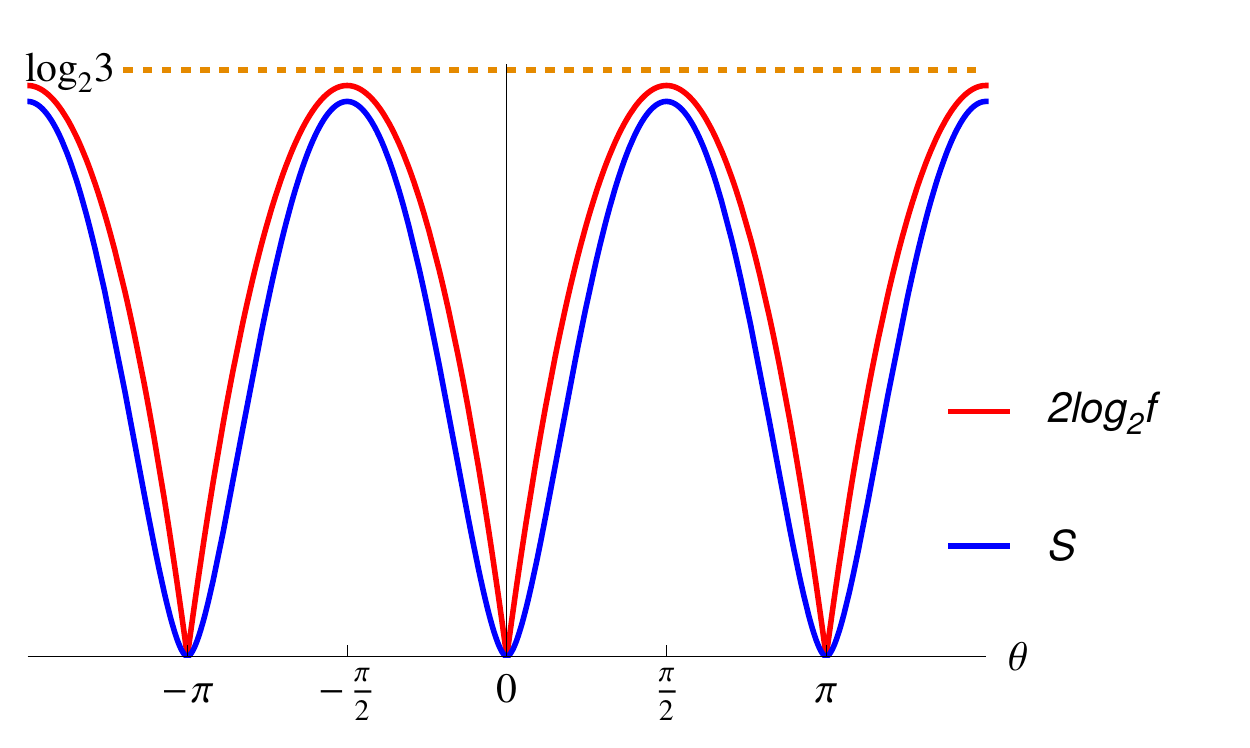}
\caption{}
\end{figure}

The explicit forms of the common minimum $\theta=\pm\pi$ and maximum $\theta=\pm\pi/2$ for both $L_1$-norm of $|d^j_{m'm}|$ and $S(\varrho_a)$ for $j=1, 3/2$ can be seen in Appendix E.

Next, the Fig.$5$ shows the derivatives of $S(\rho_a)$ and $f$ with respect to $\theta$. The zero points correspond to the extreme points of $S(\rho_a)$ and $f$. Except the two common zero points $\theta=\pi/2$ and $\pi$, other four zero points of $S(\rho_a)$ do not coincide with the zero points of $f$.
\begin{figure}[!htb]
\centering
\includegraphics[width=0.69\columnwidth,clip]{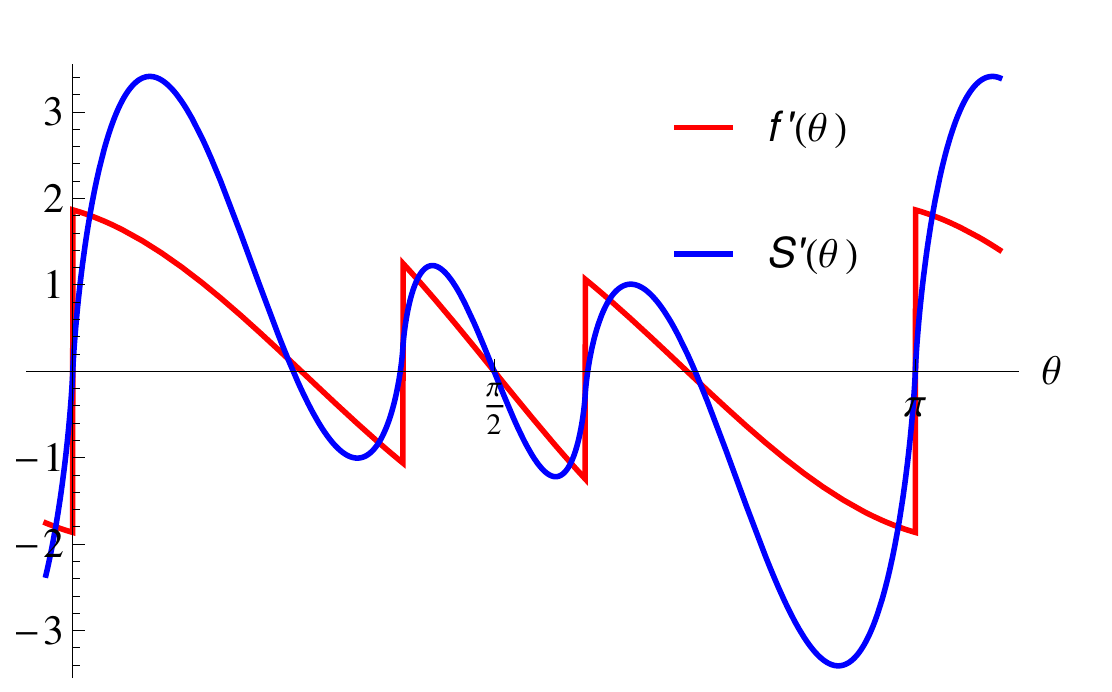}
\caption{}
\end{figure}

In general, we can prove that $S(\rho)$ and $f$ always share the same common extreme points $\theta=\pi/2$ and $\pi$ in the period $[0,\pi]$ for arbitrary $j$ and $m$ (m=0 is excluded). Firstly, let's consider the case of $\theta=\pi$. Because $d^j_{m'm}(\pi)=(-1)^{j+m'}\delta_{m'\,-m}$,
we have $d^j_{-m\,m}(\pi)=(-1)^{j-m},\ d^j_{m'm}=0\ (m'\neq-m)$ for arbitrary $m$, and it exactly denotes a separable state. Therefore $S(\rho_a)$ and $f$ both take the minimum value at $\theta=\pi$, and
\begin{equation}
    \left.S(\rho_a)\right|_{\theta=\pi}=2\log_2f(\pi)=0.
\end{equation}

When $\theta=\pi/2$, we have \cite{21}
\begin{equation}\label{eq2}
    d_{m'm}^{j}(\frac{\pi}{2})=\pm(-1)^{2m'}d_{-m'm}^{J}(\frac{\pi}{2}),
\end{equation}
and
\begin{equation}
   \left.\frac{d}{d\theta}d_{m'm}^{j}\right|_{\theta=\frac{\pi}{2}}=\mp(-1)^{2m'}\left.\frac{d}{d\theta}d_{-m'm}^{j\ \prime}\right|_{\theta=\frac{\pi}{2}}.
\end{equation}
In Ref.\cite{1}, it had proved that the $L_1$-norm of D-functions reaches the extreme value at $\theta=\pi/2$. We just need to prove $S(\rho_a)$ also to have extreme at $\theta=\pi/2$. Considering
\begin{equation}
\eqalign{
\frac{d}{d\theta}S(\rho_a)|_{\theta=\frac{\pi}{2}}&= -2\sum_{m'=-j}^{j}d_{m'm}^{j}\frac{d}{d\theta}d_{m'm}^{j}
\textrm{log}_2\left(|d_{m'm}^{j}|^2\right) \\
&=\delta_{0,(2j~\textrm{mod}~2)} d_{0m}^{j} \frac{d}{d\theta}\textrm{log}_2\left(|d_{m'm}^{j}|^2\right)}
\end{equation}
hence when $2j$ is odd, $\delta_{0,(2j\;\mathrm{mod}\;2)}=0$, thus $\left.\frac{d}{d\theta}S(\rho_a)\right|_{\theta=\frac{\pi}{2}}=0$. When $2j$ is even, according to Eq.~(\ref{eq2}), we have
\[
  d^j_{0m}\frac{d}{d\theta}d^j_{0m}=(\pm)(\mp)d^j_{0m}\frac{d}{d\theta}d^j_{0m}=-d^j_{0m}\frac{d}{d\theta}d^j_{0m},
\]
i.e. $d^j_{0m}\frac{d}{d\theta}d^j_{0m}=0$, therefore $\left.\frac{d}{d\theta}S(\rho_a)\right|_{\theta=\frac{\pi}{2}}=0$. To sum up, $\theta=\pi/2$ is the common extreme points of $S(\rho_a)$ and $f$.

\section{Conclusion Remarks}
Similar to the standard strategy of the construction of the 2-D topological basis for TQFT associated with Temperley-Lieb algebra \cite{1,2,3,8,9}, the extension has been made to construct the 3-D basis for BWM algebra.
The point is to introduce the basis (\ref{Eq1.9}) and (\ref{Eq1.10}), then all of the 3-D representations of $S_{12}$, $S_{23}$, $E_{12}$ and $E_{23}$ can be shown in terms of the graphic technique \cite{17} that yield the $3$-d representations of the corresponding braiding $A, B$ and $E$-operations.
For both $S^{\dag}=S$ and $S^{\dag}=S^{-1}$ we have given the explicit matrix forms of them in the Sec. \ref{Sec2} and Sec. \ref{Sec3}. In comparison with the case of TLA now the $E$-involved relations appear. Matching for the $S_{12}$, $S_{23}$, $E_{12}$ and $E_{23}$ in $9\times9$ matrix form the corresponding $A$, $B$, $E_A$ and $E_B$ have been found in $3\times3$ matrix forms that are nothing but the natural extension of $2\times2$ matrix forms for TLA.

The physical meaning of the TQFT associated with TLA has been well established \cite{8,9,10}. However, the physical meaning of the counterpart for BWMA deserves to be explored in the future. In ref \cite{16} a physical model was proposed, but here we present a different approach. As for the interaction model arisen from BWMA it deserves more discussions. The connection of (\ref{Eq5.14}) with BWMA has been shown for spin-$1$. How to solve the model for any N based on the 3-D representation in terms of the topological basis is an open problem.

More interesting point should be emphasized. We have shown that the $L_1$-norm of $D^j$-function and von Neumann entropy share the common maximum and minimum points of $\theta$, i.e $\pm\pi/2$ and $\pm\pi$, respectively, as least for any half-integer $j$. As was proved that $D^j(\theta,\varphi)$ are $(2j+1)-d$ solutions of YBE. The connection of the extreme points for both the $L_1$-norm of $D^j$ and von Neumann entropy explores the deep meaning of $L_1$-norm in Quantum Mechanics. It also happened in other model \cite{26}. It describes quantum entanglement in the quantum information. The reason is that braiding operation is very natural to describe entanglement for two particles. Because the entanglement only exists in the interaction between "particles" A and B themselves. An intuitive explanation of $(N\times N)$-dimensional braiding can easily be made in terms of $D(\theta,\varphi)$-function. Suppose two parallel lines along $Z$ axis (time pointed) represent A and B at the site $1$ and $2$, and form $X-Z$ plane. In the spherical coordinates an entanglement between A and B occurs by over crossing the two lines with angle $\theta$. The (\ref{Eq63}) is to fix $\varphi$ for a given $\theta$. Different $\theta$ also gives rise the rotation of $X-Z$ plane about $Z$ axis by $\varphi$. The $\theta=\pi/2$ means that the two lines are perpendicular to each other locally at different $\varphi$. Whereas $\theta=\pi$ means parallel to each other. The former corresponds with the maximum entangled state and the later with decomposable one, i.e. disentangled. Since YBE is the factorization condition of multi-body S-matrix to 2-body scatterings. We look over the facts such as factorization of S-matrix, topological basis, D-function, $L_1$-norm and von Neumann entropy. Now all of them have been connected within the frame work of YBE for quantum information, especially for the entanglement.

We appreciate the interesting discussion with Prof. Z. H. Wang and Prof. J. Birman, and X. B. Peng. This work is in part supported by NSF of China with the Grant No.11075077 and 11275024. Additional support was provided by the Ministry of Science and Technology of China, No.2011AA120101.


\appendix
\section{Graphic explanation of BWA}
For the self-contain we list the graphic expressions for the later use. The BWA reads
  \begin{equation}
    S=\cross,\quad S^{-1}=\cros,\quad E=\Un \quad (\mathrm{operators})
  \end{equation}
  \begin{equation}
    S_{i}S_{i\pm1}S_{i}=S_{i\pm1}S_{i}S_{i\pm1}\label{Eq.A2}
  \end{equation}
  \begin{equation}
    E_{i}E_{i\pm1}E_{i}=E_{i}, \quad E_{i}^{2}=dE_{i}
  \end{equation}
  \begin{equation}
    E_{i}S_{i}=S_{i}E_{i}=\sigma E_{i}
  \end{equation}
  \begin{equation}
    S_i\equiv S_{i,i+1},\quad E_i\equiv E_{i,i+1}
  \end{equation}
  \begin{equation}
    S_{12}|e_{\mu}\rangle=\lambda_{\mu}|e_{\mu}\rangle \quad(\mu=1,2,3)
  \end{equation}
The (\ref{Eq.A2}) reads graphically as
  \begin{equation}
    \SSSSijk=\sssijk
  \end{equation}
and the $E$ takes the simple graph and satisfies T-L algebra
\begin{equation}
    E_{i}=\Unii,\quad \complexijk=d\,\Unll,\quad d=\myloop,\quad (\mathrm{T-L})
  \end{equation}
All the other relations can then be expressed in terms of the graphs. Say,
  \begin{equation}
    E_{i}E_{i+1}E_{i}=E_{i}, \quad E_{i}^{2}=dE_{i}
  \end{equation}
  \begin{equation}
    S_iE_i=E_iS_i=\sigma E_i
  \end{equation}
  \begin{equation}
    \SicrossEi=\EiSicross=\sigma\Un
  \end{equation}
  \begin{equation}
    \Ncross=\sigma\Ncup,\qquad \Ucros=\sigma\Ucup,\qquad \Ncros=\sigma^{-1}\Ncup
  \end{equation}
  \begin{equation}
    E_iS_{i+1}S_i=S_{i+1}S_iE_{i+1}
  \end{equation}
  \begin{equation}
    \EiSiSj=\SiSjEj=\EiEj
  \end{equation}
  \begin{equation}
    S_{i+1}E_iS_{i+1}=S_i^{-1}E_{i+1}S_i^{-1}
  \end{equation}
  \begin{equation}
    \EiSjEj=\SihSjhEj\Rightarrow\UcrossNn=\tcrossH
  \end{equation}
  \begin{equation}
    E_{i+1}E_iE_{i+1}=S_i^{-1}E_{i+1}
  \end{equation}
  \begin{equation}
    \EiSjEjb=\SihEjb\Rightarrow\UcrossNn=\lUn
  \end{equation}
  \begin{equation}
    \EiEjSj=\SihEj
  \end{equation}
  \begin{equation}
    E_{i+1}E_iS_{i+1}=E_{i+1}S_i^{-1}
  \end{equation}
  The dependent relations $E_iS_{i\pm1}E_i=\sigma^{-1}E_iE_{i\pm1}E_i=\sigma^{-1}E_i$ can also easily be expressed in terms of the similar graphs.

It can be checked that all of $E_{A}$,$E_{B}$,$A$ and $B$ obey BWM algebra in 3-D representation:
  \begin{equation}
    ABA=BAB
  \end{equation}
  \begin{equation}
    \begin{array}{rcl}
    E_{A}E_{B}E_{A}&=&E_{A}\\
    E_{B}E_{A}E_{B}&=&E_{B}\\
    \end{array}
  \end{equation}
  \begin{equation}
    E_{A}^2=dE_{A},\quad E_{B}^2=dE_{B}
  \end{equation}
  \begin{equation}
    \begin{array}{rcl}
    A-A^{-1}&=&W(I-E_{A})\\
    B-B^{-1}&=&W(I-E_{B})\\
    \end{array}
  \end{equation}
  \begin{equation}
    \begin{array}{ccccc}
    AE_{A}&=&E_{A}A&=&\sigma E_{A}\\
    BE_{B}&=&E_{B}B&=&\sigma E_{B}\\
    \end{array}
  \end{equation}
  and other dependent relations:
  \begin{equation}
  \eqalign{
    E_{A}BA=BAE_{B}=E_{A}E_{B}\\
    E_{B}AB=ABE_{B}=E_{B}E_{A}\\
  }
  \end{equation}
  \begin{equation}
  \eqalign{
    AE_{B}A=B^{-1}E_{A}B^{-1},\quad BE_{A}B=A^{-1}E_{B}A^{-1}\\
    AE_{B}E_{A}=B^{-1}E_{A},\quad BE_{A}E_{B}=A^{-1}E_{B}\\
    E_{A}E_{B}A=E_{A}B^{-1},\quad E_{B}E_{A}B=E_{B}A^{-1}\\
  }
  \end{equation}

\section{Proof of (\ref{Eq1.12})}
The basis $|e_{\mu}\rangle$ is defined by (\ref{Eq1.7}), so
\begin{equation}
  S_{12}E_{12}|e_3\rangle=\sigma E_{12}|e_3\rangle=d \cdot S_{12}|e_3\rangle \nonumber
\end{equation}
i. e.
\begin{equation}
  S_{12}|e_3\rangle=\sigma|e_3\rangle\quad (\sigma=\lambda_3)
\end{equation}
For $i=1,2$ (no sum over repeating indices)
\begin{equation} \fl
\eqalign{
  S_{12}|e_i\rangle&=f_i\left\{\Uu+\alpha_i [\crosUu+w\Uu-w\tsep+\sigma\beta_i\tsep]\right\} \nonumber\\
  &=\alpha_{i}f_{i}\left\{\tcross+\alpha_i^{-1}(1+\alpha_{i}w)\Uu+\alpha_i^{-1}(\sigma\beta_i-\alpha_{i}w)\tsep\right\}
}
\end{equation}
In order that $|e_i\rangle$ are eigenstates for $S_{12}$ it should hold
\begin{eqnarray}
  \alpha_{i}^{-1}(1+\alpha_{i}w)=\alpha_i \nonumber\\
  \alpha_{i}^{-1}(\sigma\beta_i-\alpha_i w)=\beta_i. \nonumber
\end{eqnarray}
We then have
\begin{equation}
  \alpha_1+\beta_1 d=d_2 +\beta_2 d= -\sigma^{-1} \nonumber
\end{equation}
i. e.
\begin{equation}
 \beta_i =-d^{-1}(\sigma^{-1}+\lambda_i) \label{Eq.B2}
\end{equation}
Because of $\Ncross=\sigma\Ncup$ ($\sigma=\lambda_3$), $\Ncros=\sigma^{-1}\Ncup$, it holds ($i=1,2$) \nonumber
\begin{eqnarray}
  \langle e_3|e_i\rangle&=&d^{-1}f_i\left(\CcrossC+\alpha_i \myloop +\beta_i\tsquare\right) \qquad (\myloop=d) \nonumber \\
  &=&d^{-1}f_i\left(\sigma^{-1}d+\alpha_i d+\beta_i d^2\right)=0 \nonumber
\end{eqnarray}
where $d=1+\frac{\sigma^{-1}-\sigma}{w}$, $\alpha_i=\lambda_i$ and (\ref{Eq.B2}) have been used. Substituting (\ref{Eq.B2}) into (\ref{Eq1.10}) the basis $|e_{\mu}\rangle$ take the forms
\begin{eqnarray}
  |e_3\rangle&=&d^{-1}\tsep\\
  |e_i\rangle&=&f_i\left(\Utcross+\lambda_i \Uu+d^{-1}(\sigma^{-1}+\lambda_i)\tsep\right)
\end{eqnarray}
The relations for BWA in the Appendix A have been used.

To verify $\langle e_i|e_j\rangle=\delta_{ij}$ we have to distinguish two types of the braiding matrices from each other for $S^{\dag}=S$ (Hermitian) and $S^{\dag}=S^{-1}$ (unitary), respectively.

(a) For $S^{\dag}=S$ there are
\begin{eqnarray}
  \langle\tcross|\Utcross\rangle=d[(\sigma^{-1}-\sigma)W+d]
  \end{eqnarray}
\begin{eqnarray}
  \langle\tcross|\Uu\rangle=\sigma d \qquad\
\end{eqnarray}
After calculation we find
\begin{eqnarray}
  \langle e_1|e_2\rangle&=&d[(\sigma^{-1}-\sigma)W+d]+W\sigma d+(\beta_1 +\beta_2)\sigma^{-1}d\\
  &&+(\lambda_1 \beta_2-\beta_1 \lambda_2)d+(\lambda_1 \lambda_2+\beta_1 \beta_2)d^2\\
  &=&d\{\sigma^{-1}(W+\beta_1)+\lambda_2 \beta_1\}=0
\end{eqnarray}
The $\langle e_1|e_1\rangle=\langle e_2|e_2\rangle=1$ leads to (\ref{Eq1.13}).

(b) For $S^{\dag}=S^{-1}$ we have ($i=1,2$)
\begin{equation}
\eqalign{
  \langle e_i|=f_{i}^{*}\left\{\ntcros+\alpha_{i}^{*}\Nn+\beta_{i}^{*}\tncup\right\},\\
   \crossUu=S_{12}\Uu,  \quad \mathrm{and}\quad \langle e_1|e_2\rangle=\tloop=d^2
}
\end{equation}
Now $\alpha_i=\lambda_i$, $\beta_i$ are complex and $\lambda_i^{*}=\lambda_i^{-1}$ that gives
  \begin{equation} \fl
  \eqalign{
  \langle e_i|e_j\rangle&=f_i^*f_j\left\{d^2+\alpha_j\sigma^{-1}d+\beta_j\sigma d
    +\alpha_i^*\left(\sigma d+\alpha_j d^2+\beta_j d\right)
    +\beta_i^*\left(\sigma^{-1}d+\alpha_j d+\beta_j d^2\right)\right\}\\
    &=f_i^*f_j d\left\{d+\alpha_j \sigma^{-1}+\beta_j (\alpha_i^{*} +\sigma)+\alpha_j \alpha_i^{*} d+\alpha_i^{*}\sigma\right\}
  }
   \end{equation}
   \begin{eqnarray}
     \langle e_1|e_2\rangle =f_1^{*} f_2 d\left\{-\lambda_2W d+\lambda_2W+\lambda_2(\sigma^{-1}-\sigma)\right\}=0\\
     f_i=\left[(d-1)(\lambda_i+\lambda_i^{-1})(\sigma+\lambda_i d+\lambda_i^{-1})\right]^{-1/2}
   \end{eqnarray}

\section{Derivation of $3$-D matrix forms of braiding matrices}
For $S^{\dag}=S$ ($i=1,2$)
\begin{eqnarray}
  E_{12}|e_i\rangle&=&f_i \Unot \left\{\Utcross+\lambda_i \Uu+\beta_i \tsep\right\}\\
  &=& d f_i (\sigma^{-1}+\beta_i d+\lambda_i) |e_3\rangle
\end{eqnarray}
\begin{equation}
  \langle e_3|E_{12}|e_3\rangle=d
\end{equation}
\begin{equation} \fl
  E_{23}|e_i\rangle=f_i \left\{\sigma+\lambda_i d-d^{-1}(\sigma^{-1}+\lambda_i)\right\}(\lambda_1-\lambda_2)^{-1}\left\{f_1^{-1}|e_1\rangle-f_2^{-1}|e_2\rangle+|e_3\rangle\right\}
\end{equation}
\begin{equation}
  E_{23}|e_3\rangle=d^{-1} \left\{(\lambda_1 -\lambda_2^{-1})^{-1}(f_1^{-1}|e_1\rangle-f_2^{-1}|e_2\rangle)+|e_3\rangle\right\}
\end{equation}
\begin{equation}
\eqalign{
  S_{23}|e_i\rangle=&f_i\left\{(\lambda_1-\lambda_2^{-1})^{-1}f_1^{-1}[(\lambda_i-W)\lambda_3-(W+\beta_i)\lambda_2]|e_1\rangle\right.\\
  &+(\lambda_1-\lambda_2)^{-1}f_2^{-1}[(W+\beta_i)\lambda_1-(\lambda_i-W)\lambda_3]|e_2\rangle\\
  &\left.+[(W+\beta_i)\lambda_3^{-1}+(\lambda_i-W)\lambda_3+d]|e_3\rangle\right\}
}
\end{equation}
\begin{equation}
  S_{23}|e_3\rangle=d^{-1}\left\{(\lambda_2-\lambda_1)^{-1}[\lambda_2 f_1^{-1}|e_1\rangle-\lambda_1 f_2^{-1}|e_2\rangle]+\lambda_3^{-1}|e_3\rangle\right\}
\end{equation}
Obviously the base $|e_1\rangle$, $|e_2\rangle$ and $|e_3\rangle$ form a closed set for the operations $S$ and $E$. For $S^{\dag}=S^{-1}$ ($i=1,2$)
  \begin{equation}\fl
    E_{23}|e_i\rangle=f_i(\sigma+\beta_i+\lambda_i d)(\lambda_1+\lambda_1^{-1})^{-1}
    \left\{f_1^{-1}|e_1\rangle-f_2^{-1}|e_2\rangle+(\lambda_1+\lambda_1^{-1})|e_3\rangle\right\}
  \end{equation}
  \begin{equation} \fl
    E_{23}|e_3\rangle=d^{-1}(\lambda_1+\lambda_1^{-1})^{-1}
    \left\{f_1^{-1}|e_1\rangle-f_2^{-1}|e_2\rangle+(\lambda_1+\lambda_1^{-1})|e_3\rangle\right\}\\
  \end{equation}

\section{Acting braiding operations on the topological basis}
From (\ref{Eq1.6})-(\ref{Eq1.13}) it follows that $d=3$ and for $m=-3$ as well as $q=1$ (for $S=S^\dagger$) we get:
 \begin{eqnarray}
  f_{1}=\frac{1}{2\sqrt{3}}, f_{2}=\frac{1}{2\sqrt{3}}, \alpha_{1}=1, \alpha_{2}=-1, \beta_{2}=0,
 \end{eqnarray}
 and
\begin{eqnarray}
   \beta_{1}=\frac{q-q^{-1}}{q^{-3}-q}|_{q\rightarrow1}=-\frac{2}{3}
\end{eqnarray}
that leads to:
\begin{equation} \fl
 T_{12}|e_{1}\rangle=T_{34}|e_{1}\rangle=|e_{1}\rangle,\,
 T_{12}|e_{2}\rangle=T_{34}|e_{2}\rangle=-|e_{2}\rangle,\,
 T_{12}|e_{3}\rangle=T_{34}|e_{3}\rangle=|e_{3}\rangle
\end{equation}
\begin{equation} \fl
 M_{12}|e_{1}\rangle=M_{34}|e_{1}\rangle=0,\,
 M_{12}|e_{2}\rangle=M_{34}|e_{2}\rangle=0,\,
 M_{12}|e_{3}\rangle=M_{34}|e_{3}\rangle=3|e_{3}\rangle
\end{equation}
\begin{eqnarray}
 T_{23}|e_{1}\rangle=T_{34}|e_{1}\rangle=\frac{1}{6}(|e_{1}\rangle-\sqrt{15}|e_{2}\rangle+2\sqrt{5}|e_{3}\rangle)
 \end{eqnarray}
\begin{eqnarray}
 T_{23}|e_{2}\rangle=T_{34}|e_{2}\rangle=\frac{1}{6}(-\sqrt{15}|e_{1}\rangle+3|e_{2}\rangle+2\sqrt{3}|e_{3}\rangle)
 \end{eqnarray}
\begin{eqnarray}
 T_{23}|e_{3}\rangle=T_{34}|e_{3}\rangle=\frac{1}{6}(2\sqrt{5}|e_{1}\rangle+2\sqrt{3}|e_{2}\rangle+2|e_{3}\rangle)
 \end{eqnarray}
\begin{eqnarray}
 M_{23}|e_{1}\rangle=M_{14}|e_{1}\rangle=\frac{1}{3}(5|e_{1}\rangle-\sqrt{15}|e_{2}\rangle+\sqrt{5}|e_{3}\rangle)
 \end{eqnarray}
\begin{eqnarray}
 M_{23}|e_{2}\rangle=M_{14}|e_{2}\rangle=\frac{1}{3}(-\sqrt{15}|e_{1}\rangle+3|e_{2}\rangle-\sqrt{3}|e_{3}\rangle)
 \end{eqnarray}
\begin{eqnarray}
 M_{23}|e_{3}\rangle=M_{14}|e_{3}\rangle=\frac{1}{3}(\sqrt{5}|e_{1}\rangle-\sqrt{3}|e_{2}\rangle+|e_{3}\rangle)
 \end{eqnarray}

\section{More Examples for the coincidence of extremes of $D^j$ and von Neumann entropy}
Example 3: {$j=1,\ m=0$}
When $\theta=\pm\pi$, the state is separable, and $S(\rho_a)=2\log_2f$ arrive at their minimum value simultaneously. The point $\theta=\pm\pi/2$ is both the local minimum point of $S(\rho_a)$ and $f$, however, $S(\rho_a)<2\log_2f$. In addition, $S(\rho_a)$ and $f$ shares another two common local maximum points in the period $(0,\pi]$, and the two common maximum points both correspond to the maximally entangled state, therefore $S(\rho_a)=2\log_2f=2\log_23$ at the two points. See Fig.E1 and Fig.E2.
\begin{figure}[!htb]
\centering
\includegraphics[width=0.5\columnwidth,clip]{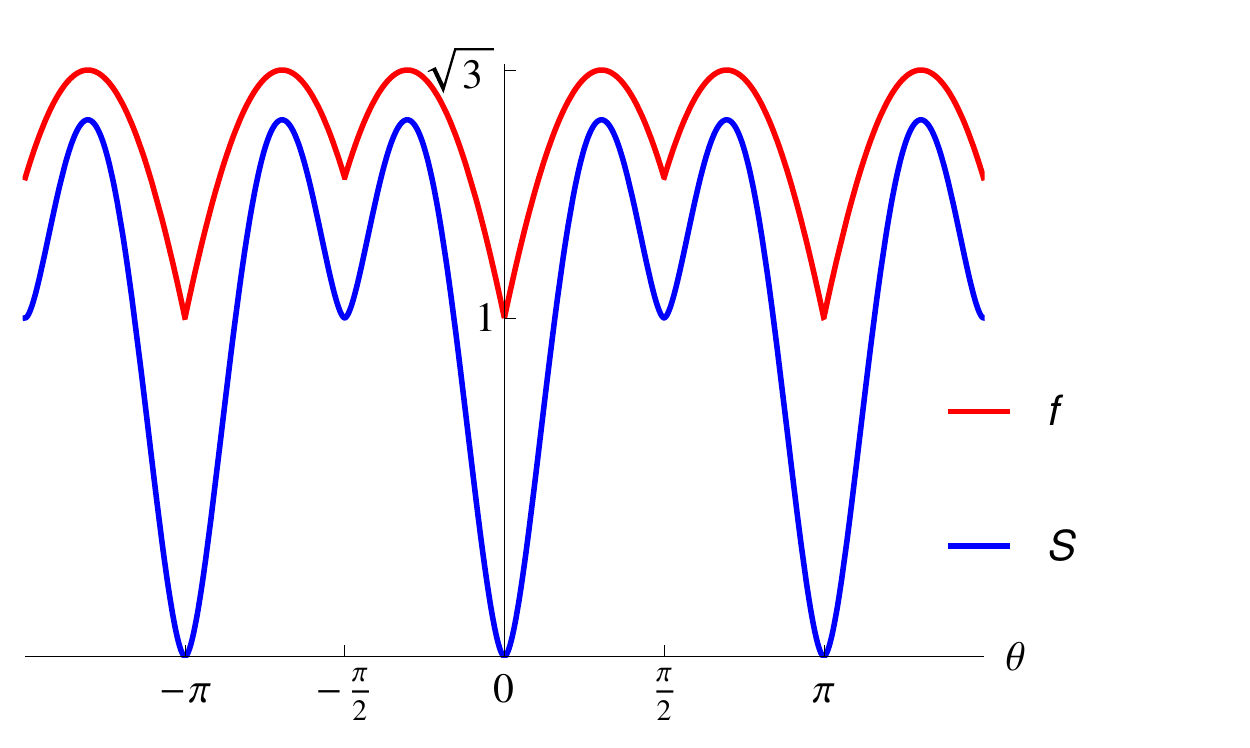}
\caption{}
\end{figure}

\begin{figure}[!htb]
\centering
\includegraphics[width=0.5\columnwidth,clip]{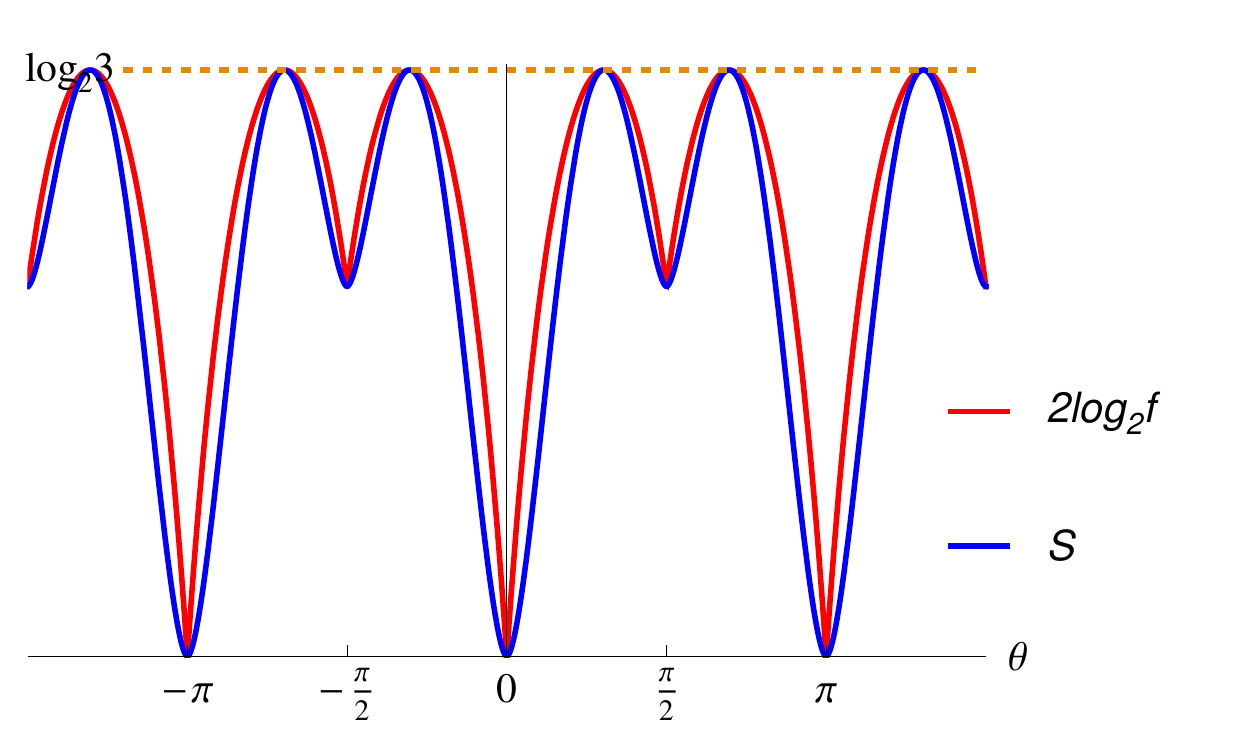}
\caption{}
\end{figure}

Example 4: {$j=3/2,\ m=\pm3/2$}


When $\theta=\pm\pi$, the state is a separable sate and $S(\rho_a)=2\log_2f$ reach their minimum value simultaneously. When $\theta=\pm\pi/2$, $S(\rho_a)$ and $f$ reach their  maximum values simultaneously, however both the values are not the same because $S(\rho_a)<2\log_2f<\log_24=2$. See Fig.E3 and Fig.E4.
\begin{figure}[!htb]
\centering
\includegraphics[width=0.5\columnwidth,clip]{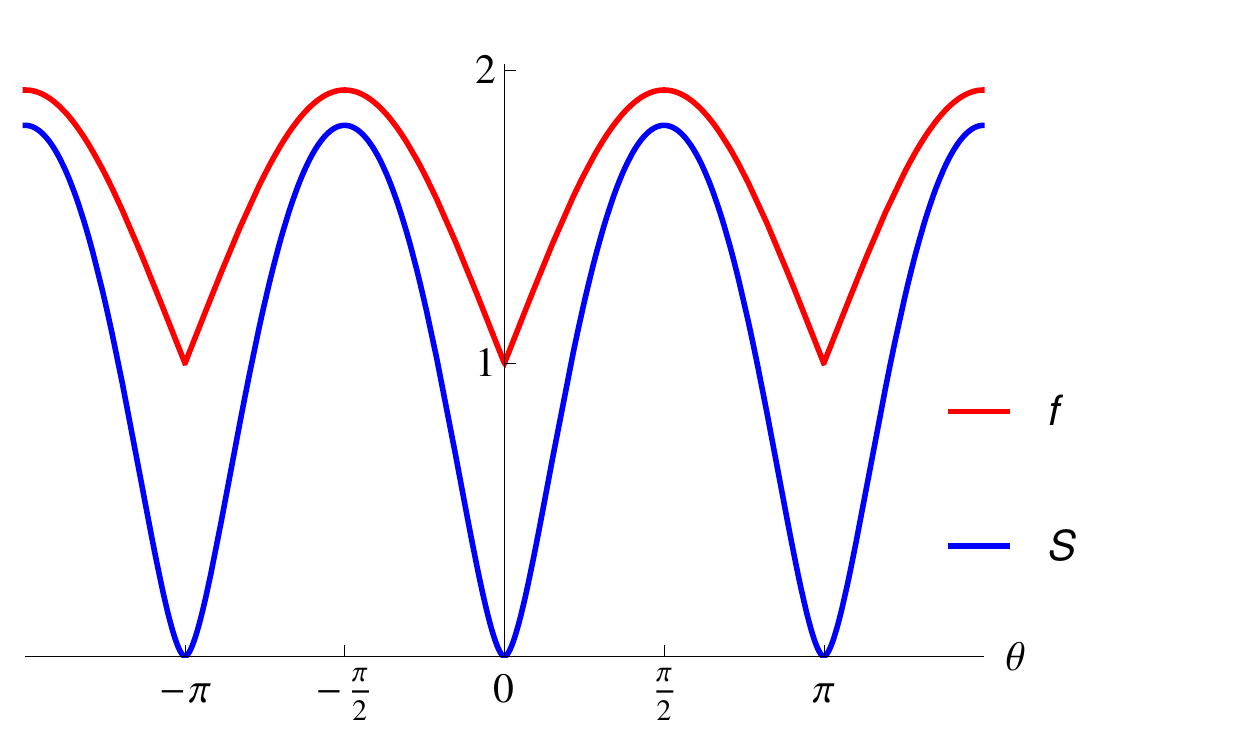}
\caption{}
\end{figure}

\begin{figure}
\centering
\includegraphics[width=0.5\columnwidth,clip]{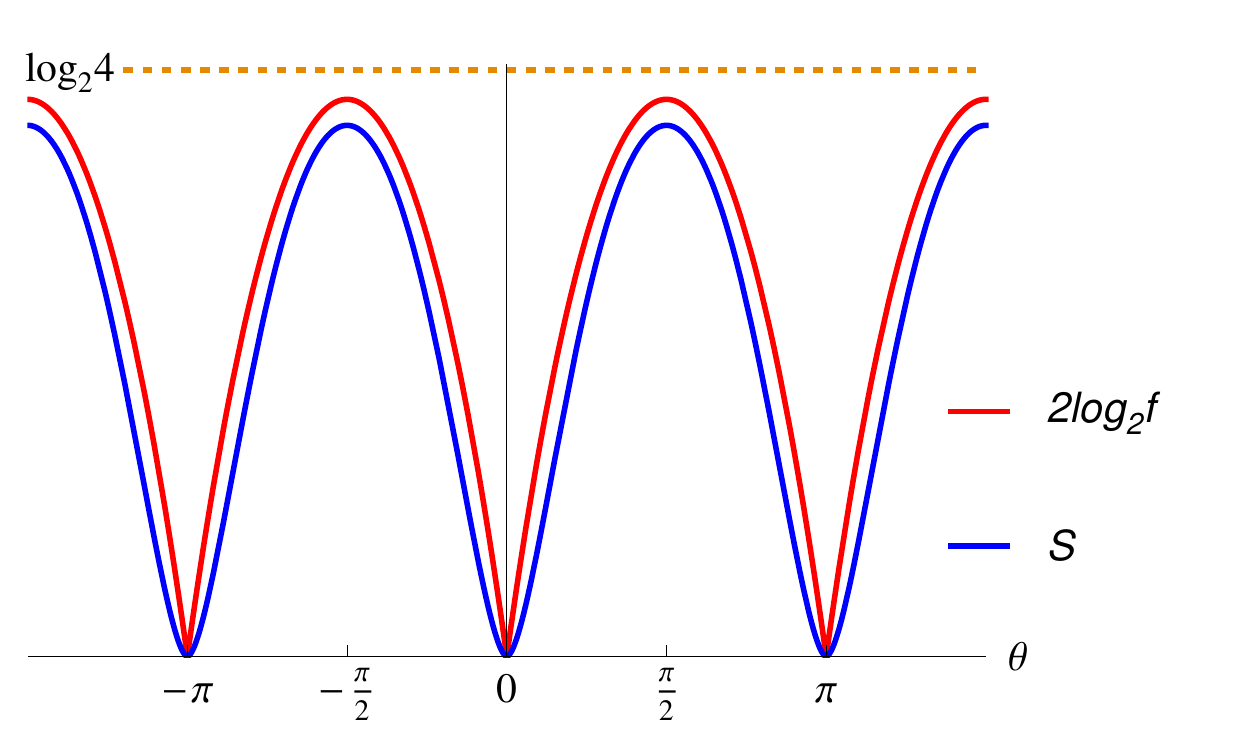}
\caption{}
\end{figure}

Example 5: {$j=3/2,\ m=\pm1/2$}.
When $\theta=\pm\pi$, the state is separable, and $S(\rho_a)=2\log_2f$ arrives at the minimum value. The point $\theta=\pm\pi/2$ is both the local minimum point of $S(\rho_a)$ and $f$, with $S(\rho_a)<2\log_2f$. It is worth noting that $S(\rho_a)$ and $\ell_1$ norm $f$ both have other four local extreme points, in the period $(0, \pi)$. It is shown in Fig. $E5$ and Fig.$E6$.
\begin{figure}[!htb]
\centering
\includegraphics[width=0.5\columnwidth,clip]{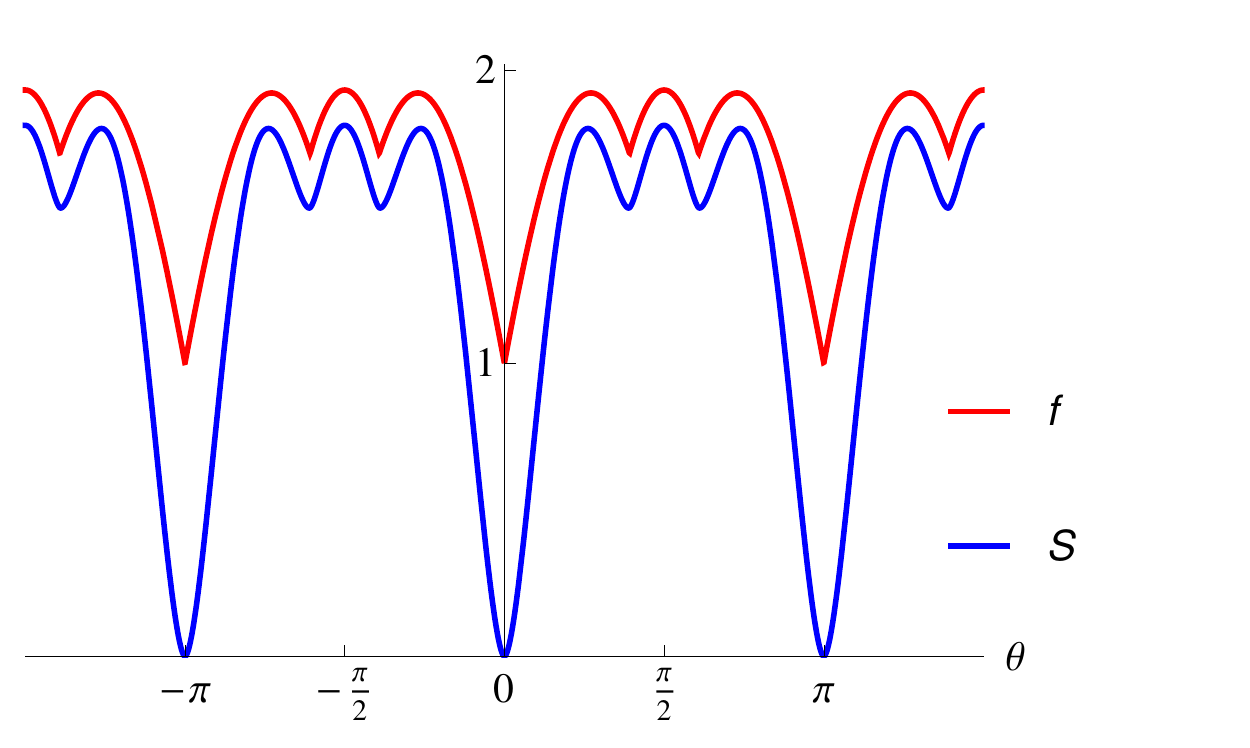}
\caption{}
\end{figure}

\begin{figure}[!htb]
\centering
\includegraphics[width=0.49\columnwidth,clip]{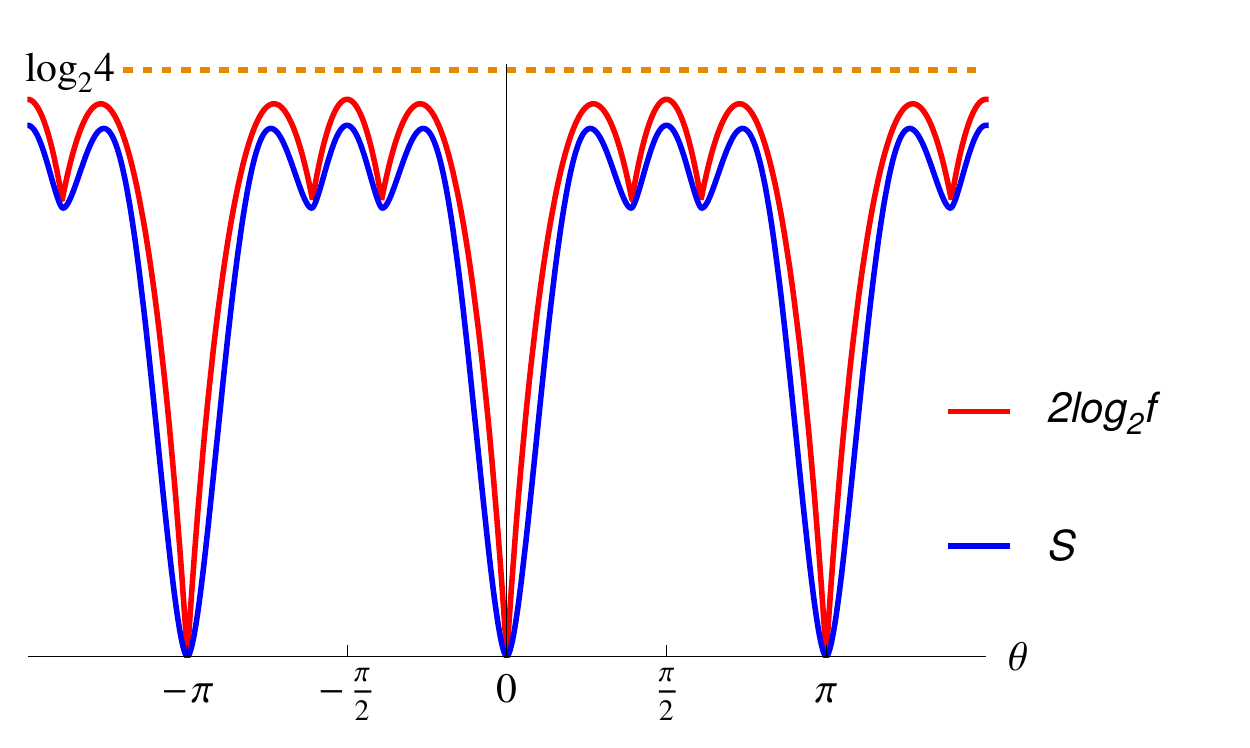}
\caption{}
\end{figure}

\end{document}